# Surface modulation of metal-organic frameworks for on-demand photochromism in the solid state

Samraj Mollick,[a] Yang Zhang,[a] Waqas Kamal,[b] Michele Tricarico,[a] Annika F. Möslein,[a] Vishal Kachwal,[a] Nader Amin,[c] Alfonso A. Castrejón-Pita,[b] Stephen M. Morris,[b] and Jin-Chong Tan[a,*]

[a]Multifunctional Materials & Composites (MMC) Laboratory, Department of Engineering Science, University of Oxford, Parks Road, Oxford OX1, United Kingdom.

[b]Department of Engineering Science, University of Oxford, Parks Road, Oxford, OX1 3PJ, United Kingdom.

[c]Department of Chemistry, University of Oxford, Mansfield Road, Oxford OX1 3TA, United Kingdom.

*Corresponding author's e-mail: jin-chong.tan@eng.ox.ac.uk

## Abstract

Organic photoswitchable molecules have struggled in solid state form to fulfill their remarkable potential, in terms of photoswitching performance and long-term stability when compared to their inorganic counterparts. We report the concept of non-electron deficient host's surface with optimal porosity and hydrophobicity, as *a priori* strategy to design photoefficient organic solid-state photochromic materials with outstanding mechanical robustness. When exposed to a light stimulus including natural sunlight, the photoswitchable nanocomposite changes color promptly and reversibly, in a matter of seconds along with excellent photo-fatigue resistance, which are on a par with inorganic photochromes. Exemplars of commercially viable prototypes that are optically clear, comprising smart windows, complex photochromic sculptures, and self-erasing rewritable devices, were engineered by direct blending with resilient polymers; particularly, the use of high-stiffness polymer (>2 GPa) is no longer an insurmountable challenge. Finally, photochromic films with anticounterfeiting features could be manufactured through precision inkjet printing of nanocrystals.

## Introduction

Inspired by nature's large repertoire of biological processes, scientists have developed stimuli-responsive smart systems constructed from unique photochromic switches that can reversibly change color; this field has recently attracted considerable scientific interest (*1-3*). Conversion of such photochromic molecules into continuum materials will enable translation of stimuli-controlled nanoscopic changes to the macroscopic scale (*4-6*). In 1964, Corning (USA) marketed the first photochromic ophthalmic lenses based on silver halide crystals (*7*), and ever since, photochromism has been exploited to evolve into a global multimillion-dollar industry (*8*). Nonetheless, organic photochromic materials have appeared far less in commercial applications and industrial sectors despite their prevalence in scientific research publications. For example, spiropyran (SP) molecules are amongst the most versatile organic photochromic materials, whose geometry transforms from a closed-ring (nonpolar) form to an opened-ring (highly polar merocyanine) form upon exposure to ultraviolet (UV) light, while the reverse process can be triggered by visible light as well as by a variety of other stimuli (*9, 10*). This intriguing reversible visual color-switching ability provides numerous opportunities for a variety of different fields, such as optoelectronic devices, data storage, fiber optics, bioadhesive mediators, optical delay generators, and much more (*11*). However, the low efficiency and poor long-term stability of such multifunctional SP molecules in the solid state have hindered their commercial and industrial proliferations (*8, 12*). Recently, encapsulation into a confined microenvironment of a porous solid state material has been recognized as an effective strategy to markedly improve the chemical and physical properties of organic photoswitchable molecules (*13, 14*). Nevertheless, the photoisomerization process of such solid-state photoswitchable materials has specific challenges, and this significantly hampers their capability due to geometrical constraints imposed by the solid-state environment (*15, 16*). To this end, the primary endeavor in the current research is to accomplish efficient bulk photoresponsivity integrated in a robust solid matrix of nanoporous materials, without compromising the remarkable molecular photoisomerization properties.

Metal-organic frameworks (MOFs) have recently become a considerable hot topic for photoresponsive applications, owing to their vastly tunable properties and porosity by design (*17-20*). Photoresponsive MOFs are primarily tailored for light modulated guest separation/storage and conductivity applications (*21, 22*), and increasingly for guest-induced luminescent sensing (*23*). Despite the fact that multiple attempts have been made to improve the overall performance efficiency and durability of photoswitchable SP molecules through encapsulation inside the MOF cavity, the majority have suffered from a lack of intense color changing features which are critical for real-world applications (*16, 24*). Additionally, none of them satisfies or is nowhere close to the inorganic commercial photochromic materials in terms of both performance and long-term stability. Hence, a rational design strategy for the selection of suitable MOF hosts is urgently needed to meet the commercial/industrial standards set for photochromic materials, but this has remained elusive.

In this study, we have judiciously engineered and harnessed the microenvironment of a series of MOFs, where non electron deficient surface with intermediate hydrophobicity and optimal porosity offers efficient solid state photoswitchable materials. The newly designed photochromic fluorescent nanocomposites were systematically characterized by a wide range of spectroscopic techniques to gain insights into the materials in the bulk powder phase, as single crystals, membranes and thin films. We demonstrate that intense color switching, outstanding long-term stability, enhanced photo-fatigue resistance, production scalability, along with efficient solution-like photochromic response in the solid state are achievable using the proposed approach. Specifically, we show that the optimal porosity provides enough room for the geometrical transformation of the photochromic guest confined in the MOF pore, while intermediate hydrophobicity enables fast reversible switching and enhanced stability. Together, these criteria yield a highly efficient photoswitchable material, which not only markedly outperforms the existing reported photoresponsive materials, but also satisfies the stringent commercial standards (*8*). Finally, to mimic practical applications, alongside an assortment of prototype photochromic sculptures, self-erasing rewritable surfaces, bare photochromic materials were inkjet printed at the nanometer scale on a wide range of technical substrates for fabricating thin-film photoswitchable devices.

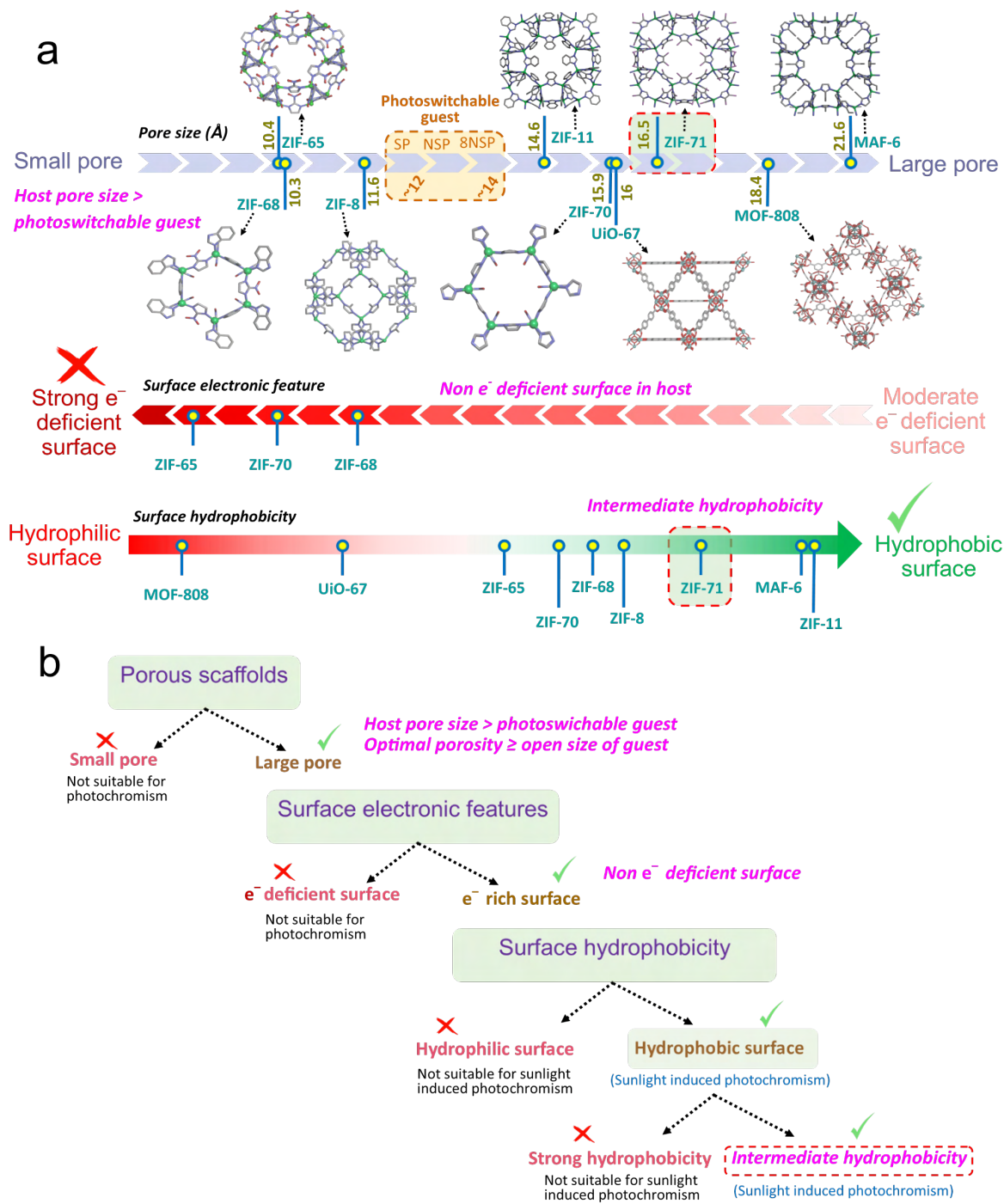

**Fig. 1. Schematic of a general design strategy for high-performance photochromic materials.** (a) MOF structures with different pore size and surface functionalization selected as a host matrix in this study for the encapsulation of organic photochromic guests, namely SP, NSP, and 8NSP. Here, porous frameworks with a cavity size larger than the spiropyran guest are classified as 'large pore', while frameworks with a pore cavity smaller than the spiropyran guest are termed as 'small pore'. (b) Selection rules for engineering a highly sensitive photochromic guest@MOF material for solid-state applications.

# Results and discussion

A series of efficient photochromic MOF nanocomposites were designed and fabricated through a facile, single-step *in situ* guest confinement protocol at room temperature (see Supplementary Information (SI) for details on the synthesis protocol). We have identified three photochromes: spiropyran (SP), nitro spiropyran (NSP), and 8-methoxy nitrospiropyran (8NSP) as promising spiropyran molecules for the organic photochromic guest. A variety of MOF structures with different topologies and surface features was screened as a host matrix, in accordance with the selection rules depicted in Figure 1, giving the best performing photoisomerization of the encapsulated guest (Figures S1-S8). In this work, UiO-67 and MOF-808 for hydrophilic pore surface; ZIF-65, ZIF-68 and ZIF-70 for nitro-functional group containing pore surface; ZIF-8, ZIF-71, ZIF-11, and MAF-6 for hydrophobic pore surface, were interrogated as a prototypal host matrix to construct photochromic guest@MOF composites (Figure 1a). Figure 2a shows the behavior of an efficient photochromic composite comprising the ZIF-71 host matrix, which has an ideal cage size and intermediate hydrophobicity to enable fast and reversible color switching of NSP upon UV irradiation or sunlight exposure (Figure S8); the latter requires a greater sensitivity photochromic material. A maximum loading of 30.4 wt.% of NSP confined in ZIF-71 has been achieved for the nanocomposite termed NSP@ZIF-71(20) (denoting 20 mM NSP guest used during synthesis of the composite), revealed by $^1$H nuclear magnetic resonance (NMR) spectroscopy (see Figure 2b, Figures S9-S11 and Table S1).

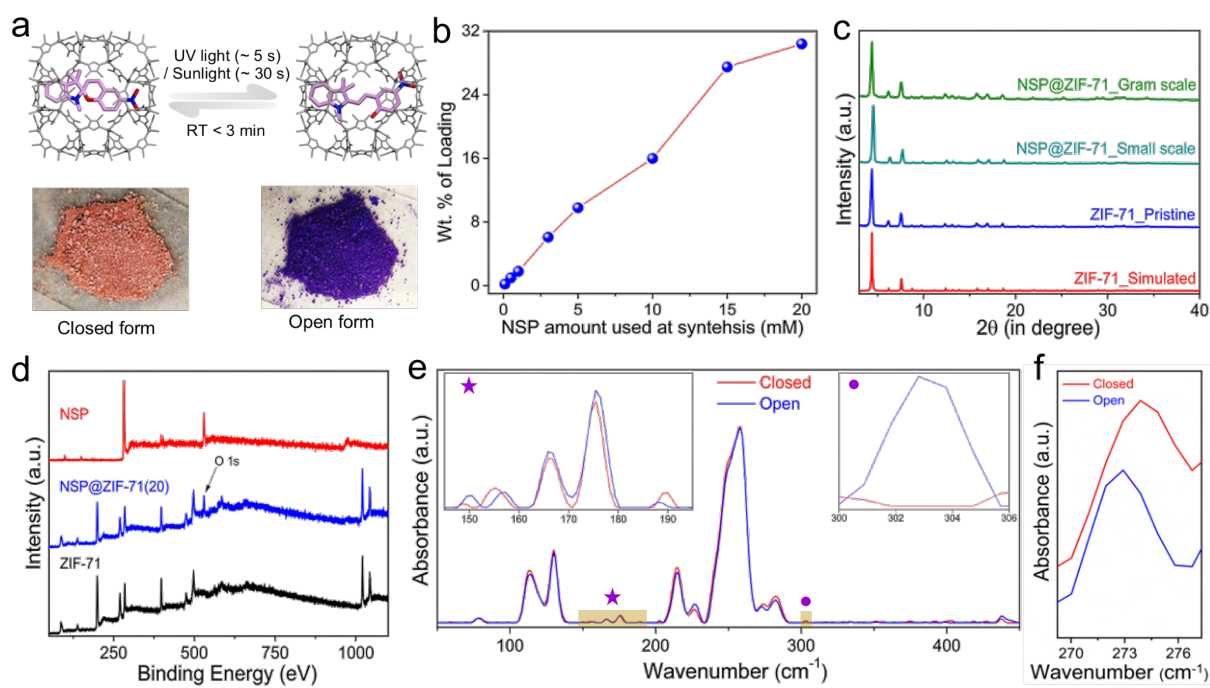

**Fig. 2.** Powder sample characterizations of the NSP@ZIF-71(20) nanocomposite. (a) Schematic diagrams and reversible color switching of the open ⇌ closed forms of the solid state photochromic material. (b) Different loading (wt.%) of NSP into the ZIF-71 host matrix. (c) PXRD patterns of ZIF-71 and composites, derived from the small- (~50 mg) and gram-scale (~4 g) production of NSP@ZIF-71(20). (d) XPS survey spectra of ZIF-71, NSP@ZIF-71(20) and NSP. (e) Synchrotron-radiation far-infrared (SR-FIR) vibrational spectra for the open and closed form of NSP@ZIF-71(20); insets show a magnified view of the highlighted regions. (f) An enlarged view of Figure 1e showing a peak shift at the strongest absorption band due to photochromic switching.

Notably, gram-scale production of NSP@ZIF-71(20) has been achieved in a facile single-step reaction protocol at room temperature: such scalability is crucial for commercial and industrial applications (Figure 2c). The XRD patterns of the composites matched the pristine MOF structure, indicating the formation of a highly crystalline phase. Raman scattering and Fourier transform infrared (FTIR) measurements were performed to determine the local microenvironment of the composites (Figures S12-S16). The existence of the characteristic Raman peak for NSP at 1577 cm$^{-1}$ in the composite confirms that NSP is embedded within the microenvironment of the ZIF-71 matrix (Figure S12). The characteristic FTIR vibrational bands corresponding to NSP appear at 748.6 cm$^{-1}$ and 1383.1 cm$^{-1}$, revealing encapsulation of NSP inside the ZIF-71 host matrix (Figures S13-S16). Further, to gain detailed surface chemical composition of the composite, X-ray photoelectron spectroscopy (XPS) measurements were performed. Figure 2d shows the appearance of a new O1s signal in the XPS spectra, which can be assigned to the NSP molecules in the composite material. The morphological features of the materials were examined by atomic force microscopy (AFM), see Figure S17. Depending upon the reaction conditions used, either faceted micron-sized crystals (*ca*. 1–2 μm) or small nanocrystals (*ca*. 50 nm) may be obtained. Systematic controlled experiments revealed that adequate porosity is essential for facilitating the photoisomerization process (*25*) of NSP guests confined in the ZIF-71 pores (Figures S18-S20).

To correlate the (bulk) powder phase photoisomerization process, high-resolution synchrotron far-infrared (FIR) vibrational spectra of nanocomposites were recorded (Figure 2e,f and Figures S21).The shifting of the terahertz (THz) collective modes of NSP@ZIF-71(20) below 200 cm$^{-1}$ (< 6 THz) upon ring opening of the NSP guests revealed the dynamic nature of the host matrix (inset Figure 2e) (*26*). Of note, a new characteristic vibrational band appears at 303.1 cm$^{-1}$ (~10 THz) due to the opened configuration of the NSP molecules, while the existing vibrational band at 273.9 cm$^{-1}$ (8.21 THz) of the ZIF-71 framework is red-shifted to 272.8 cm$^{-1}$ (8.18 THz) due to the strong interaction between the host framework and the polar opened form of the NSP guest (Figure 2f). The experimental determination of new vibrational bands at 202.5 cm$^{-1}$, 445 cm$^{-1}$, and 501 cm$^{-1}$ after UV illumination is further evidence of the opened-ring merocyanine form of the NSP guests (Figure S21). Near-field infrared nanospectroscopy (nanoFTIR) technique was employed as a local probe to interrogate the single-crystal photoisomerization process of NSP@ZIF-71(20) (Figure 3a, Figures S22-S25).

Here, nanoFTIR and scattering-type scanning near-field optical microscopy (s-SNOM) imaging provided us with the unique opportunity to chemically pinpoint the host-guest interaction and reversible photoisomerization process of the photochromic NSP molecules at the nanoscale, within the sub-micron crystal of the composite (*27*). The obtained s-SNOM optical phase images (Figure S22) revealed that NSP is homogeneously distributed throughout the ZIF-71 crystals, without agglomeration of the guests on the surface (*28*). Upon UV irradiation of NSP@ZIF-71(20), the nanoFTIR spectra show the appearance of three new vibrational bands at 1391 cm$^{-1}$, 1607 cm$^{-1}$ and 1672 cm$^{-1}$, while the intensity of the vibrational peaks at 1105 cm$^{-1}$, 1638 cm$^{-1}$ and 1662 cm$^{-1}$ decreased due to a transformation of the closed-form of NSP into an opened merocyanine form which is polar (Figures S23-S25). The vibrational band at 1130 cm$^{-1}$ splits into two distinct peaks at 1139 cm$^{-1}$ and 1118 cm$^{-1}$ upon UV exposure; an observation that is attributed to the collective bond vibration of the opened form which is attained by cleaving off the oxygen-containing six-membered ring of the closed form (inset Figure 3a). The characteristic vibrational peak of ZIF-71 at 1469 cm$^{-1}$ is red-shifted to 1466.5 cm$^{-1}$ upon UV irradiation, most likely due to either the open merocyanine form of NSP strongly interacting with the host framework or the host framework being slightly distorted due to the dynamic nature of the framework (Figure S25). These findings are supported by the terahertz vibrational response of the collective modes for the closed- versus opened-ring forms of NSP@ZIF-71(20) (Figure 2e).

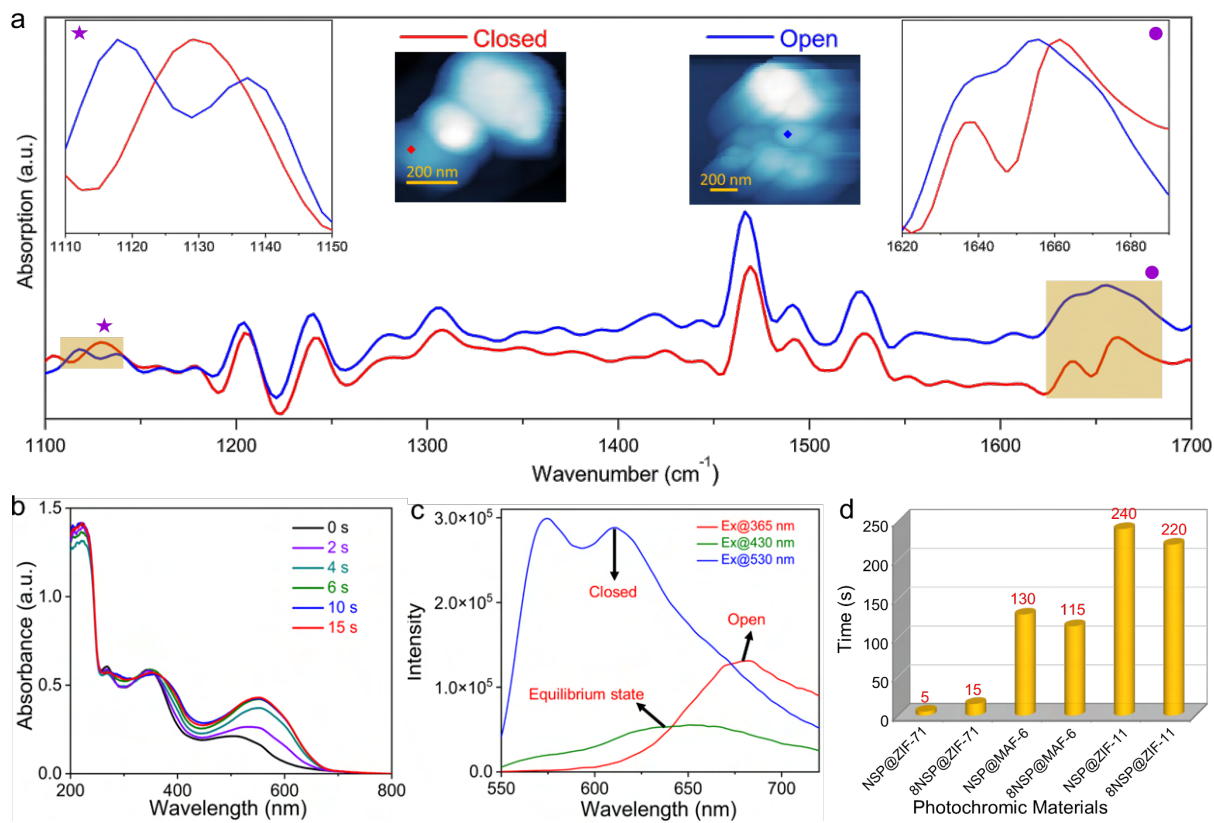

**Fig. 3.** Single crystal and optical characterizations of the photochromic materials. (a) Near-field nanoFTIR spectra of the closed and open forms of individual NSP@ZIF-71(20) nanocrystals at the designated locations (probe size < 20 nm) marked on the inset AFM images (red = before UV; blue = after UV irradiation). Inset graphs show the highlighted parts of the figure. (b) UV-Vis DRS of the NSP@ZIF-71(20) powder for different exposure times (0–15 s) subject to a 365-nm UV irradiation. (c) Emission spectra of the NSP@ZIF-71(20) powder at different excitation wavelengths. (d) A comparison of the photochromic transition times for various photochromic materials in powder form, each of which was synthesized using the same guest concentration of 20 mM.

UV-Vis diffused reflectance spectroscopy (DRS) revealed bathochromic shifts in the presence of 365 nm UV light or sunlight, and hypsochromic shifts occurring in the dark under ambient conditions (Figures S26-S27, Table S2) (14). The distinct visual color change was clearly noticed for the NSP@ZIF-71(20) powder within 5 s of UV illumination or after 30 s of sunlight exposure, before reverting back to its original color within ~2.5 min under ambient conditions (or within 5 s at 50 °C). This photoswitchable behavior indicates facile reversible photoisomerization of the NSP confined in the ZIF-71 cavity (Figure 3b). Using fluorescence spectroscopy, the open and closed forms of the NSP@ZIF-71(20) nanocomposite were registered upon excitation at 365 nm and 530 nm, respectively, while these two forms remained in an equilibrium state upon excitation at 430 nm (Figure 3c, Figures S28-S33 and Table S3). The absolute photoluminescence quantum yield (QY) of all the photochromic materials decreases upon UV irradiation (Tables S4-S7). The NSP@ZIF-71(20) nanocomposite powder exhibits the fastest switching time in its coloration and fading response, compared with other photochromic nanocomposites synthesized in this work (Figure 3d); these results demonstrate the general applicability of the selection rules outlined in Figure 1.

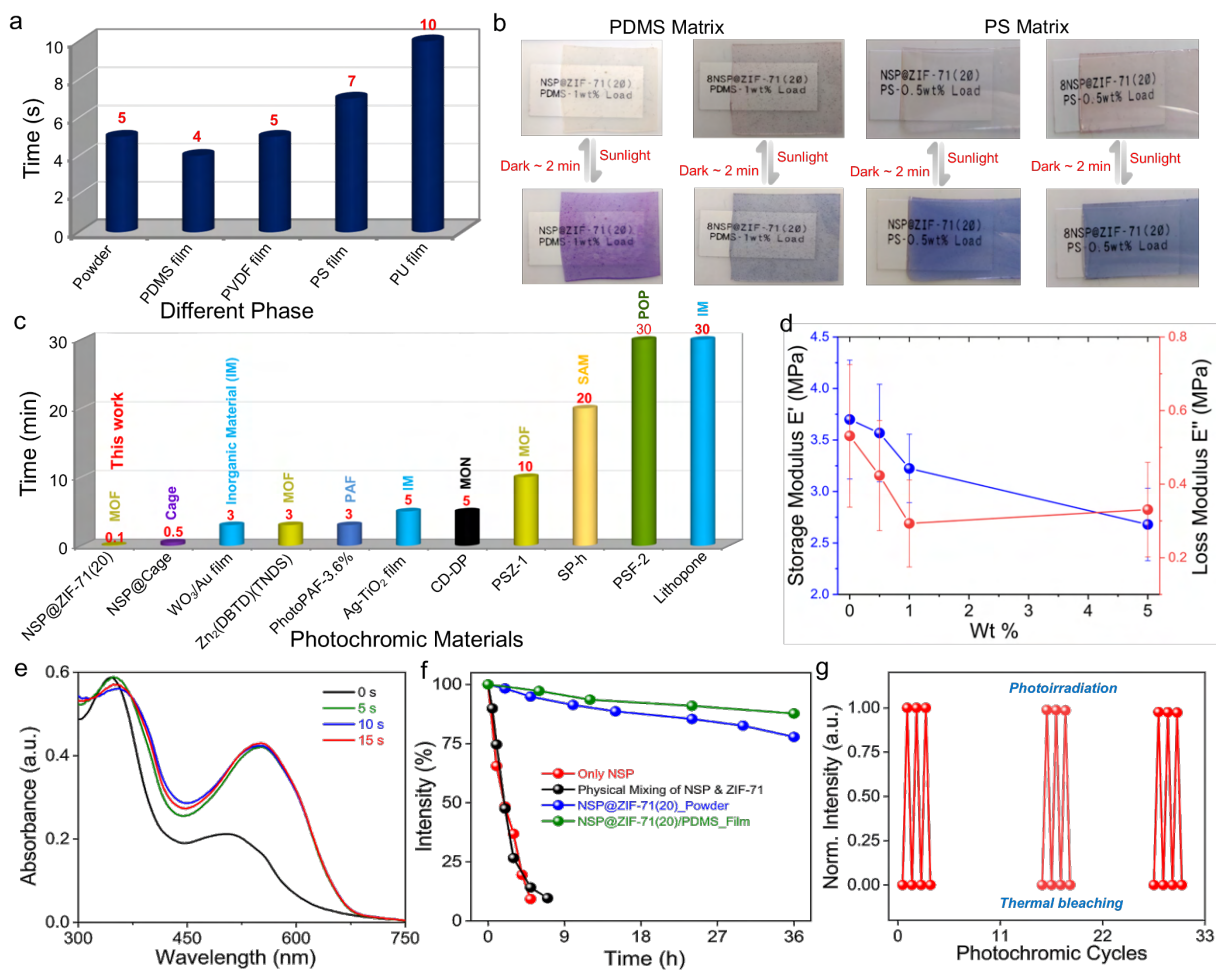

**Fig. 4.** Film characterization and photostability of photochromic materials. (a) Time taken for visual color change of the different photochromic nanocomposite films under 365 nm UV light. (b) Various thin films were fabricated by blending NSP@ZIF-71(20) and 8NSP@ZIF-71(20) with PDMS and PS polymer matrices, respectively. These films switch color in the presence of either 365 nm UV light (6 W lamp) or natural sunlight, where the nominal thickness of the film is *ca*. 190 μm. (c) A comparison of the transition times for photochromic switching of NSP@ZIF-71(20) versus other competing photochromic materials (see details in Table S8). (d) Probe-based dynamic mechanical analysis (nanoDMA) curves for the different loadings of NSP@ZIF-71(20) into PDMS films. (e) UV-Vis DRS of NSP@ZIF-71(20) powder subject to 0–15 s of 365 nm UV irradiation; samples tested after eight months of production. (f) Photostability tests for different photochromic materials. (g) Reversible photochromic cycles showing the photo-induced fatigue resistance of the NSP@ZIF-71(20)/PDMS film.

Pursuing real-world applications with these photochromic materials, detailed photochromic studies on thin films were carried out using a wide range of polymer matrices, such as PVDF (polyvinylidene difluoride), PDMS (polydimethyl siloxane), PS (polystyrene), PU (polyurethane), acrylate, and PMMA (polymethyl methacrylate) (Figure 4, Figures S34-S42). In this work, we have performed in-depth photochromic investigations of the NSP@ZIF-71(20)/PDMS system due to its merits in the form of transparent films, efficient photoswitching ability, and superior mechanical properties compared to the other polymer matrices being tested herein (Figure S34). We fabricated several films with different loadings of photochromic material as a filler to yield various film thicknesses, ranging from 12 to 2000 μm utilizing the doctor-blade technique (Figure S35-S36). Remarkably, we established that a filler content as low as 1 wt.% is already sufficient to achieve an intense visual color change in the presence of UV or natural sunlight (Figures S37-S40). Upon UV irradiation, the characteristics of the merocyanine absorption maximum appear at 576 nm in UV-vis DRS (Figure 4e), further confirming the open form of NSP molecules confined in the ZIF-71 host. Visual color switching was observed within 5 s upon UV irradiation before reverting to its original state under ambient conditions within 120 s, which is faster than the transition time observed for the bulk powder samples (Figure 4a, Figure S41).

Significantly, the NSP@ZIF-71(20)/PDMS films exhibit the quickest reversible photoisomerization in contrast with existing reported photochromic materials. Moreover, it also fulfils the commercial standards required for photochromic materials (Figure 4b-c, Tables S8-S9) (*24*). The mechanical properties of the photochromic films were further examined by the nanoindentation technique, so as to determine basic properties key to commercial thin films (Figures S43-S44). The flat punch and Berkovich indenters were used for the measurement of the mechanical properties of NSP@ZIF-71(20) incorporated into PDMS and PS polymer matrices, respectively. The elastic modulus ($E$) of the PS polymer decreases with an incremental amount of filler loading, where the stiffness value fell from 4.6 GPa (neat PS) to around 4.1 GPa at 15 wt.% filler content (Figure S43). This is a significant result because at present mechanically stiff polymers with $E > 2$ GPa will yield a poor photochromic response (*8*) (*vide infra*). In terms of the viscoelastic (time-dependent) mechanical response, the storage modulus ($E'$) of the polymer matrix decreases with increasing amount of NSP@ZIF-71(20) filler, while the loss modulus ($E''$) initially decreases with the incremental amount of filler and reaches a minimum for 1 wt.% (Figure 4d). Likewise for the 1 wt.% composite, the loss tangent values stay the lowest across all tested frequencies (Figure S44).

Despite the success in the synthesis of various photochromic materials, the insufficient long-term stability, photobleaching, and the low recycling ability of organic photochromic materials are major barriers that seriously impede their potential in real-world applications. Figure 4e depicts the similar coloring rate as pristine materials over a period of several months, demonstrating the long-term environmental stability. Photobleaching tests were conducted by continuous exposure to an intense 365 nm UV light. Both the NSP@ZIF-71(20) powder and films have retained more than 75% of their initial emission intensity even after 36 hours of continuous exposure to high-intensity UV irradiation, while the bare NSP and physical mixing of the NSP and ZIF-71 samples rapidly degraded within 5 hours (Figure 4f, Figure S45). Furthermore, no detectable photodegradation was observed in the absorption intensity even after several spontaneous colorations and fading cycles for both the photochromic powders and films (Figure 4g, Figure S46), validating the exceptional photostability and robustness of the composite materials to photo-induced fatigue (Figure 4f). The combination of quick reversible photoisomerization process, long-term stability, and outstanding fatigue features render the NSP@ZIF-71 material a strong contender for deployment in commercial and industrial settings (*24*).

A variety of well-defined smooth and crack-free prototype photochromic sculptures were successfully fabricated by blending 1 wt.% of NSP@ZIF-71(20) into a PDMS or PS polymer matrix to imitate the production of commercially viable photochromic components (Figure 5 and Figures S47-S49). The solid sculptures visually switch color within 5 s and 30 s upon irradiation by 365 nm UV light and sunlight, respectively. Subsequently, these materials could revert to their original color in the dark within 3 min at room temperature (Figure 5a), and this transformation time is as good as commercial-grade inorganic photochromic materials (*24*). Furthermore, prototypes of a large optically clear photochromic window and rewritable device were fabricated by blending 1 wt.% of NSP@ZIF-71(20) with the PS polymer. The photochromic window changes color instantly when exposed to natural sunlight for a few seconds and will return to its original color in less than a minute in darkness (Figure 5b, Movie S1). In another exemplar case, by illuminating a collimated 365-nm laser pointer on the rewritable device (see Movie S2), this instantly generates high-contrast handwriting or drawings that are swiftly erasable within minutes. Consequently, it turns out that high stiffness engineering polymers, such as polystyrenes ($E \sim 4$ GPa), can be employed as a polymer matrix for the manufacture of photochromic films/devices if the photochromic guests are encapsulated in porous scaffolds, which was previously unattainable because mechanically stiff polymers typically hinder the molecular switching of the photochromes (*8*). Additionally, the devices made from the PS films are readily re-dissolved and recycled, which may reduce polymer waste and expand the life cycle of engineered components.

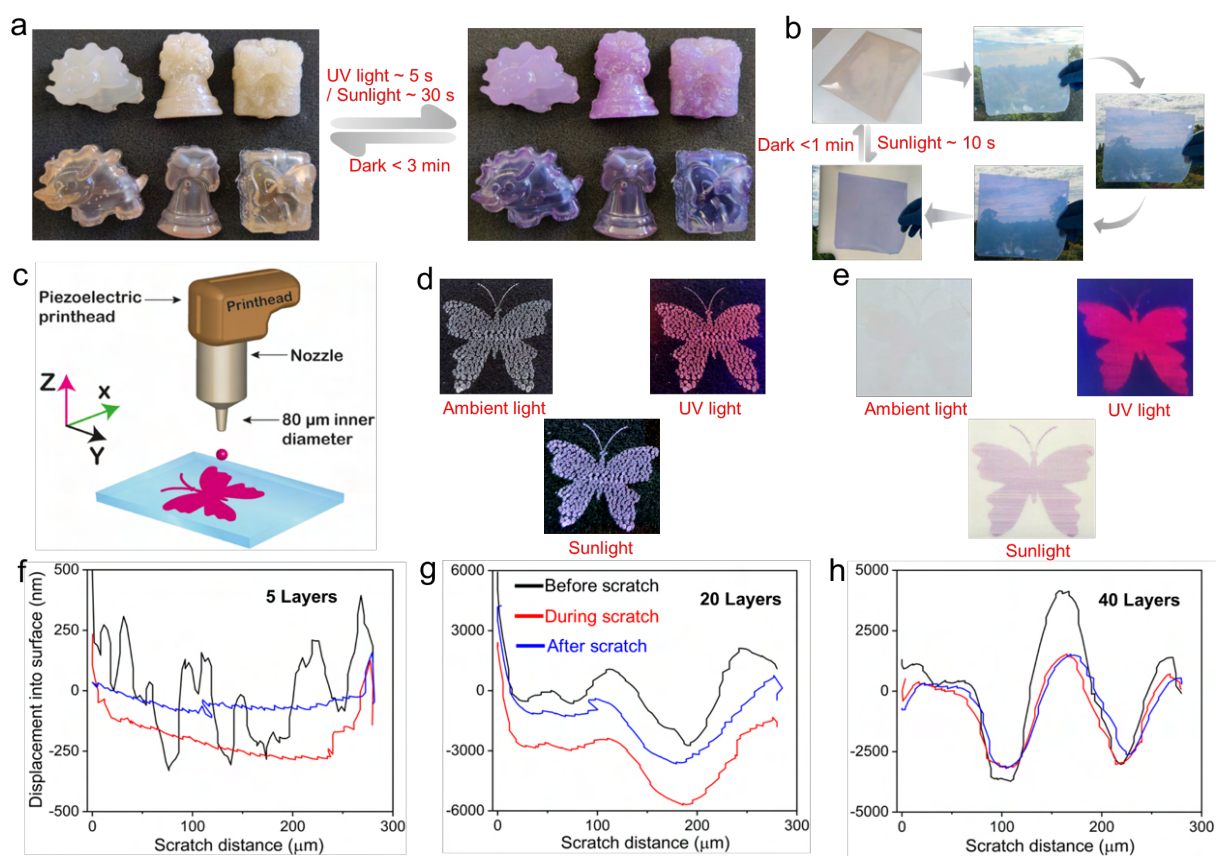

**Fig. 5.** Fabrication of different photochromic sculptures, transparent window, and emblems using the NSP@ZIF-71(20) nanocomposite. (a) Various photochromic sculptures were fabricated by blending 1 wt.% of NSP@ZIF-71(20) with PDMS (top) and PS (bottom) polymer matrices. These sculptures readily switch their color in the presence of either 365 nm UV light or natural sunlight. (b) 20 × 15 cm$^2$ NSP@ZIF-71(20)/PS based photochromic window with a nominal thickness of *ca*. 350 μm. (c) An illustration of the inkjet printing technique. (d) Photographs of 2 layers of a butterfly image printed on a glass substrate. (e) Photographs of 2 layers of butterfly image printed on a PET substrate. When exposed to UV and sunlight, the printed image on both the glass and PET substrates changes color within just 5–10 s. Displacement and scratch distance data from a 1 mN nanoscratch ramp-load test by the plough mode over a scratch distance of 200 μm, on (f) 5 layers, (g) 20 layers, and (h) 40 layers of inkjet-printed NSP@ZIF-71(20) film on a glass substrate. Black, red, and blue traces corresponding to the initial surface scan, linear ramp scratch, and residual profile scan after tip unloading, respectively.

Finally, piezoelectric-driven drop-on-demand inkjet printing was used for the precision coating of the bare nanocomposite materials on different substrates to create prototype photoswitchable printed films (Figures 5c-e and Figures S50-S66, see SI for details of the deposition method) (*29*). The photochromic nanocomposites were homogeneously printed on a wide range of engineering substrates, such as silicon wafer, indium tin oxide (ITO)-coated flexible acetate sheets, glass substrates, PET substrates, and ITO-coated glass substrates, ranging from 50 nm to 2 μm in film thickness (Figures S51-S66). AFM characterization revealed that the thickness of each layer of bare NSP@ZIF-71(20) film was around 30-50 nm, which is in good agreement with the height of each nanocrystal of the composites (Figures S17, S57-S60). Figure 5d and 5e demonstrates the 2-layer thickness of a butterfly emblem inkjet-printed on glass and PET substrates, respectively, showing their ability to spontaneously switch color subject to either UV irradiation or sunlight (Figures S61-S63).

The printed photochromic film is invisible when viewed against the white PET substrate in ambient light, but it becomes visible promptly under UV/sunlight, demonstrating its potential use in anticounterfeiting technology (Figure 5e, Figures S64-S65). The materials adhesion to the substrate is a crucial parameter for commercial applications, and it was investigated in these inkjet-printed samples on a glass substrate by systematic nanoscratch experiments accompanied by topographic characterizations (Figure 5f-h, Figures S66-S68, see the characterizations and physical measurements section in SI for more details) (*30*). The scratch test revealed that as the inkjet-printed film thickness increases, the roughness of the film decreases while the material cohesion interaction is improved (Figures S67-S68). In the scratch test for the 5-layer film, the indenter has already reached the substrate and the elastic recovery is attributed to the substrate; while for 40 layers, there is no elastic recovery because the indenter has just scratched the surface of the films (Figure S67). Additionally, we also observed that chipping is the failure mechanism, and this is related to the cohesion of the nanocrystals of the photochromic materials (Figure S68). The results suggest there is scope to tailor the mechanical resilience of the photochromic films and its fracture toughness by tuning the number of printed layers and thus modifying the film-to-substrate interactions for specific applications.

## Conclusions

In summary, we have developed a conceptual methodology for engineering exceptionally stable and efficient solid-state photoswitchable materials that exhibit a solution-like photochromic response. Structural, optical, chemical, and mechanical characterizations have been systematically performed using a suite of techniques to elucidate the detailed behavior of the newly developed photochromic materials, from bulk powders to single crystals, and from polymer membranes to nanocrystalline thin films. The commercially viable prototype photochromic films and photoresponsive devices fabricated from these composites show high optical transparency, quick fading kinetics, and excellent fatigue resistance, making them a strong contender for deployment in practical applications. It is envisaged that the straightforward photochrome@host methodology outlined can be extended to other solid-state nanoporous host matrices, including but not limited to MOFs, covalent organic frameworks (COFs), and metal-organic polyhedra (MOPs), which will leapfrog the development of durable organic photoswitchable materials supported in mechanically robust engineering polymers fit for real-world applications.

## Acknowledgements

S.M., M.T., A.F.M., V.K. and J.C.T. thank the ERC consolidator Grant PROMOFS (grant agreement 771575) for funding the research. J.C.T. acknowledges the EPSRC IAA award (EP/R511742/1) for additional support. W.K. acknowledges the financial support provided by the Punjab Education Endowment Fund (PEEF), Pakistan. A.F.M. thanks the Oxford Ashton Memorial Scholarship for a D.Phil. studentship award. A.A.C-P was supported by The Royal Society through a University Research Fellowship (URF\R\180016) and the John Fell Fund, Oxford University Press, via a Pump-Priming grant (0005176). We acknowledge the Diamond Light Source (Harwell, Oxford) for the award of beamtime SM25407; we thank Dr Mark Frogley and Dr Gianfelice Cinque for their assistance during the B22 MIRIAM beamline. We thank the Research Complex at Harwell (RCaH) for the provision of materials characterization facilities. We thank Dr Sudarshan Narayan (Oxford Materials) for the XPS measurements.

## Author Contributions

S.M. conceived, designed, and developed the project under the supervision of J.C.T. S.M. conducted all the material synthesis and characterizations. S.M. analyzed all the data with the help of Y.Z., V.K. and J.C.T. M.T. performed the mechanical testing measurements. W.K. performed the inkjet printing experiments under the supervision of A.A.C.-P and S.M. A.F.M. and S.M. performed the nanoFTIR measurements. N.A. performed the NMR experiments and data analysis. S.M. wrote the original draft of the manuscript with input from J.C.T. All the authors contributed to the final version of the manuscript.

# Supplementary Information

**Surface modulation of metal-organic frameworks for on-demand photochromism in the solid state**

Samraj Mollick,[a] Yang Zhang,[a] Waqas Kamal,[b] Michele Tricarico,[a] Annika F. Moslein,[a] Vishal Kachwal,[a] Nader Amin,[c] Alfonso A. Castrejón-Pita,[b] Stephen M. Morris,[b] and Jin-Chong Tan [a],*

[a]Multifunctional Materials & Composites (MMC) Laboratory, Department of Engineering Science, University of Oxford, Parks Road, Oxford OX1, UK.

[b]Department of Engineering Science University of Oxford, Parks Road, Oxford, OX1 3PJ, United Kingdom.

[c]Department of Chemistry, University of Oxford, Mansfield Road, Oxford OX1 3TA, UK.

Corresponding author. E-mail: jin-chong.tan@eng.ox.ac.uk

# Table of Contents

# Materials

All the pristine MOFs such as ZIF-68, ZIF-65, ZIF-8, ZIF-70, MOF-808, UiO-67, ZIF-71, MAF-6 and ZIF-11 were synthesized using a previously reported literature method (*1-3*). All the reagents, solvents, and photochromic guests were commercially available and used as received. All of the commercially available materials were purchased from Sigma-Aldrich. Fisher Scientific, Fluorochem, and Alfa Aesar depending on their availability. The 100 mM stock solution of different photochromic guests such as SP, NSP, and 8NSP were prepared in *N,N*-dimethyl formamide (DMF) solutions.

# Methods

**Materials Synthesis**

**Photochromic ZIF-71 composites.** 1 mmol zinc acetate was soluble in 5 mL of methanol and rapidly mixed into a 5 mL methanol solution of 4 mmol of 4,5-dichloroimidazole (dcIm) under stirring. The mixed solution transformed from a clear to turbid appearance after a few minutes. After 24 hours of stirring at room temperature, the sample was centrifuged at 10,000 rpm to collect the product. The product was further washed six times with a copious amount of methanol to remove the excess reactants and dried at 60 °C overnight. The SP@ZIF-8, NSP@ZIF-8 and 8NSP@ZIF-8 composites were prepared following the aforementioned same synthesis protocol where different guests such as SP, NSP and 8NSP were mixed into the linker solution prior to the addition of the metal source of the MOF building blocks. Various NSP@ZIF-71 composites were also prepared following the aforementioned method where the amount of NSP stock solution gradually increased from 0.1 mM, 0.5 mM, 1 mM, 3 mM, 5 mM, 10 mM, 15 mM, to 20 mM, and keeping the fixed amount of MOF components. The resultant guest@MOF composite materials are denoted as NSP@ZIF-71(0.1), NSP@ZIF-71(0.5), NSP@ZIF-71(1), NSP@ZIF-71(3), NSP@ZIF-71(5), NSP@ZIF-71(10), NSP@ZIF-71(15), NSP@ZIF-71(20), and NSP@ZIF-71(30), respectively.

**Different nanosized photochromic ZIF-71 composites.** The synthesis of different nanosized ZIF-71 particles was conducted using methanol as a solvent in the presence of triethyl amine (TEA). The dcIm : TEA : Zn(OAC)$_2$ · 2H$_2$O molar ratio was varied from 1:0:1, 1:1:1, 2:2:1, 3:3:1 and 4:4:1 at room temperature in a sealed vial and left to stand at room temperature for 24 hours in stirring conditions. The different nanosized NSP@ZIF-71 composites were prepared following the aforementioned synthesis protocol, keeping the NSP amount fixed and mixed prior to the addition of the metal source of MOF building blocks. After 24 hours of stirring under room temperature, the sample was centrifuged at 10,000 rpm to collect the product. The product was further washed six times with a copious amount of methanol to remove the excess reactants and dried at 70 °C overnight. A large single crystal of NSP@ZIF-71(20) was obtained by following the same reaction conditions except the stirring process continued for 72 hours at room temperature.

**Gram scale production of photochromic ZIF-71 composites.** 40 mmol of 4,5-dichloroimidazole (dcIm) and 20 mL of 100 mM stock solution of guests were mixed properly in 50 mL of methanolic solution. Another 10 mmol of zinc acetate was dissolved in another 30 mL of methanolic solution, and then these two solutions were rapidly combined under vigorous stirring. After 24 hours of stirring at room temperature, the sample was centrifuged at 10,000 rpm to collect the product. The product was further washed six times with a copious amount of methanol to remove the excess reactants and dried at 70 °C overnight to obtain the desired product on a gram scale (yield ~ 3.6 g).

**ZIF-72 composite.** The solid-state nonporous ZIF-72 phase was synthesized by following the reported protocol in the literature with slight modification (*4*). Briefly, a physical mixture of ZnO, HdcIm and NSP (82 mg: 411 mg: 32.2 mg; 1: 3: 0.1 molar ratio) was prepared and transferred to a 50 mL Schott bottle. A glass vial containing the additive (H$_2$O, 5 mL) was placed into the Schott bottle next to the powders. The capped Schott bottle was placed in a preheated oven at 130 °C for 24 hours. After a reaction of 1 day, the powder samples were collected and washed with a copious amount of methanol to remove the excess reactants and dried at 70 °C overnight.

**Transformation of photochromic ZIF-71 to ZIF-72 composite.** The porous ZIF-71 phase transformed into nonporous ZIF-72 by exposing the vapor of the dcIm linker at high temperatures. The NSP@ZIF-71(20) was transformed into NSP@ZIF-72 in the presence of dcIm vapor at 130 °C for 24 hours. Finally, the product was washed with a copious amount of methanol to remove the excess reactants and dried at 70 °C overnight.

**Fabrication of various photochromic sculptures.** A variety of desired geometrical shapes and different-sized photochromic sculptures were prepared using polydimethylsiloxane (PDMS) and polystyrene (PS) polymer matrices, where NSP@ZIF-71(20) was used as a filler. Toluene solvent was used to dissolve the solid PS polymer and the NSP@ZIF-71(20) composite was homogeneously dispersed into the PS polymer matrix by stirring for 24 hours under ambient conditions. The obtained jelly-type materials were poured into different shaped molds and kept for another 12 hours at 60 °C to obtain the various-shaped photochromic sculptures. For the creation of PDMS-based photochromic sculptures, PDMS and NSP@ZIF-71(20) were first combined for 24 hours before Sylgard 184, a curing agent, was applied. The mixture of NSP@ZIF-71(20)/PDMS and curing agent was kept 90 °C for 12 hours to obtain the PDMS based sculptures. A PTFE (polytetrafluoroethylene) spray was used before pouring the mixtures of NSP@ZIF-71(20)/PDMS and curing agent into the molds to facilitate the detachment of the cured sculptures from the molds.

**Fabrication of photochromic thin films.** A variety of photochromic thin films were prepared by using a wide range of polymer matrices such as polyvinylidene difluoride (PVDF), PS, PDMS, polyurethane (PU), acrylate polymer and poly(methyl methacrylate) (PMMA). First, photochromic nanocomposites were dispersed homogeneously into different polymer matrices by a combination of sonication (30 min) and magnetic stirring (24 hours). Next, homogeneously mixed polymer-nanocomposites were cast onto a glass substrate by using the doctor blade technique. The thickness of the thin-film membranes can be varied from 3 to 2200 μm using different thickness gaps of the blade during casting, and the casting speed was set to 12 mm/s.

**Fabrication of photochromic window and rewritable device.** The PS polymer matrix was selected for the fabrication of a photochromic window and rewritable device. First, NSP@ZIF-71(20) was well-dispersed in a toluene solvent by sonication and then solid PS polymeric beads (molecular weight = 192,000) were added to this mixture and left stirring at 300 rpm for 24 hours at room temperature. The concentration of the photochromic materials in the PS matrix was fixed to 0.1 wt.%. The obtained homogenous mixture was degassed at a reduced pressure to remove the air bubbles. Afterwards, the mixture was cast onto a clean large glass substrate using the doctor blade applicator at room temperature and a casting speed set to 12 mm/s. Finally, the casted films were kept in an oven at 60 °C for another 24 h, and next the film was peeled off to obtain the desired photochromic window. The rewritable device was also fabricated following the same protocol.

**Photochromic MAF-6 composites.** Pristine MAF-6 was synthesized using a rapid solution mixing protocol, where a mixture of 9 mL ethanol solution of 2-ethylimidazole (1 mmol) and 2 mL cyclohexane was rapidly combined with 1 mL of a concentrated ammonia hydroxide (35%) solution of ZnO (0.5 mmol). The resultant reaction mixture was stirred at room temperature for the next 2 h, and then filtered, washed with a copious amount of methanol, and dried at 70 °C under a vacuum for 12 h to obtain the desired product. The NSP@MAF-6 and 8NSP@MAF-6 composites were prepared following the same synthesis protocol where different guests such as NSP and 8NSP were dissolved with the linker solution mixture prior to the addition of the metal source of the MOF building blocks. Composites with different guest concentrations were also prepared by using 100 mM stock solution of guests.

**Photochromic ZIF-11 composites.** Pristine ZIF-11 was synthesized using a rapid solution mixing protocol where a mixture of 6 mL ethanol solution of benzimidazole (1 mmol) and 3 mL toluene was mixed rapidly with 1 mL of concentrated ammonia hydroxide (35%) solution of zinc acetate dehydrate (0.5 mmol). The resultant reaction mixture was stirred at room temperature for the next 3 h, and then filtered, washed with a copious amount of ethanol, and dried at room temperature overnight to obtain the desired product. The NSP@ZIF-11 and 8NSP@ZIF-11 composites were prepared following the same synthesis protocol where different guests such as NSP and 8NSP were dissolved with benzimidazole prior addition of zinc acetate dihydrate. Composites with different guest concentrations were also prepared by using a 100 mM stock solution of guests.

**Photochromic ZIF-8 composites.** The ZIF-8 was synthesized according to the literature reported protocol *(1)*. Two precursor solutions were prepared by dissolving 1 mmol of $Zn(NO_3)_2·6H_2O$ and 8 mmol of 2-methylimidazole (mIm) in 10 mL of methanol, respectively. A colloidal mixture was obtained within a few minutes by combining the two precursor solutions, which were stirred for the next 12 h at room temperature. Afterwards, the colloidal solution mixtures were centrifugation at 8000 rpm for 10 min, followed by solvent exchange with a copious amount of fresh methanol, and sonication for 30 s and the same procedure was repeated 5 times. After repeated washing, the product was dried at 70 °C under a vacuum for 12 h to obtain the desired product. The SP@ZIF-8, NSP@ZIF-8 and 8NSP@ZIF-8 composites were prepared following the aforementioned synthesis protocol where different guests such as SP, NSP and 8NSP were mixed in prior to the addition of the metal source of the MOF building blocks.

**Photochromic ZIF-65 composites.** 0.5 mmol zinc acetate was soluble in 5 mL of DMF and rapidly mixed into 1 mmol of 2-nitroimidazole in 5 mL methanol under vigorous stirring. After 24 h of stirring at room temperature, the sample was centrifuged at 10,000 rpm to collect the product. The product was further washed three times with a copious amount of N, N-dimethylformamide (DMF) and methanol to remove the excess reactants and dried at 90 °C overnight. The SP@ZIF-65, NSP@ZIF-65 and 8NSP@ZIF-65 composites were prepared following the aforementioned same synthesis protocol where different guests such as SP, NSP and 8NSP were dissolved with 2-nitroimidazole solution prior addition of zinc acetate dehydrate solution.

**Photochromic ZIF-68 composites.** 1 mmol of 2-nitroimidazole, 0.32 mmol of benzimidazole and 1 mmol of $Zn(NO_3)_2·6H_2O$ were mixed in 10 mL of different DMF solution separately. After that, three different solutions were combined and heated in a capped vial at 130 °C for 96 h and left to cool for 12 h. The mother liquor was decanted and the products were further washed three times with a copious amount of DMF and methanol to remove the excess reactants and dried at 70 °C overnight. The SP@ZIF-68, NSP@ZIF-68 and 8NSP@ZIF-68 composites were prepared following the aforementioned synthesis protocol where different guests such as SP, NSP and 8NSP were dissolved with 2-nitroimidazole and benzimidazole solutions prior addition of $Zn(NO_3)_2·6H_2O$ solutions.

**Photochromic ZIF-70 composites.** 1 mmol of 2-nitroimidazole, 1 mmol of imidazole and 1 mmol of $Zn(NO_3)_2·6H_2O$ were mixed in 10 mL of different DMF solutions separately. After that, three different solutions were combined and heated in a capped vial at 130 °C for 96 h and left to cool for 12 h. The mother liquor was decanted and the products were further washed three times with a copious amount of DMF and methanol to remove the excess reactants and dried at 70 °C overnight. The SP@ZIF-70, NSP@ZIF-70 and 8NSP@ZIF-70 composites were prepared following the aforementioned synthesis protocol where different guests such as SP, NSP and 8NSP were dissolve with 2-nitroimidazole and imidazole solutions prior addition of $Zn(NO_3)_2·6H_2O$ solution.

**Photochromic UiO-67 composites.** A mixture of 1 mmol $ZrCl_4$ (83 mg, 0.35 mmol) and 1 mmol 4,4′-biphenyldicarboxylic acid was dissolved in 10 mL of DMF, and the resulting solution was placed in a Teflon-lined Parr stainless steel vessel (17 mL) and sealed. The sealed vessel was then placed in the oven and heated to 120 °C for 24 hours. After slow cooling to room temperature, crystalline powders were isolated by filtration and washed with DMF three times. The obtained MOF was immersed in MeOH for 4 days to exchange occluded solvent molecules with MeOH and MeOH was exchanged every 12 hours. The MeOH exchanged sample was then evacuated at room temperature for 24 h and at 100 °C for 12 h to yield an activated UiO-67 powder sample. The SP@UiO-67, NSP@UiO-67 and 8NSP@UiO-67 composites were prepared following the same synthesis protocol where different guests such as SP, NSP and 8NSP were mixed with the MOF building blocks. Composites with different guest concentrations were also prepared by using different amounts of stock solutions of guests where the total solution mixture remained fixed at 10 mL.

**Photochromic MOF-808 composites.** The photochromic MOF-808 composites were prepared followed by an *ex situ* synthetic process where the first pristine MOF-808 powder was synthesized and afterwards the photochromic guests were exchanged into the MOF-808 powder by infiltration. In a typical MOF-808 synthesis protocol, trimesic acid (210 mg, 1 mmol) and zirconyl chloride octahydrate (970 mg, 3 mmol) were dissolved into DMF/formic acid (30 mL/30 mL) and placed in a large screw-capped glass jar, which was heated to 130 °C for two days. A white precipitate of MOF-808 was collected by filtration and washed four times with 400 mL of fresh DMF. The DMF-washed compound was then immersed in 100 mL of acetone for four days and during this time the acetone was replaced two times per day to facilitate the solvent exchange process. The acetone-exchanged sample was then evacuated at room temperature for 24 h and at 150 °C for 12 h to yield an activated sample. For the synthesis of photochromic composites, photochromic guests were added to a suspension of MOF-808 in a 10 mL methanol solution. The solution was stirred for 12 h at room temperature. The product was then washed sequentially with DMF (1 × 20 mL) and methanol (2 × 25 mL) and dried in a vacuum to yield a pink powder.

**Inkjet drop-on-demand printing.** The MOF 'ink', comprising an isopropyl alcohol (IPA) suspension of fine NSP@MOF nanoparticles without any additives were printed using a commercially available inkjet printing system (Jetlab-II, MicroFab Technologies Inc.) which can deposit droplets with a placement accuracy of ±5 μm. To print MOF droplets without the appearance of any satellite droplets, a piezoelectric nozzle with an 80-μm orifice diameter was employed. The dispenser was connected to a pneumatic pressure control system to control the back pressure so as to reduce the isopropanol solvent evaporation at the tip of the nozzle (the fast evaporation at the tip could lead to nozzle blockage). Figure S51 shows the dynamic shadowgraph images of the smooth MOF droplet formation process. The in-flight diameter of the falling droplet was ~50 μm. The glass and PTFE substrate were placed on the bed of the printer. During printing, the substrate and print head were maintained at a temperature of 21 °C. A rectangular array of droplets, according to the area of the substrates, were printed at a droplet spacing of 20 μm. A substrate with a range of film thicknesses was produced by changing the print pass of the print head for repeated deposition.

**Analysis of photochromic guest loading in the composites.** Samples for $^1$H nuclear magnetic resonance (NMR) characterization were dissolved into a solution composed of 500 μL methanol-d4 and 50 μL DCl / D$_2$O (35 wt.%). Data were processed using Bruker Topspin with a line broadening of 0.3 Hz and 2 rounds of zero-filling. Peaks were integrated using a global spectral deconvolution in the MestReNova software package. The ratio of NSP in NSP@ZIF-71 composites was calculated from the integral ratio of the peak corresponding to the 6 protons of the dimethyl group attached to the pyrrole ring of NSP and the peak corresponding to the single proton of the dcIm imidazole ring.

# Materials characterization and physical measurement techniques

**Powder X-ray diffraction (PXRD).** Powder X-ray diffraction (PXRD) patterns were performed on a Rigaku MiniFlex diffractometer with a Cu Kα source (1.541 Å) at a scan speed of 0.5°/min and a step size of 0.01° in $2\theta$.

**Thermogravimetric analysis (TGA).** Thermogravimetric analysis (TGA) was performed using a TA Instruments Q50 TGA machine equipped with a platinum sample holder under an $N_2$ inert atmosphere at a heating rate of 10 °C/min from 50 to 800 °C.

**Attenuated total reflectance Fourier transform infrared spectroscopy (ATR-FTIR).** ATR-FTIR spectra were acquired at room temperature with a Nicolet iS10 FTIR spectrometer with an ATR sample holder.

**Raman.** Raman spectra were recorded by using a MultiRam FT-Raman spectrometer (Bruker).

**Diffuse reflected spectroscopy (DRS).** A 2600 UV-Vis spectrophotometer (Shimadzu) was used to measure the absorption spectra and calculate the Kubelka-Munk (KM) function. The time-dependent UV-irradiation was conducted using a 365-nm handheld UV lamp (UVA, 6W).

**Solid-state photoluminescence spectra.** An FS-5 spectrofluorometer (Edinburgh Instruments) was used to characterize the steady-state emission, excitation spectra, quantum yield (QY), CIE 1931, and lifetime measurements.

**Nuclear magnetic resonance (NMR).** NMR spectroscopy was done at 298 K using a Bruker Avance NEO spectrometer operating at 600 MHz, equipped with a BBO cryoprobe. Data was collected using a relaxation delay of the 20 s, with 64 k points and a sweep width of 20 ppm, giving a digital resolution of 0.37 Hz. Data were processed using Bruker Topspin with a line broadening of 0.3 Hz and 2 rounds of zero-filling. Peaks were integrated using global spectral deconvolution in the MestReNova software package.

**Synchrotron radiation (SR) infrared spectroscopy.** SR FTIR spectra were recorded at the Multimode InfraRed Imaging and Microspectroscopy (MIRIAM) Beamline B22 at the Diamond Light Source (Oxfordshire, UK). IR spectroscopy was performed in vacuum via a Bruker Vertex 80 V Fourier Transform IR (FTIR) with an Attenuated Total Reflection (ATR) accessory (Bruker Optics, Germany). The mid-IR spectra were collected using a standard DLaDTGS detector. For the far-infrared spectral range below 700 cm$^{-1}$, a bolometer cooled by liquid helium was used for the detection of terahertz signals. All spectra were acquired with a resolution of 4 cm$^{-1}$ and a scanner velocity of 20 kHz. Pre-processing of spectral data was performed using the OPUS software version 7.2 (Bruker Optics). "Concave rubberband correction" from the OPUS software was applied with 4000 points followed by max-min normalization in the range of 50-680 cm$^{-1}$.

**nanoFTIR.** The morphologies, optical phase images, and nanoFTIR results were examined using the s-SNOM instrument (Neaspec GmbH), where a platinum-coated AFM probe (Arrow-NCPt, tip radius < 25 nm, 285 kHz) under the tapping mode is illuminated by a broadband mid-infrared (MIR) laser source (Toptica). To suppress background contributions, the signal was modulated at the third harmonic of the tip frequency for optical phase images and at the second harmonic for nanoFTIR absorption spectra. Each spectrum was obtained from averaging over 12 individual measurements with an integration time of 12 s, and subsequently normalized to the spectrum of the silicon substrate. In order to eliminate the instrument noise, we removed 200 cm$^{-1}$ of data from both sides of each nanoFTIR plot during the plotting process.

**X-ray photoelectron spectroscopy (XPS).** X-ray photoelectron spectroscopy (XPS) was performed using a PHI VersaProbe III system generating monochromatic Al X-rays at 1486 eV. Powder samples were mounted on adhesive carbon tape inside a glovebox and transferred into the XPS *via* a vacuum transfer vessel to avoid contamination and ambient exposure. Survey scans and high-resolution elemental scans were acquired at pass energies of 224 eV and 55 eV, respectively.

**Density functional theory (DFT) of NSP molecules.** Materials Studio software suite 2017 (Accelrys) was used to perform the density functional theory (DFT) calculations. The molecular structures of the NSP (open and closed forms) were fully relaxed using the DMol3 code. Infrared spectra were calculated using the B3LYP and PBE exchange-correlation functionals.

**Nanoindentation test.** Nanoindentation tests were conducted using an iMicro (KLA-Tencor) nanoindenter. Berkovich and flat punch diamond indenter tips were used. A 50-µm diameter flat punch was used for the nanoDMA measurements to determine the storage and loss moduli of PDMS films. A Berkovich tip was employed for characterizing the Young's modulus and hardness of the PS films, and for conducting scratch tests on inkjet-printed NSP@ZIF-71(20) films on glass substrate.

**Scratch test.** Nanoindenter scratch experiments were performed on an MTS Nanoindenter XP equipped with a Berkovich diamond indenter tip. The substrate with coated surface was laterally translated under the indenter tip for a total scratch length of 280 μm at a speed of 5 μm s$^{-1}$ (0° or 180° tip orientation). The normal load was progressively increased from 0 to 1 mN at a loading rate of 0.01 mN s$^{-1}$ during a ramp load run. Each scratch test was composed of three sequential stages: (I) a tiny load is applied to the surface to track the surface morphology (before scratch); (II) the same path is followed during the scratch test and the specified load is applied (during scratch); (III) again very tiny load is applied along the scratch path to measure the residual surface deformation (after scratch). The area established between the 'during scratch' and 'after scratch' graphs is related to elastic recovery, while the area between 'after scratch' and 'before scratch' corresponds to plastic deformation.

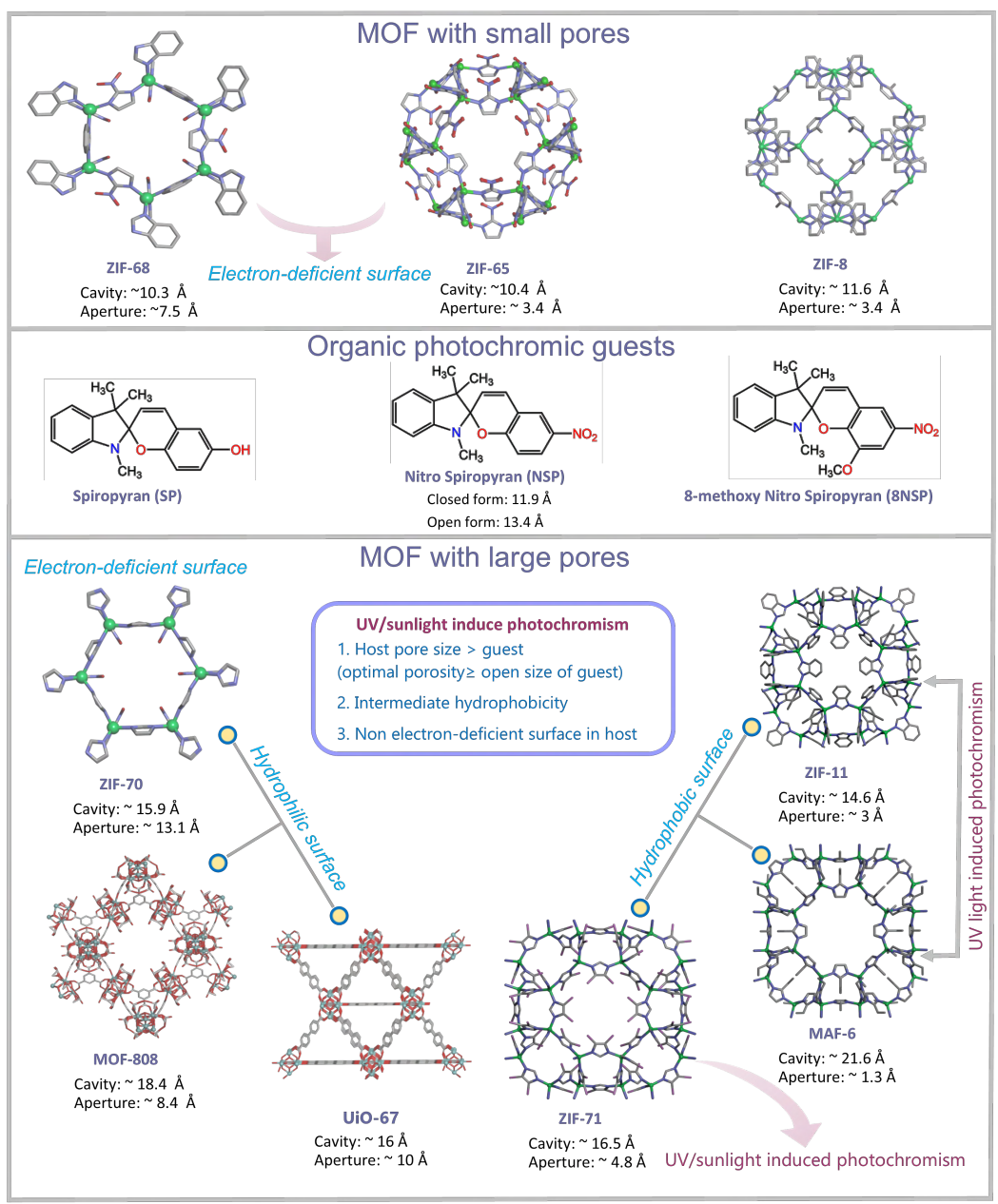

**Fig. S1.** Different MOFs were used as host matrices for the encapsulation of organic photochromic spiropyran guest molecules. Here, the pore cavity larger than spiropyran guest is termed as MOF with a small pore while the pore cavity smaller than the spiropyran guest is termed as MOF with a large pore. Three different organic photochromic guests coined as SP {1',3',3'-Trimethylspiro[chromene-2,2'-indolin]-6-ol}, NSP {1′,3′-Dihydro-1′,3′,3′-trimethyl-6-nitrospiro[2H-1-benzopyran-2,2′-(2H)-indole]} and 8NSP { 1′,3′-Dihydro-8-methoxy-1′,3′,3′-trimethyl-6-nitrospiro[2H-1-benzopyran-2,2′-(2H)-indole] respectively. Photographs portraying the photochromic color change upon irradiation of 365 nm UV light (6W) of NSP@MOF(20) composites.

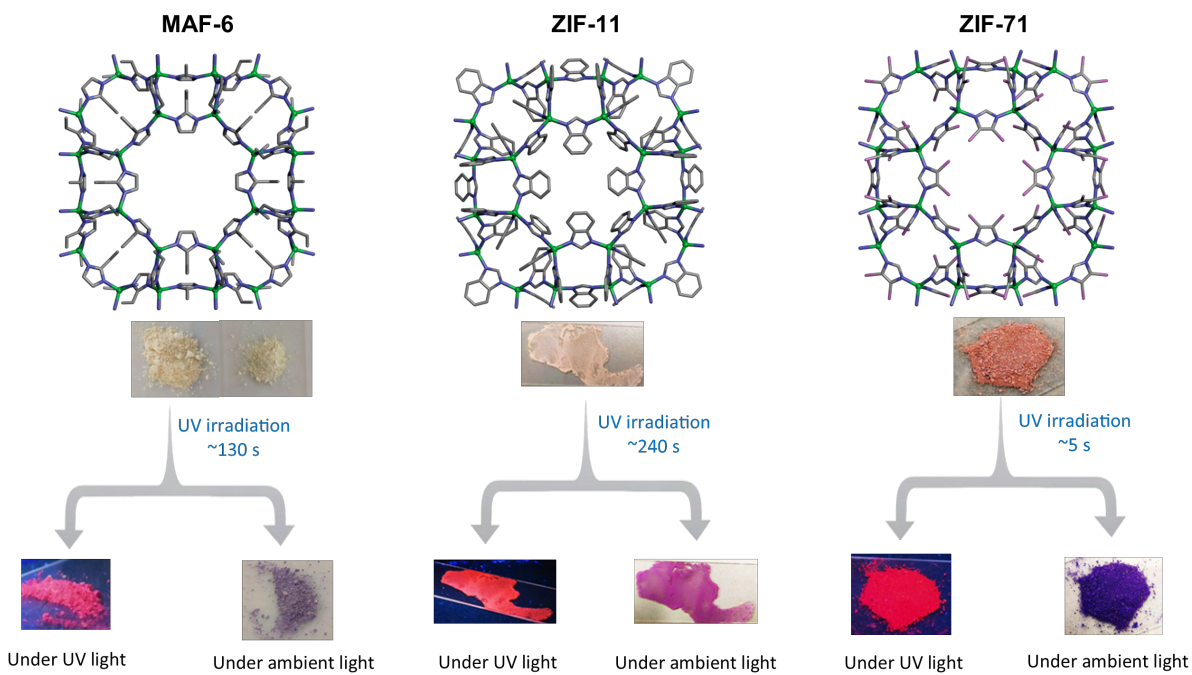

**Fig. S2.** Photographs portraying the photochromic color change upon irradiation of 365 nm UV light (6W) of NSP@MOF(20) composites.

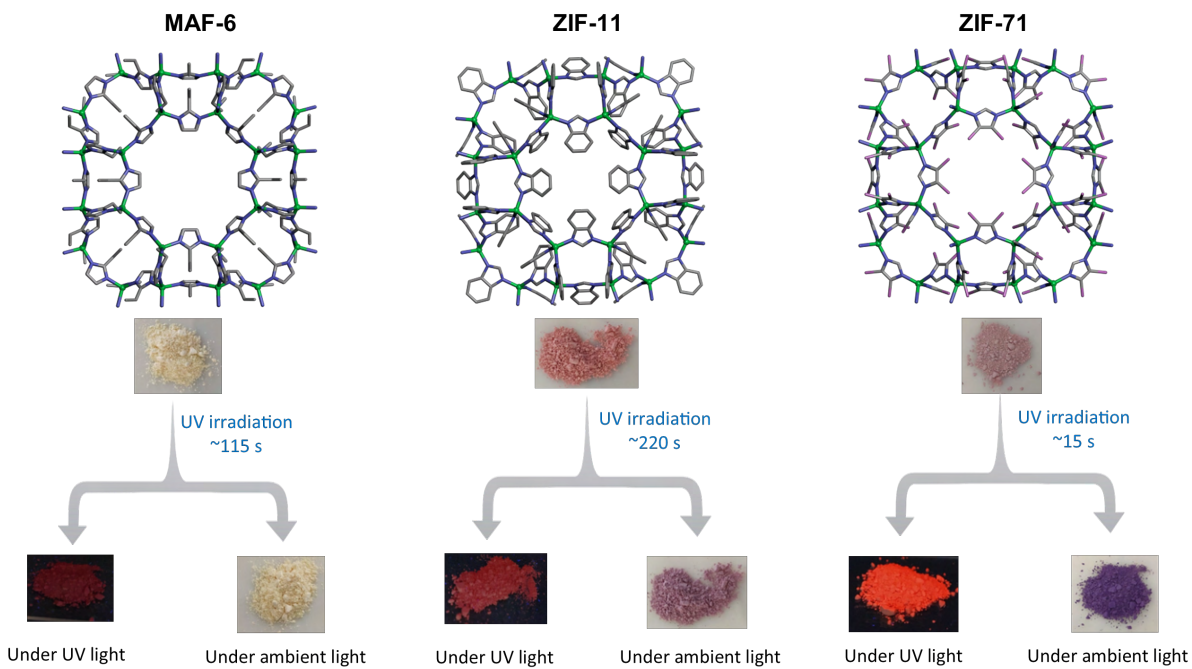

**Fig. S3.** Photographs portraying the photochromic color change upon irradiation of 365 nm UV light of 8NSP@MOF(20) composites.

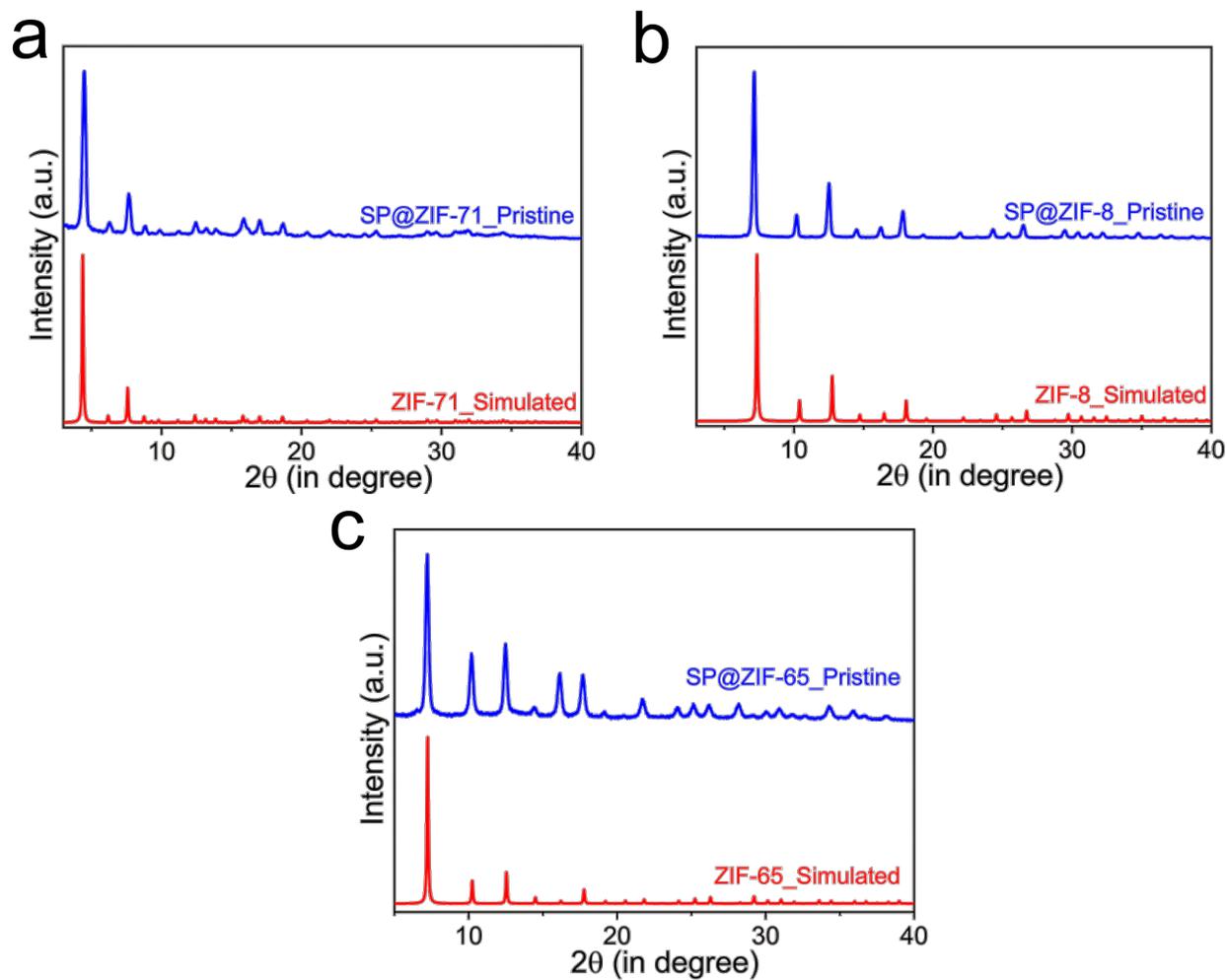

**Fig. S4. (a)** PXRD patterns of pristine ZIF-71 and SP@ZIF-71 composites. **(b)** PXRD patterns of pristine ZIF-8 and SP@ZIF-8 composite. **(c)** PXRD patterns of pristine ZIF-65 and SP@ZIF-65 composites.

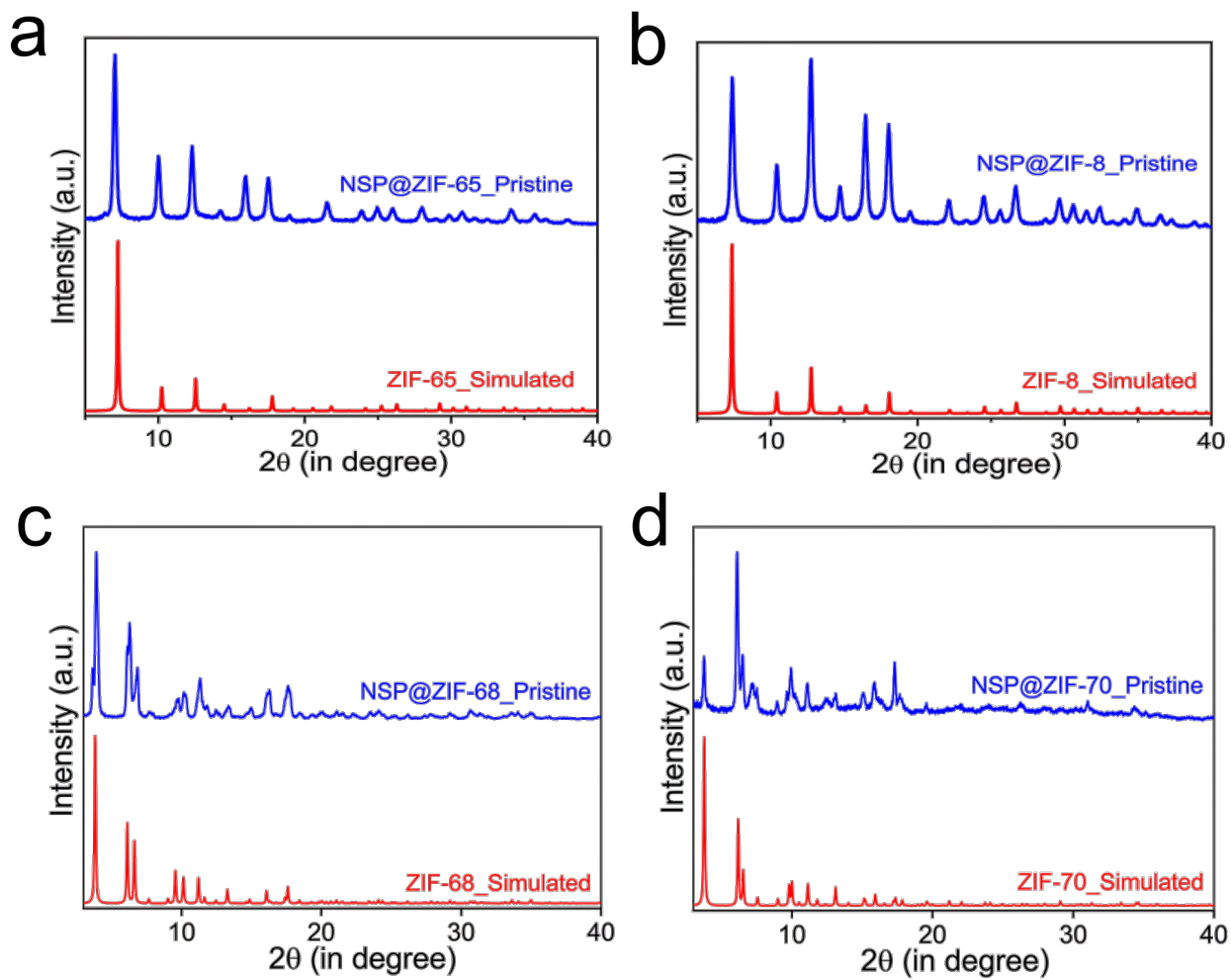

**Fig. S5. (a)** PXRD patterns of pristine ZIF-65 and NSP@ZIF-65 composites. **(b)** PXRD patterns of pristine ZIF-8 and NSP@ZIF-8 composites. **(c)** PXRD patterns of pristine ZIF-68 and SP@ZIF-68 composites. **(d)** PXRD patterns of pristine ZIF-70 and NSP@ZIF-70 composites.

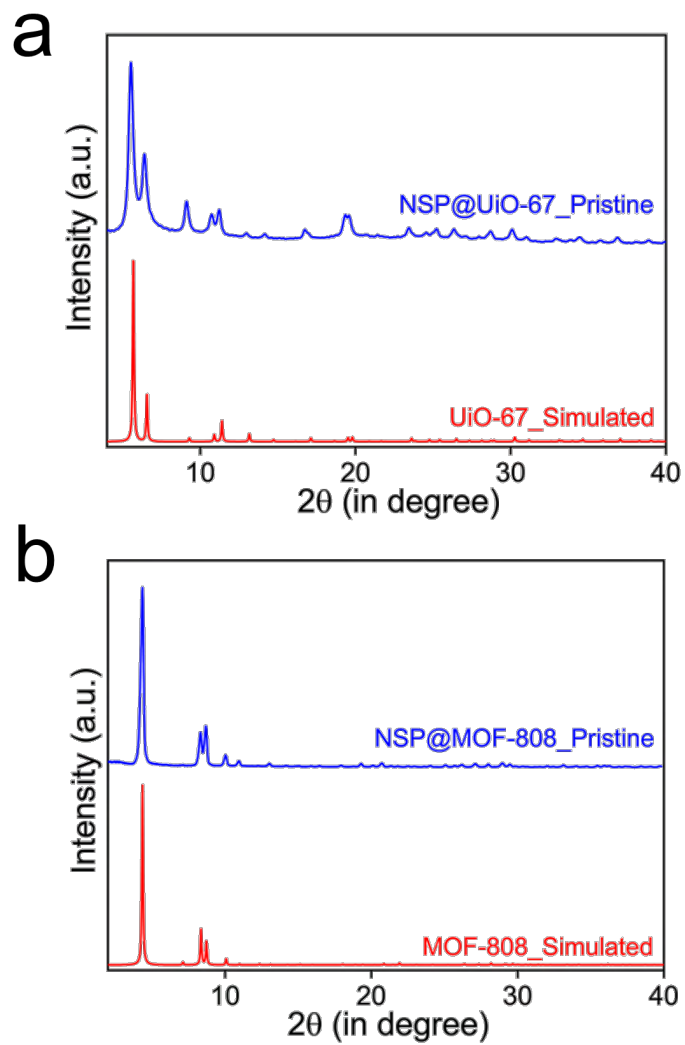

**Fig. S6. (a)** PXRD patterns of pristine UiO-67 and NSP@UiO-67 composites. **(b)** PXRD patterns of pristine MOF-808 and NSP@MOF-808 composites.

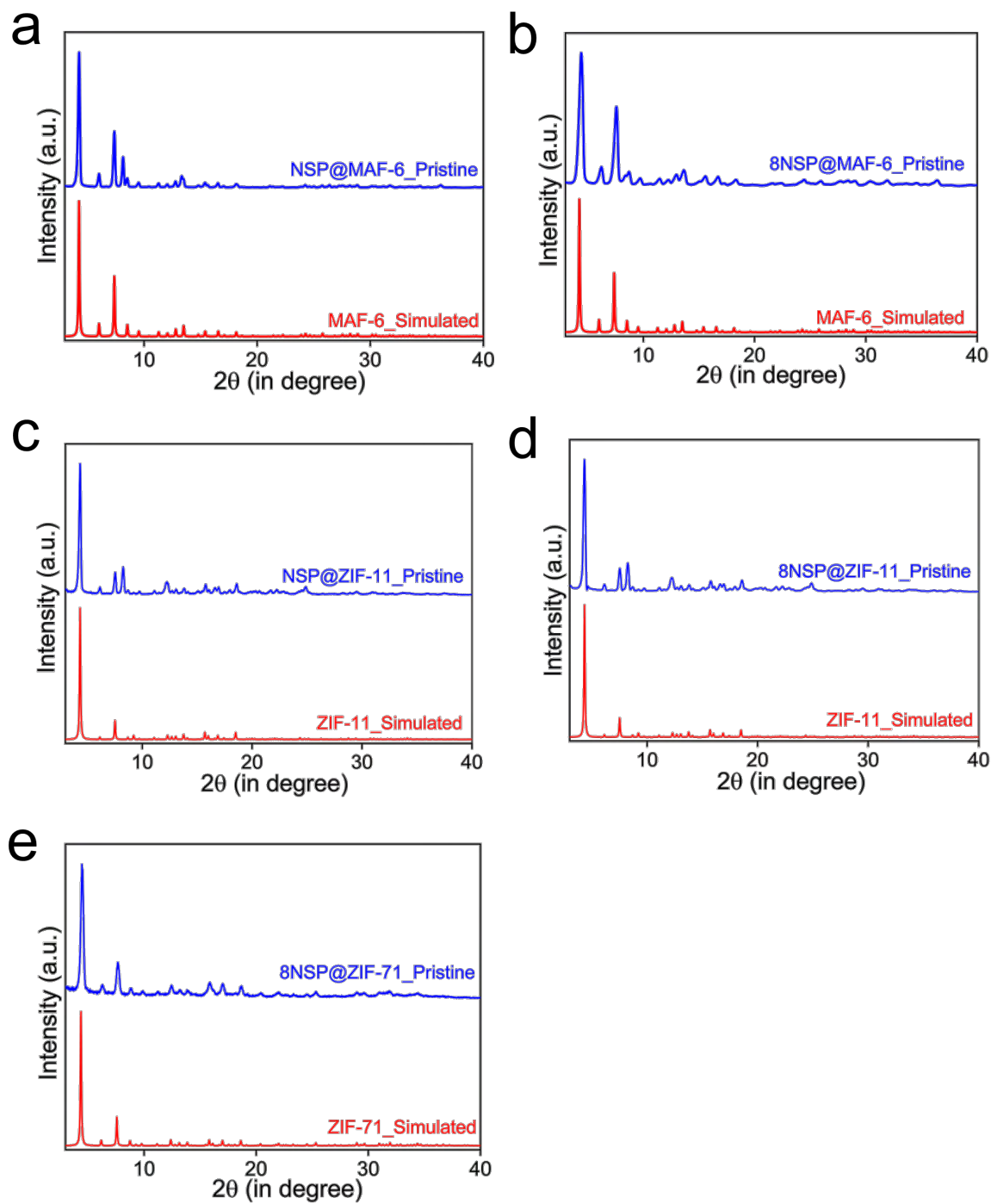

**Fig. S7. (a)** PXRD patterns of pristine MAF-6 and NSP@MAF-6 composites. **(b)** PXRD patterns of pristine MAF-6 and 8NSP@MAF-6 composites. **(c)** PXRD patterns of pristine ZIF-11 and NSP@ZIF-11 composites. **(d)** PXRD patterns of pristine ZIF-11 and 8NSP@ZIF-11 composites. **(e)** PXRD patterns of pristine ZIF-71 and NSP@ZIF-71 composites. **(e)** PXRD patterns of pristine ZIF-71 and 8NSP@ZIF-71 composites.

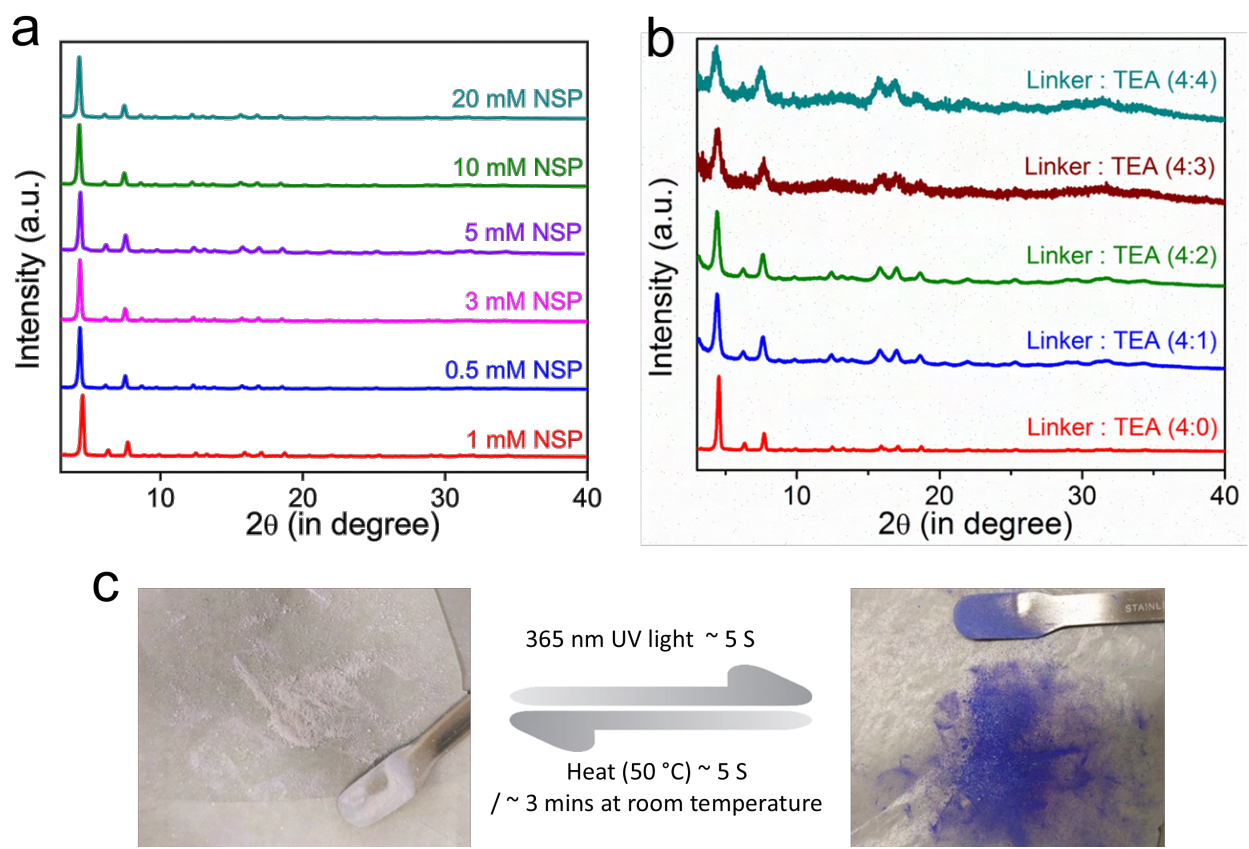

**Fig. S8. (a)** PXRD patterns of different loading of NSP into the ZIF-71 cavity. **(b)** PXRD patterns of nanosized NSP@ZIF-71(20) composites. **(c)** Photographs portraying the photochromic color change upon irradiation by 365 nm UV light of NSP@MOF(20) composites, before returning to its original color under room temperature (3 mins) or when heated at 50 °C (5 s).

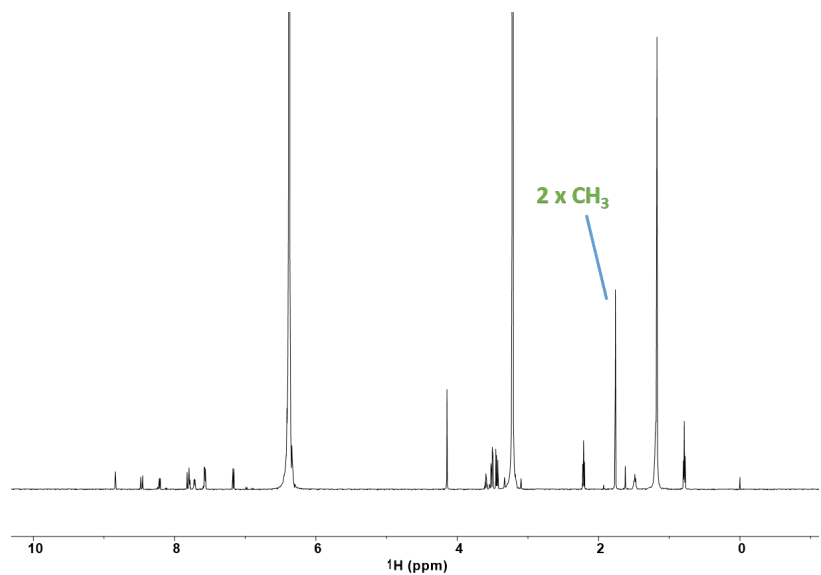

**Fig. S9.** ¹H NMR of the NSP molecule.

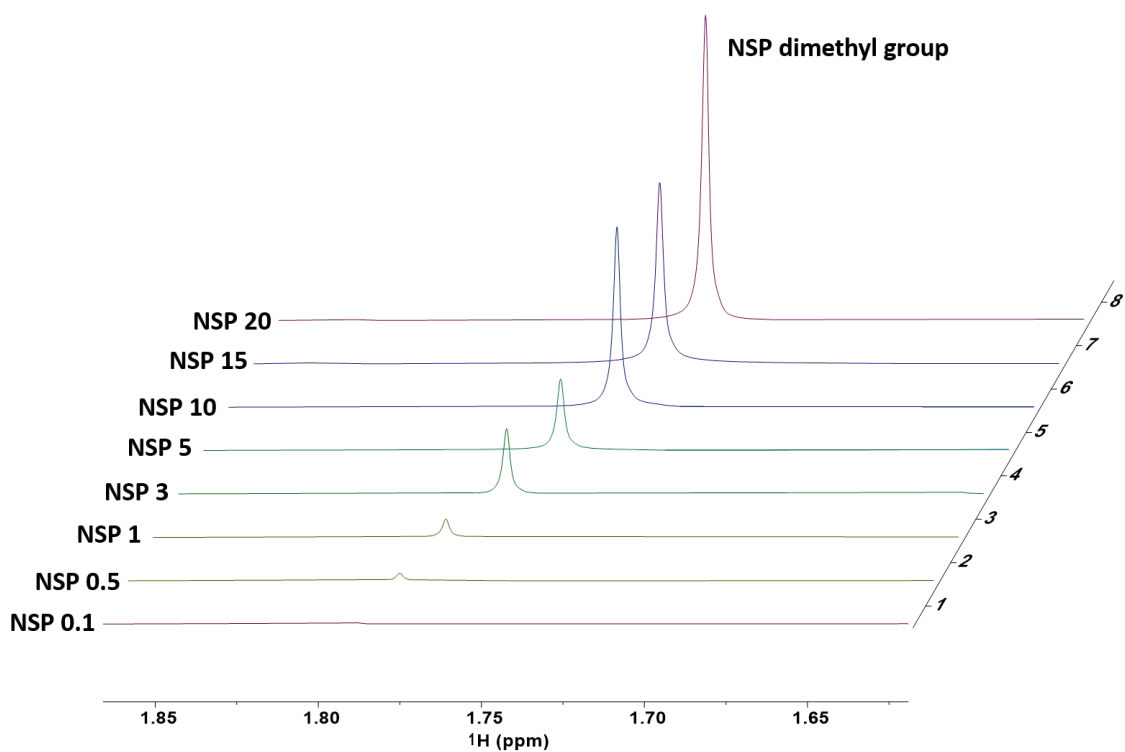

**Fig. S10.** Solution ¹H NMR of various NSP@ZIF-71 composites where the guest loading of the NSP molecules was different. The intensity of the characteristic peak for NSP increases with increasing the amount of guest loading used in the synthesis.

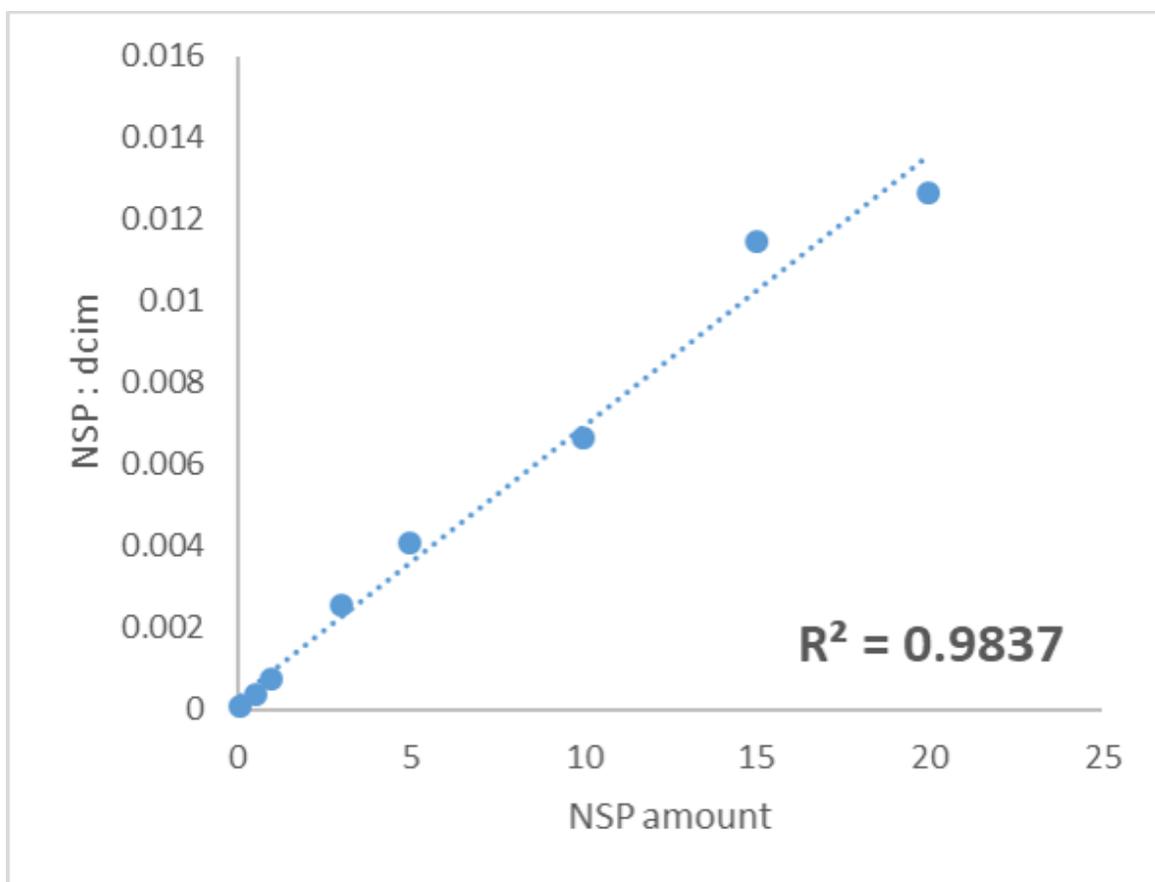

**Fig. S11.** Ratio of NSP : dcIm increases with increasing amount of NSP introduced during the synthesis step. The loading of NSP guest into the ZIF-71 MOF host follows a linear fit with concentration of NSP.

**Table S1.** Calculation of loading % of NSP in the ZIF-71 host. The ratio of NSP : ZIF-71 was calculated from the integral ratio of the peak corresponding to the 6 protons of the dimethyl group (NSP) attached to the pyrrole ring and the peak corresponding to the single proton of the dcIm imidazole ring. Considering one cage contains 24 dcIm ligands, the loading amount was obtained by NSP : dcIm multiplied by 24. The amount of NSP per cage of ZIF-71 is calculated by considering one NSP is present in one cage of the ZIF-71 cavity.

| NSP amount (mmol) | dcIm Area | NSP (2x $CH_3$ Area) | Normalised NSP area | dcIm : NSP | NSP : dcIm | Number of ZIF-71 cages per NSP guest | Loading of NSP in ZIF-71 (wt.%) |
|---|---|---|---|---|---|---|---|
| 20 | 1535831 | 116660.4 | 19443.4 | 79 | 0.01265986 | 13.2 | 30.4 |
| 15 | 1903780.2 | 130920.67 | 21820.11167 | 87 | 0.01146146 | 14.5 | 27.5 |
| 10 | 3193010.95 | 127580.57 | 21263.42833 | 150 | 0.00665936 | 25.0 | 16.0 |
| 5 | 2020881.47 | 49393.57 | 8232.261667 | 245 | 0.00407359 | 40.8 | 9.8 |
| 3 | 3180480.56 | 48477.64 | 8079.606667 | 394 | 0.00254037 | 65.7 | 6.1 |
| 1 | 2832369.68 | 12526.81 | 2087.801667 | 1357 | 0.00073712 | 226.2 | 1.8 |
| 0.5 | 2276661.77 | 5233.35 | 872.225 | 2610 | 0.00038316 | 435.0 | 1.0 |
| 0.1 | 2942809.51 | 1365.84 | 227.64 | 12927 | 7.73547E-05 | 2154.5 | 0.2 |

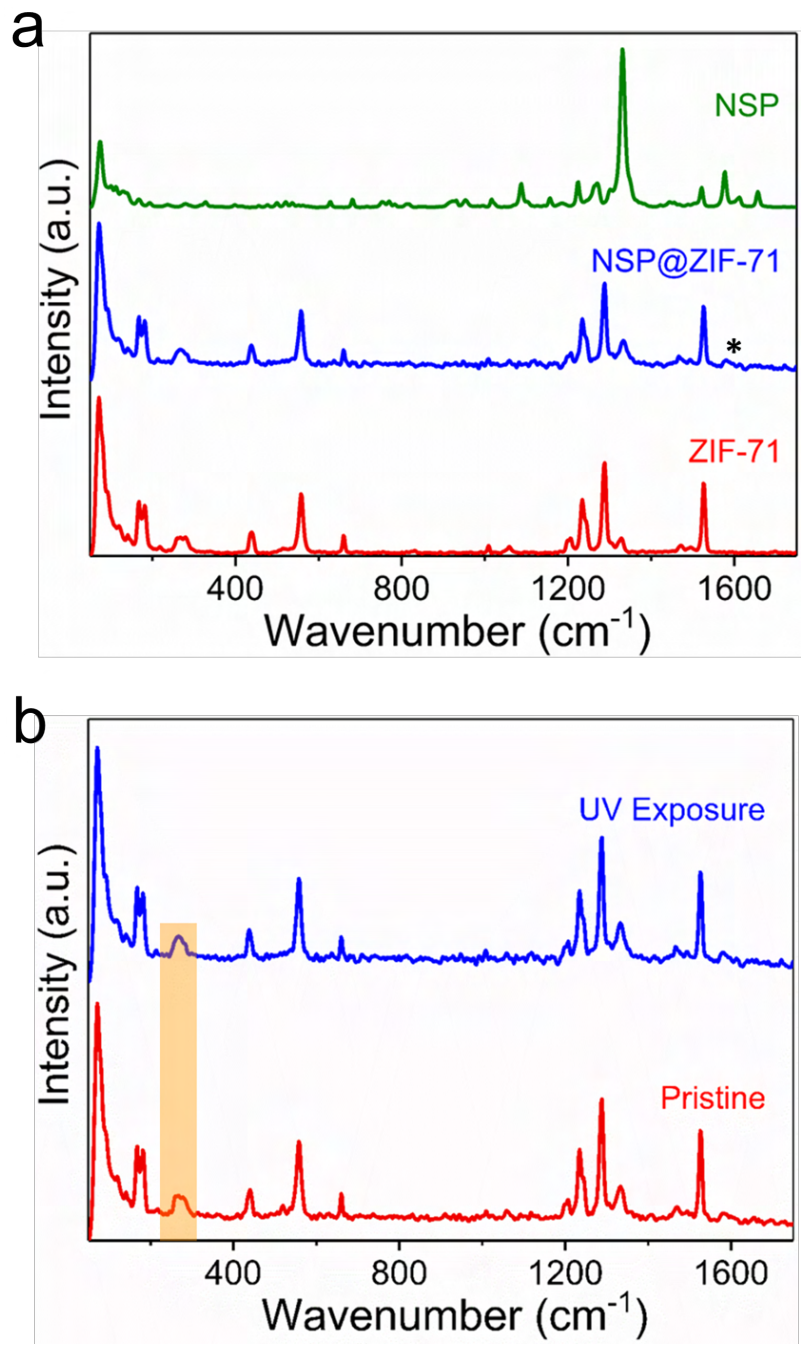

**Fig. S12. (a)** Raman spectra of ZIF-71, photochromic NSP@ZIF-71(20) composites and NSP. The (*) represents the signature peaks for the NSP molecule. **(b)** Raman spectra of NSP@ZIF-71(20) for pristine state and after UV exposure. The highlighted portion shows that there is little change in the Raman spectra upon UV exposure.

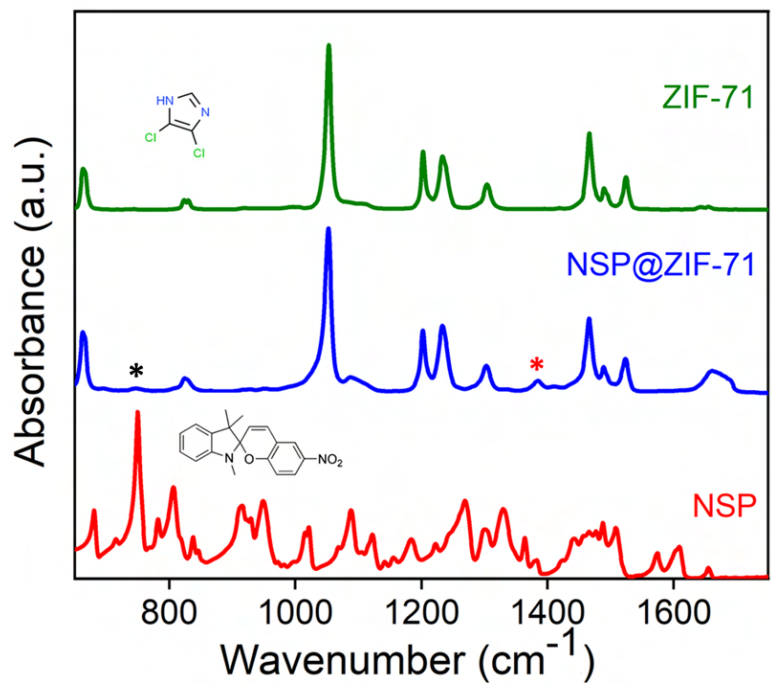

**Fig. S13.** ATR-FTIR spectra of NSP, photochromic NSP@ZIF-71(20) composite and ZIF-71. (*) represents the signature vibrational band for the NSP guest.

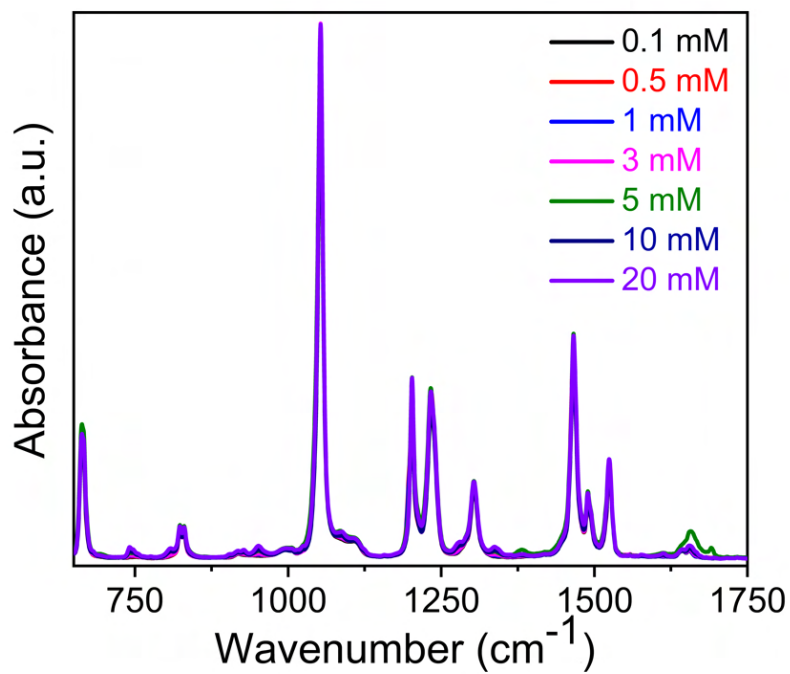

**Fig. S14.** ATR-FTIR spectra of different loading of NSP inside the ZIF-71 MOF.

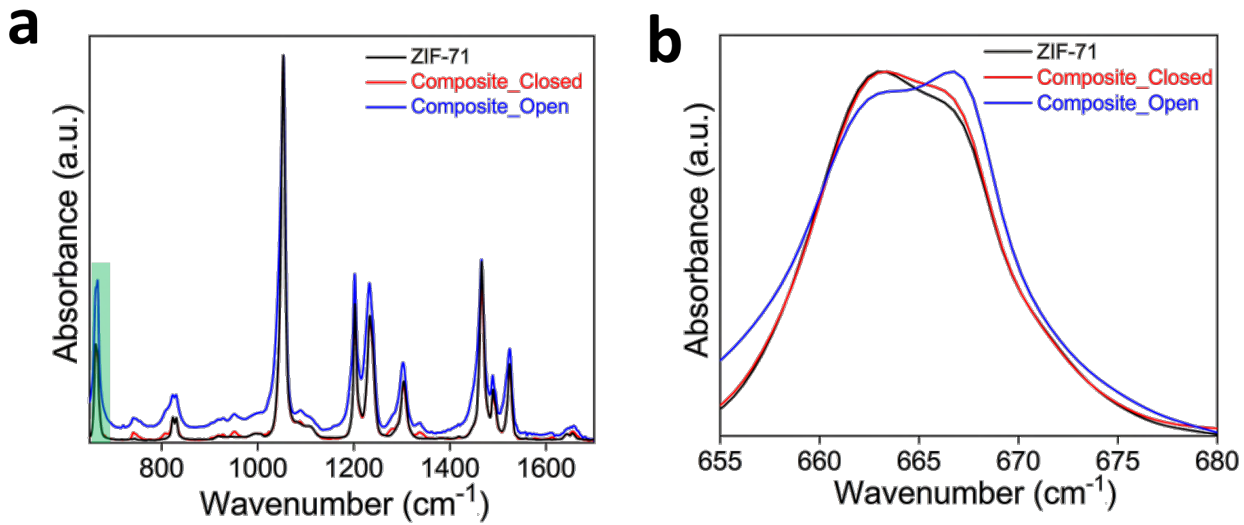

**Fig. S15. (a)** ATR-FTIR spectra of ZIF-71, closed and open form of NSP@ZIF-71(20). **(b)** Enlarged view of highlighted portion of figure a.

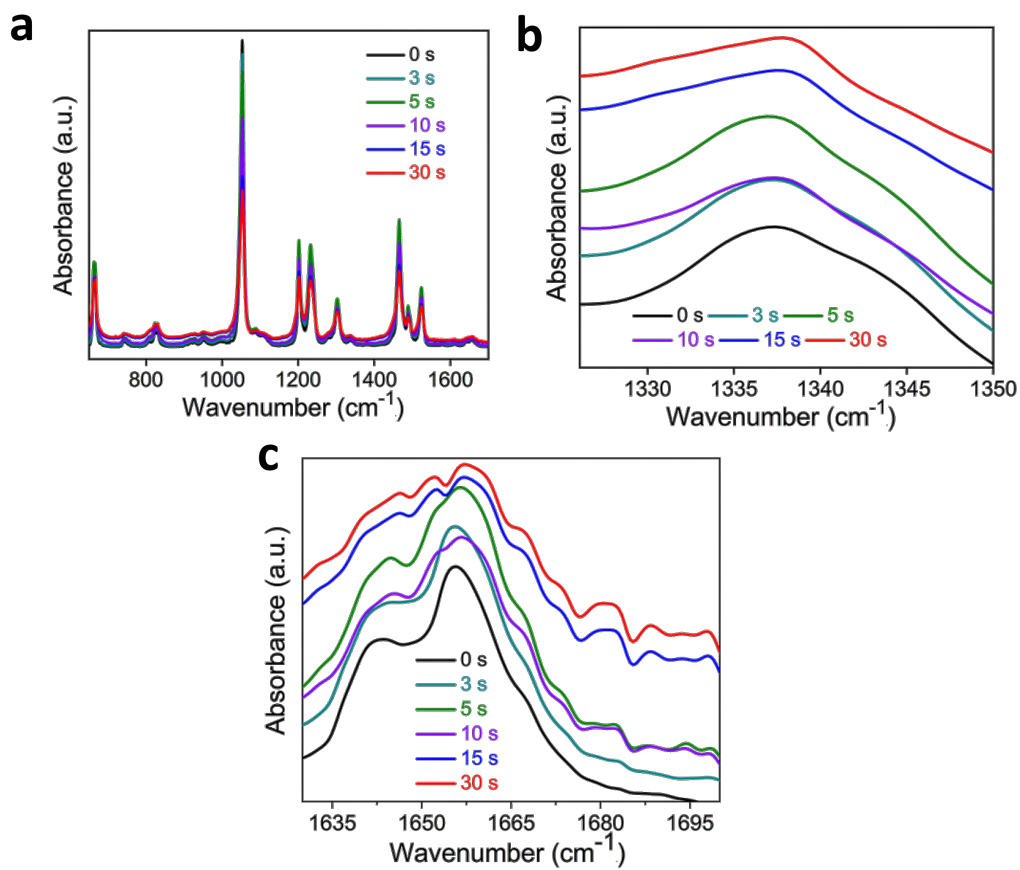

**Fig. S16. (a)** ATR-FTIR spectra of the NSP@ZIF-71(20) powder over time upon exposure to the 365 nm UV lamp under room temperature. **(b)** Enlarged view of 1337 cm$^{-1}$ peak of NSP@ZIF-71(20). **(c)** Enlarged view of 1655 cm$^{-1}$ peak of NSP@ZIF-71(20).

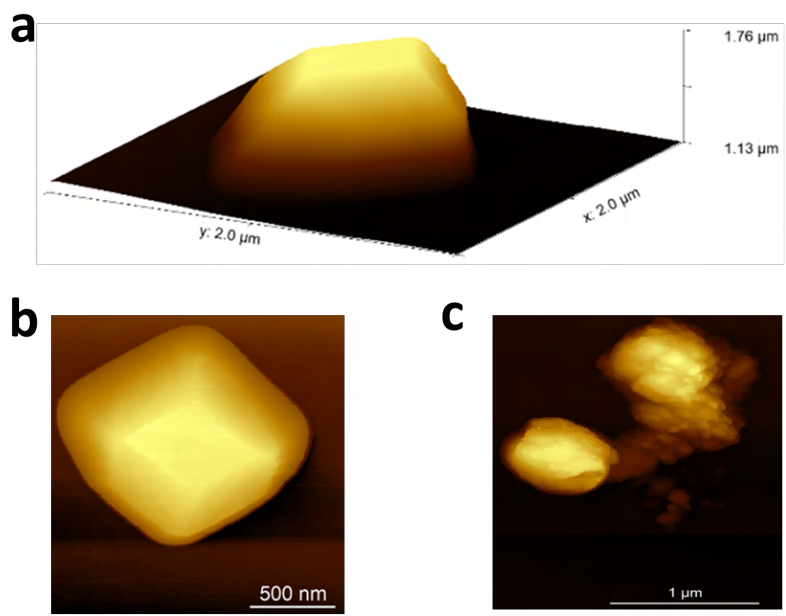

**Fig. S17.** AFM images of different crystal morphologies of the NSP@ZIF-71(20) composites. **(a, b)** Height topography of a micron-sized crystal of NSP@ZIF-71(20), designated as NSP@ZIF-71(20)_SC. The nominal size is ~ 1400-1700 nm. **(c)** Size of the nanocrystal of NSP@ZIF-71(20), designated as NSP@ZIF-71(20)_NC. The nominal size ~ 30-50 nm. The AFM images show the particle size of NSP@ZIF-71(20) may vary from 30 nm to 1700 nm.

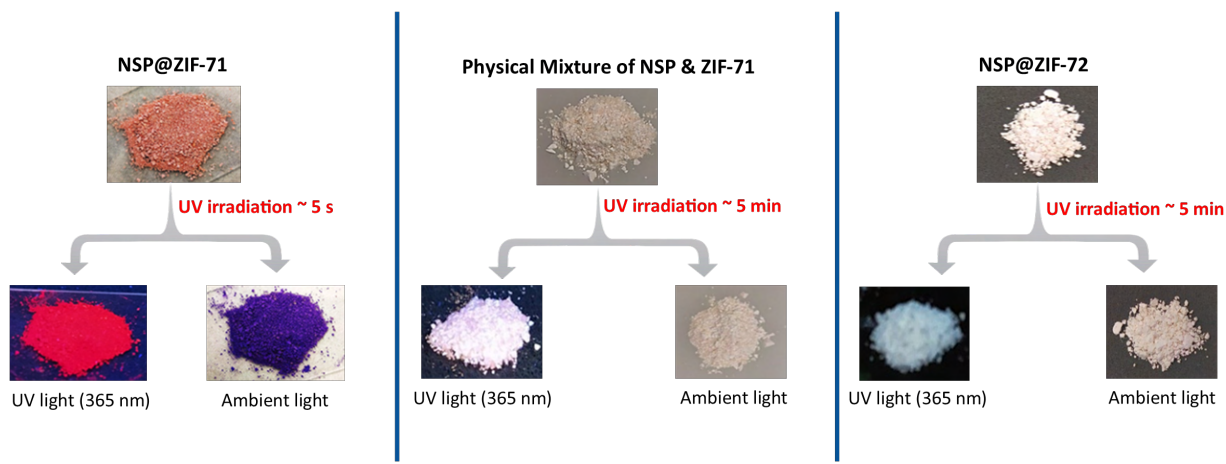

**Fig. S18.** Photographs of NSP@ZIF-71(20), physical mixing of the NSP and ZIF-71, and NSP@ZIF-72(20). The physically-mixed powders of NSP and ZIF-71, as a control sample, did not exhibit color changing phenomena upon UV irradiation. This result indirectly demonstrates that NSP is not present on the outer surface of the frameworks of NSP@ZIF-71(20). The photochromic composite did not exhibit any photoisomerization upon transforming porous the ZIF-71 to nonporous ZIF-72 host (NSP@ZIF-72), confirming that the pore within the host MOF framework is crucial for the facile isomerization of NSP.

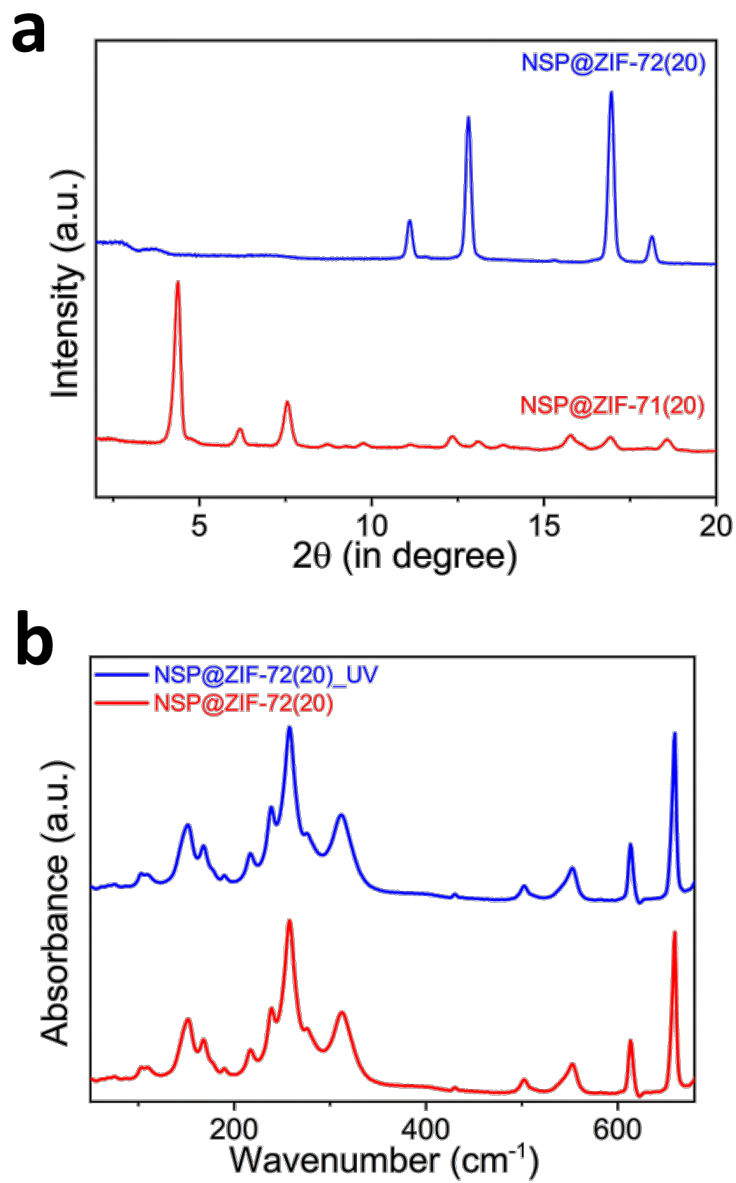

**Fig. S19. (a)** PXRD of NSP@ZIF-71(20) and NSP@ZIF-72(20). **(b)** Far-infrared (farIR) spectra of NSP@ZIF-72(20) before and after UV irradiation. The absence of any signature vibrational band in the terahertz region for an open form of NSP upon UV irradiation confirmed that the pore is crucial for the photoisomerization process.

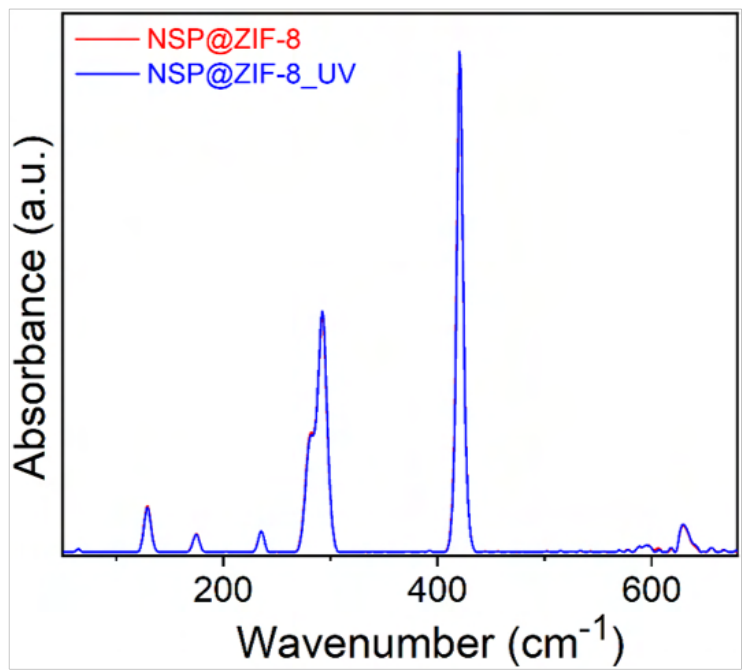

**Fig. S20.** FarIR spectra before and after 365 nm UV exposure of NSP@ZIF-8(20). The matching vibrational peaks indicate that the NSP guest is not switching in the ZIF-8 cavity, due to the smaller pore size of the host matrix.

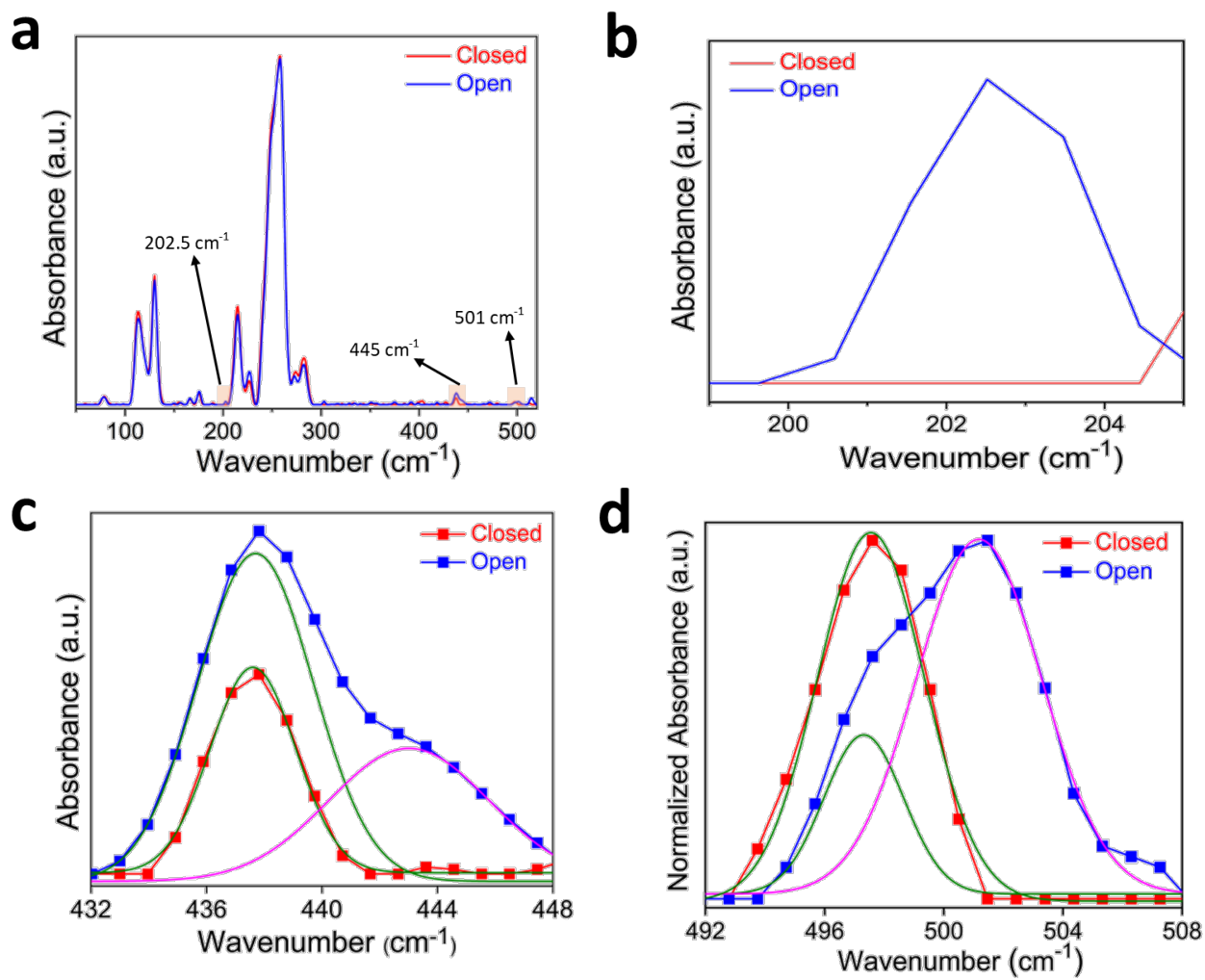

**Fig. S21.** FarIR spectra measured with synchrotron radiation for the open and closed forms of NSP@ZIF-71(20). The highlighted portions are presented as enlarged views in panels (b-d). **(a)** Full spectra, **(b)** 202.5 cm$^{-1}$ band, **(c)** deconvolution of 445 cm$^{-1}$ band, and **(d)** 501 cm$^{-1}$ band.

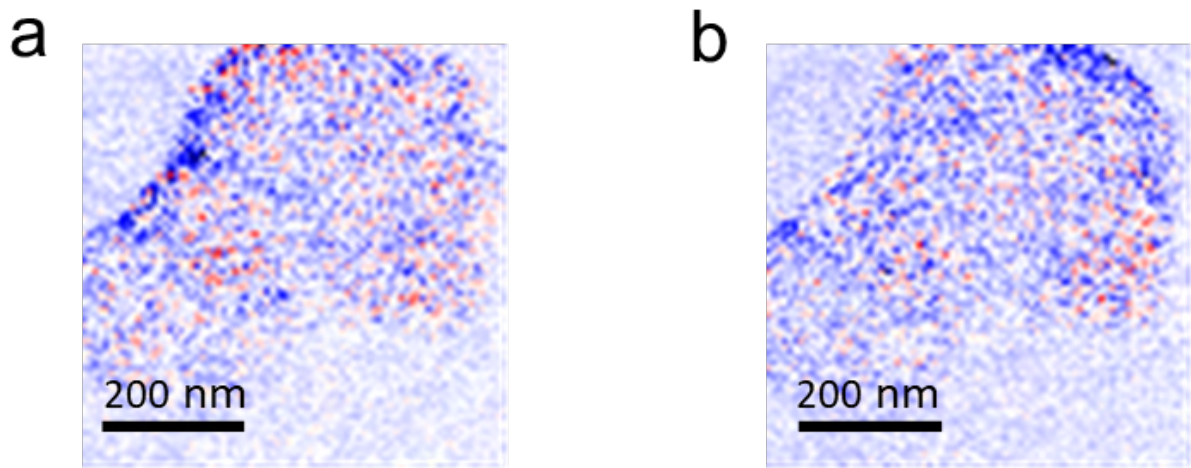

**Fig. S22.** The scattering-type scanning near-field optical microscopy (s-SNOM) optical phase images of pristine NSP@ZIF-71(20) for **(a)** forward scan and **(b)** backward scan.

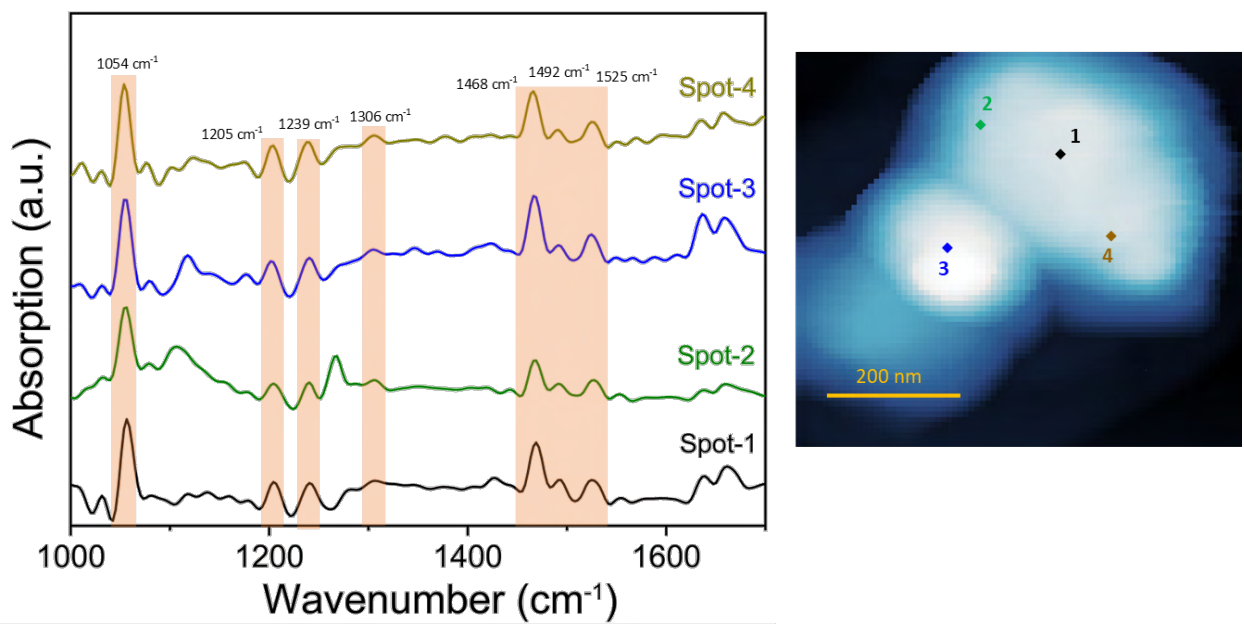

**Fig. S23.** Nano-FTIR spectra of the pristine NSP@ZIF-71(20) nanocrystals at the designated locations on the AFM image. The presence of signature vibrational peaks for the ZIF-71 host corresponds to 1054 cm$^{-1}$, 1205 cm$^{-1}$, 1239 cm$^{-1}$, 1306 cm$^{-1}$, 1468 cm$^{-1}$, 1492 cm$^{-1}$, 1525 cm$^{-1}$, confirming the retention of the host framework after NSP guest confinement (*5*).

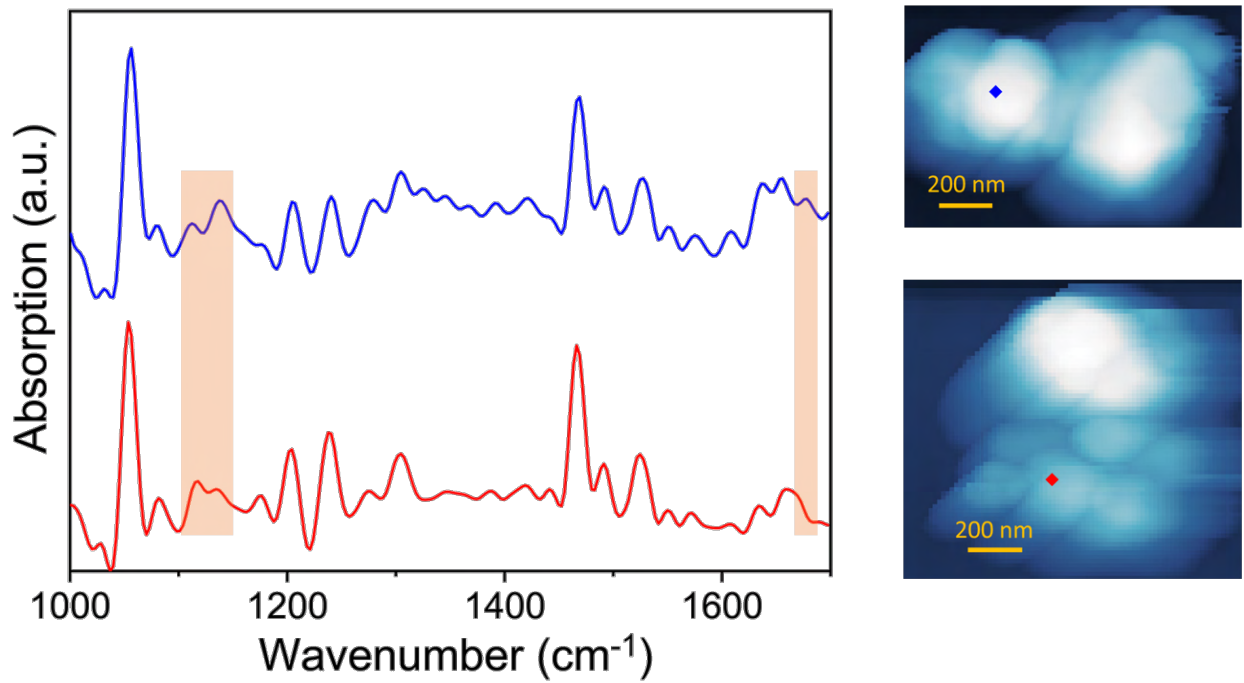

**Fig. S24.** Nano-FTIR spectra of the NSP@ZIF-71(20) nanocrystals at the designated locations on the AFM images upon exposure to UV light. The presence of characteristic vibrational peaks for the open form of NSP in different randomly selected nanocrystals of NSP@ZIF-71(20) confirms that the NSP guest is embedded into the ZIF-71 host matrix.

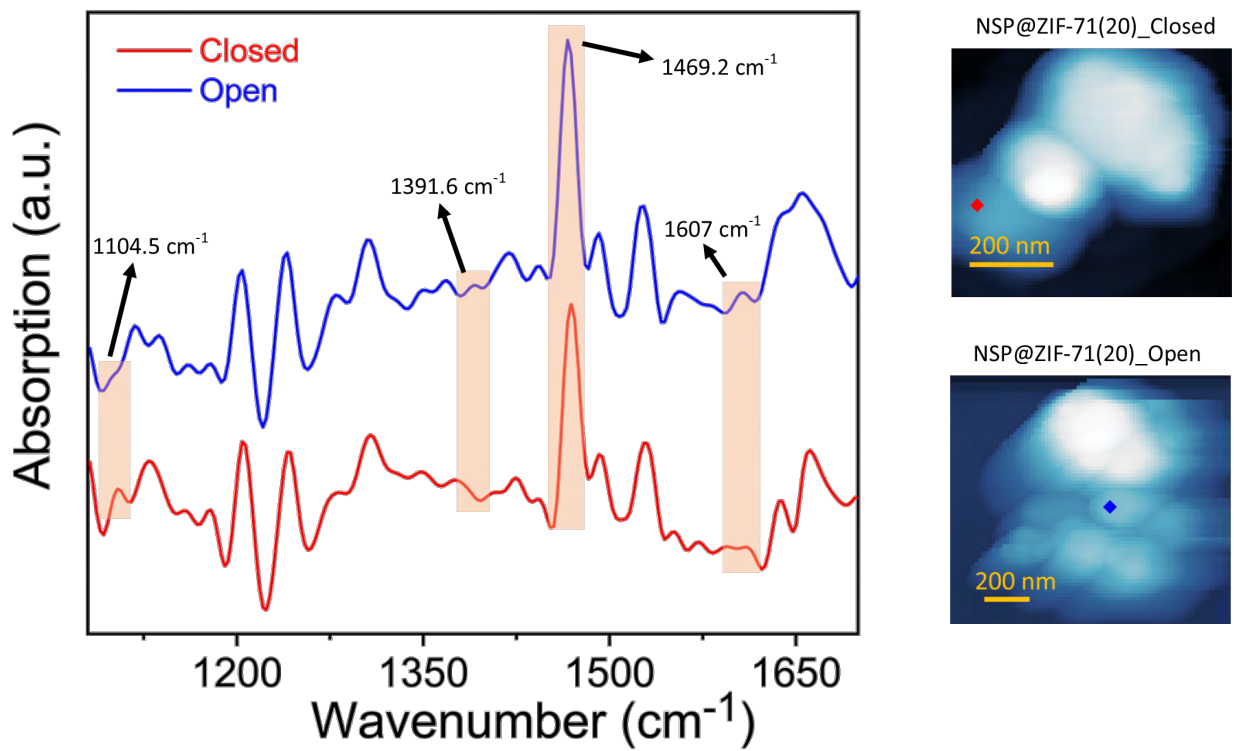

**Fig. S25.** Nano FTIR spectra of the closed and open form of NSP@ZIF-71(20) nanocrystals at the designated locations on the AFM images. The intensity of the vibrational peak at 1104.5 cm$^{-1}$ decreases in the open form due to cleavage of oxygen containing six-membered ring. Two new vibrational bands corresponding to 1391.6 cm$^{-1}$ and 1607 cm$^{-1}$ appear in the open form of the NSP@ZIF-71(20) composite, which is attributed to collective breathing of the open form of the NSP molecules. The characteristic vibrational band for host ZIF-71 at 1469.2 cm$^{-1}$ has redshifted to 1466.5 cm$^{-1}$ in the open form most probably due to the strong interaction between the polar zwitterionic form of NSP and the framework of host ZIF-71.

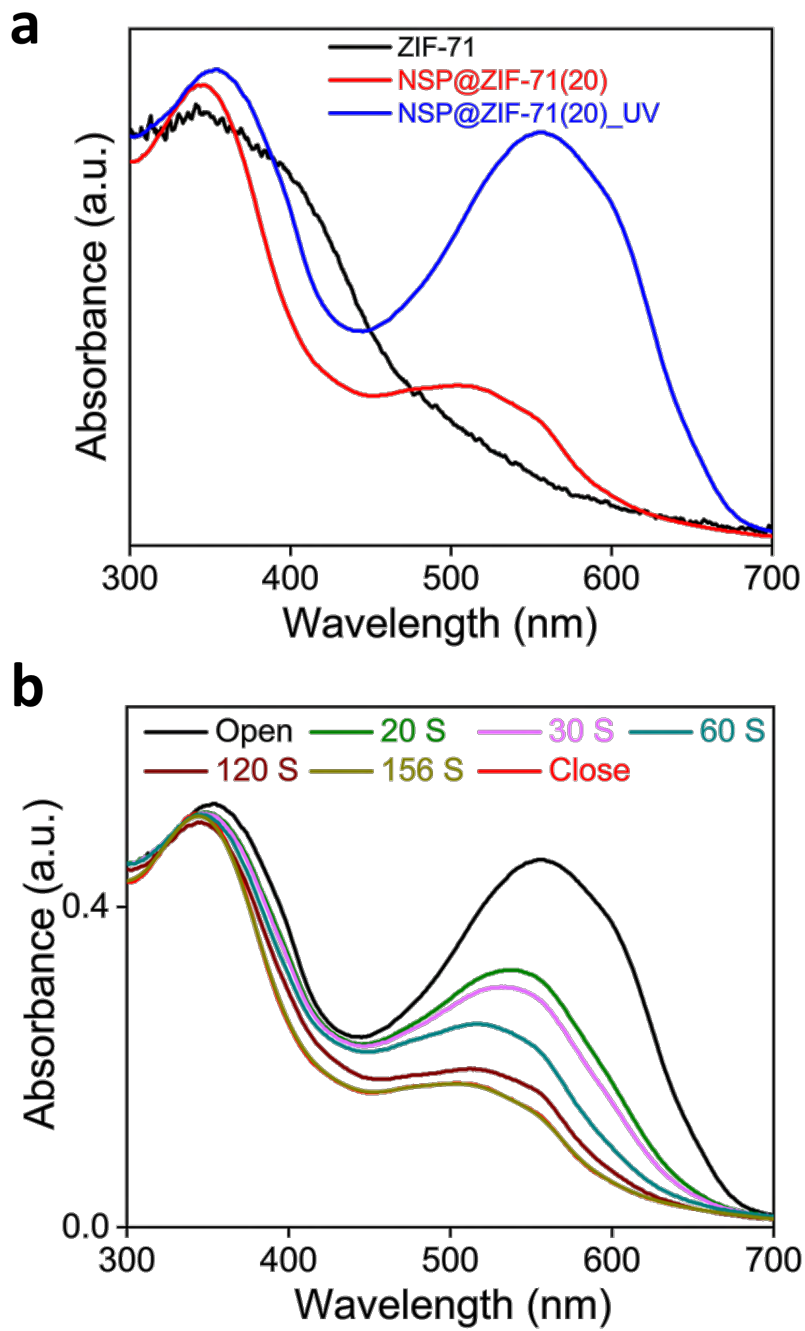

**Fig. S26. (a)** Diffused reflectance spectra (DRS) of ZIF-71 powder, before and after UV exposure of NSP@ZIF-71(20) powder. **(b)** DRS of NSP@ZIF-71(20) powder over time to determine the time required to revert back to its (pristine) closed form.

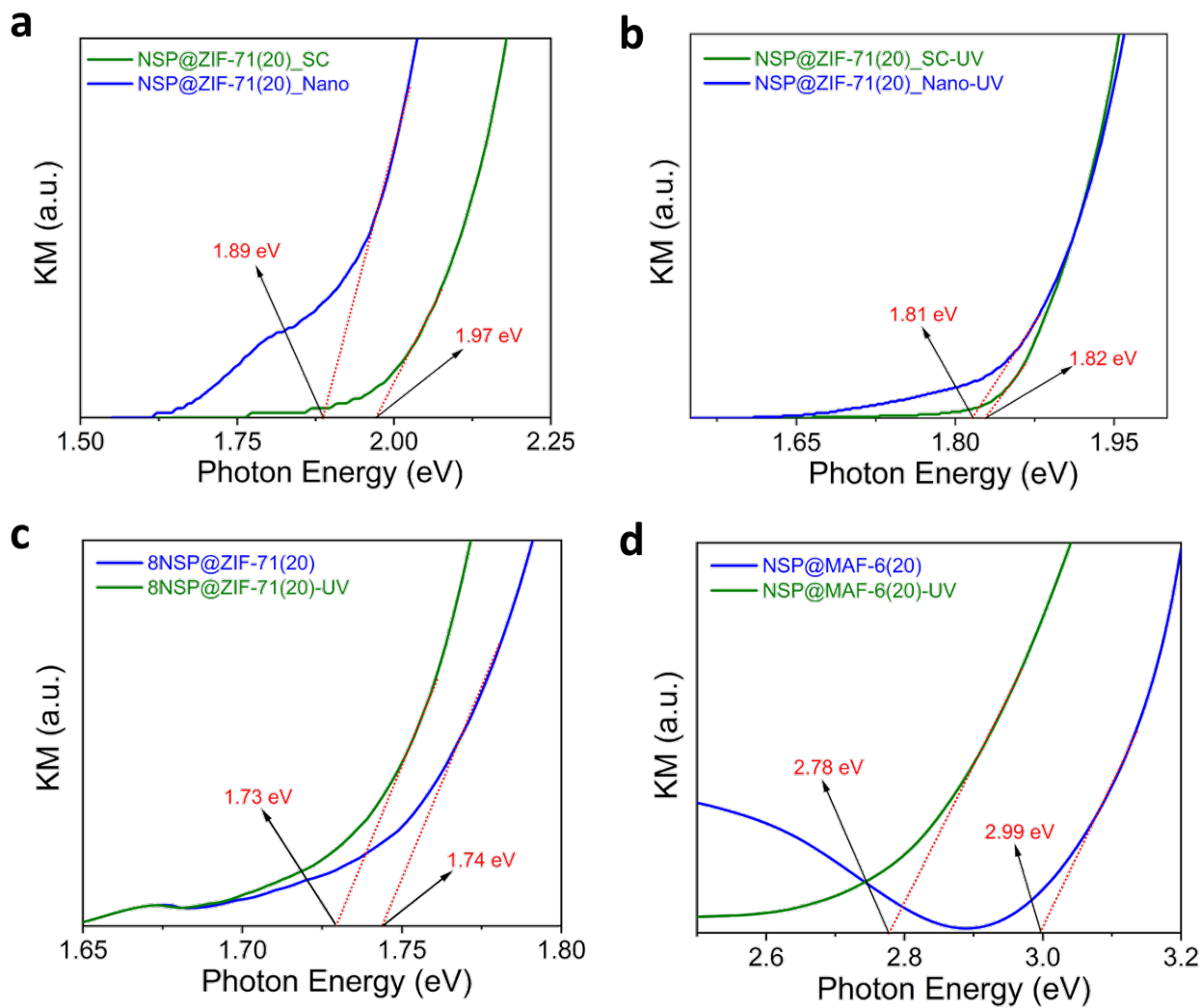

**Fig. S27.** Kubelka-Munk (KM) functions for estimating the optical band gaps based on the photon energy intercepts for NSP@ZIF-71(20)_SC and NSP@ZIF-71(20)_Nano **(a)** before UV exposure and **(b)** after UV exposure. Band gap for NSP@ZIF-71(20) and NSP@MAF-6(20) **(c)** before and **(d)** after UV exposure.

**Table S2.** The band gaps based of NSP@ZIF-71 composites in powder forms before and after UV exposure.

| Materials | Band gap (eV) | |
|---|---|---|
| | Before UV | After UV |
| NSP@ZIF-71(1) | 3.42 | 3.35 |
| NSP@ZIF-71(3) | 3.27 | 3.12 |
| NSP@ZIF-71(5) | 3.19 | 302 |
| NSP@ZIF-71(10) | 3.07 | 2.55 |
| NSP@ZIF-71(15) | 3.01 | 2.32 |
| NSP@ZIF-71(20) | 1.97 | 1.82 |

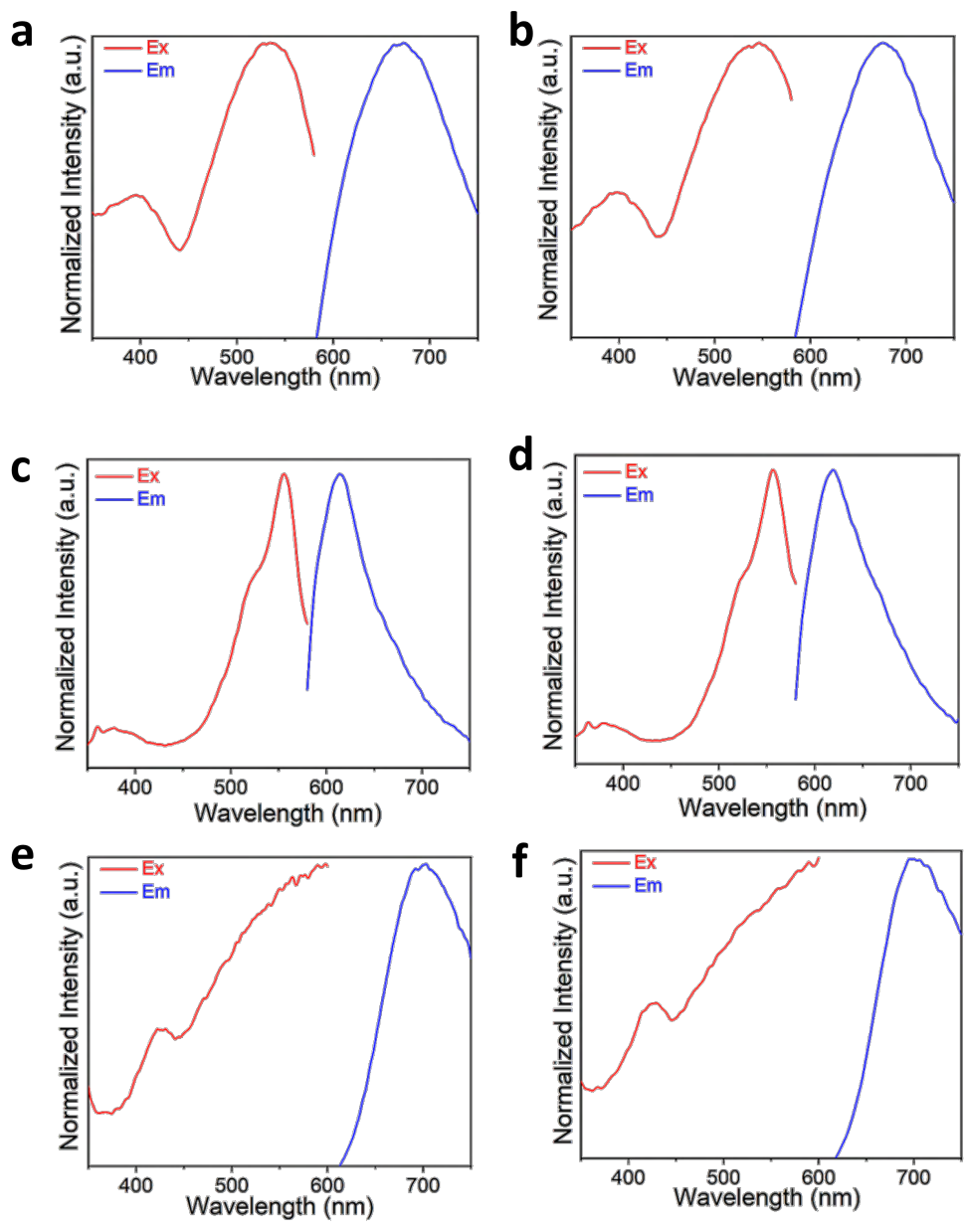

**Fig. S28.** The emission and excitation peaks for different composites before and after UV irradiation. **(a)** NSP@ZIF-11(20) before UV **(b)** NSP@ZIF-11(20) after UV **(c)** NSP@ZIF-11(20) before UV **(d)** NSP@MAF-6(20) after UV **(e)** 8NSP@ZIF-71(20) before UV **(f)** 8NSP@ZIF-71(20) after UV.

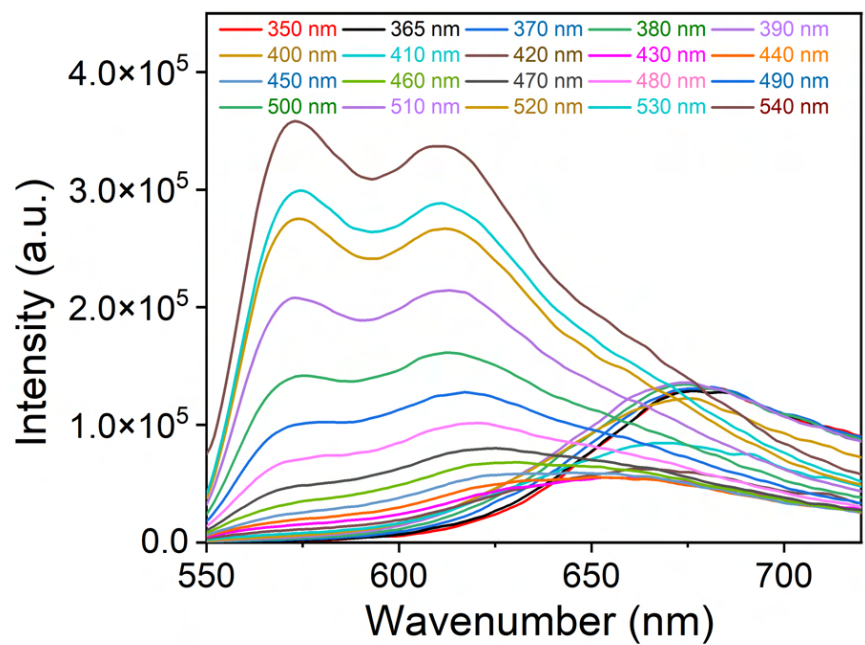

**Fig. S29.** Fluorescence emission spectra of NSP@ZIF-71(20) with different excitation wavelengths.

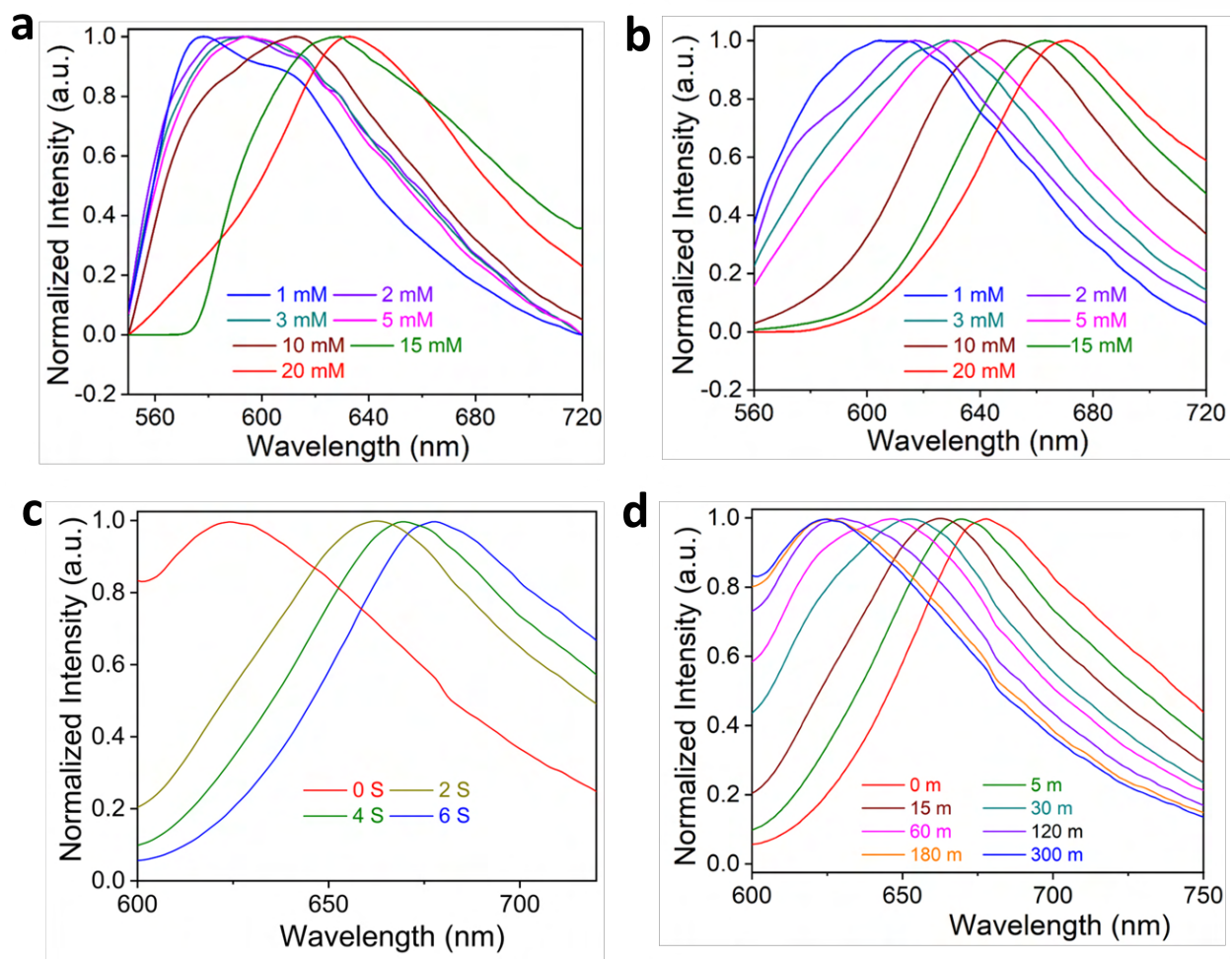

**Fig. S30. (a)** The emission spectra of different loading of NSP@ZIF-71 composites before UV exposure. **(b)** The emission spectra of different loading of NSP@ZIF-71 composites before UV exposure. **(c)** The time-dependent emission spectra of NSP@ZIF-71(20) composite powder over time subject to a 365-nm UV irradiation (6W) at room temperature**. (d)** Emission spectra of NSP@ZIF-71(20) powder to measure the time taken to return back to its pristine closed form at room temperature.

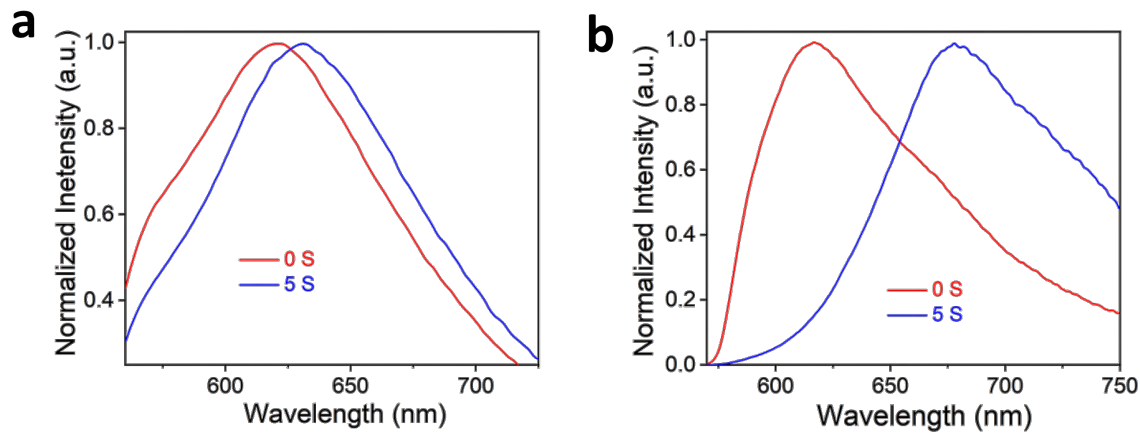

**Fig. S31. (a)** Emission spectra of nanocrystals of NSP@ZIF-71(20)_Nano composite before and after external 365 nm UV irradiation. **(b)** Emission spectra of micron-sized crystals of NSP@ZIF-71(20)_SC composite before and after external UV irradiation.

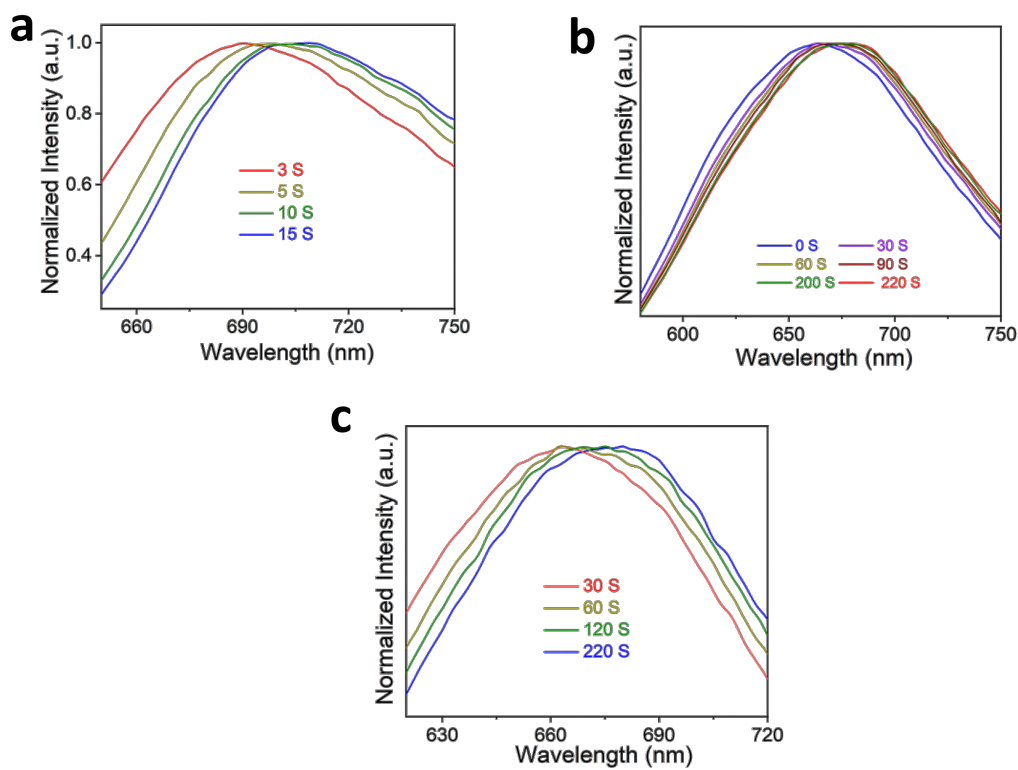

**Fig. S32. (a)** Time-dependent emission spectra of 8NSP@ZIF-71(20) composite powder upon the external 365 nm UV light (6W) irradiation at room temperature. **(b)** The time-dependent emission spectra of 8NSP@MAF-6(20) composite powder upon the external 365 nm UV light (6W) irradiation at room temperature. **(c)** The time-dependent emission spectra of 8NSP@ZIF-11(20) composite powder upon the external 365 nm UV light (6W) irradiation at room temperature.

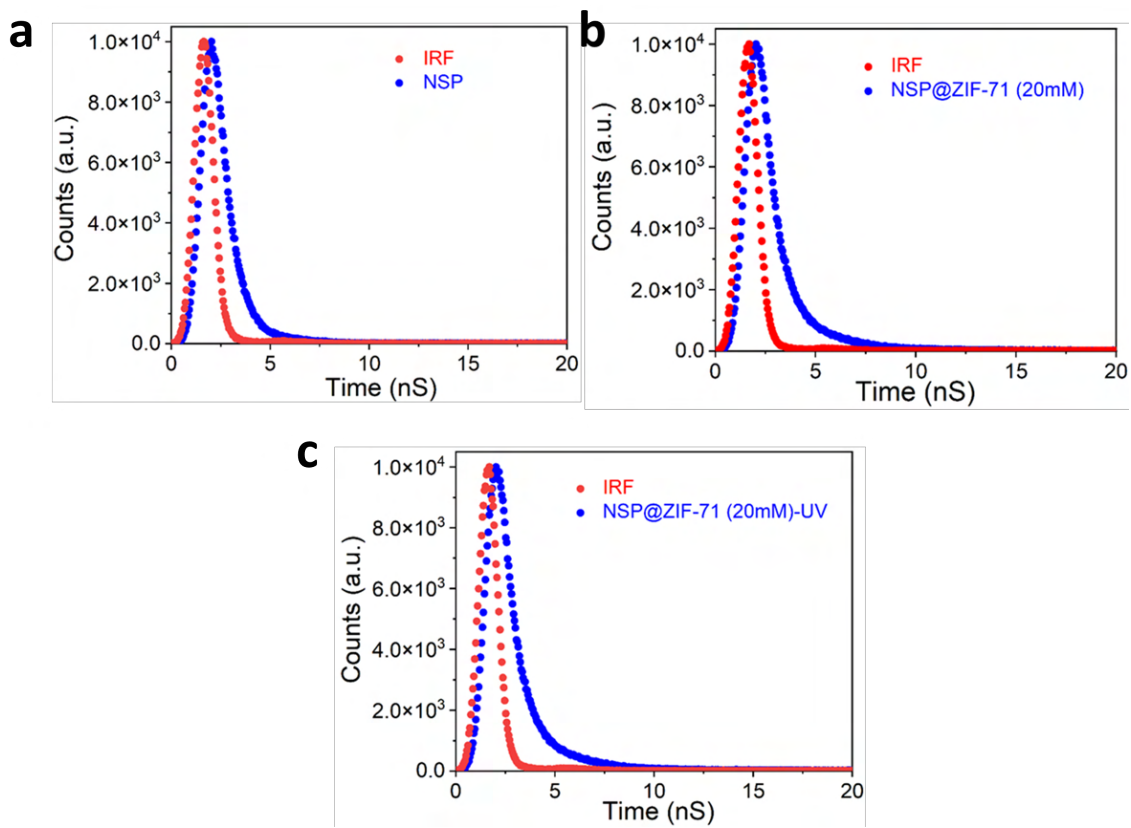

**Fig. S33.** Lifetime emission spectra showing the decays of **(a)** NSP, **(b)** NSP@ZIF-71(20) before UV and **(c)** NSP@ZIF-71(20) after UV irradiation for 5 seconds.

**Table S3.** Fitted fluorescence lifetime data for before and after UV irradiation of NSP and NSP@ZIF-71(20) composite, where UV irradiation time for NSP is 5 min while NSP@ZIF-71(20) is 5 s.

| Sample | Component 1 | | | Component 2 | | | Component 3 | | | $\chi^2$ |
|---|---|---|---|---|---|---|---|---|---|---|
| | $\tau_1$ | $\alpha_1$ | $c_1$ | $\tau_2$ | $\alpha_2$ | $c_2$ | $\tau_3$ | $\alpha_3$ | $c_3$ | |
| NSP | 0.25 | 0.045 | 16.27 | 0.76 | 0.075 | 82.65 | 4.19 | 0.0002 | 1.08 | 1.114 |
| NSP_UV | 0.284 | 0.050 | 20.56 | 0.75 | 0.072 | 78.24 | 3.5 | 0.0002 | 1.2 | 1.161 |
| NSP@ZIF-71(20) | 0.54 | 0.050 | 26.06 | 1.63 | 0.035 | 54.05 | 3.9 | 0.005 | 19.89 | 1.116 |
| NSP@ZIF-71(20)_UV | 0.43 | 0.079 | 42.71 | 1.23 | 0.032 | 49.1 | 3.4 | 0.001 | 8.19 | 1.042 |

**Table S4.** The absolute quantum yield (QY) of NSP@ZIF-71 composite with different NSP concentrations in powder forms.

| Materials | QY | |
|---|---|---|
| | Before UV | After UV |
| 0.5 mM | 42.33 | 14.68 |
| 1 mM | 37.43 | 11.98 |
| 3 mM | 33.41 | 6.78 |
| 5 mM | 30.41 | 5.63 |
| 10 mM | 7.86 | 5.37 |
| 15 mM | 7.83 | 4.85 |
| 20 mM | 4.76 | 2.32 |
| 30 mM | 2.41 | 1.03 |

**Table S5.** The absolute QY of NSP@ZIF-71(20) composite (powder samples) with different reaction time intervals.

| NSP@ZIF-71(20) | QY | |
|---|---|---|
| | **Before UV** | **After UV** |
| 24 h | 5.53 | 2.54 |
| 36 h | 5.33 | 2.24 |
| 48 h | 5.07 | 2.17 |
| 72 h | 4.76 | 2.32 |

**Table S6.** The absolute QY of different crystal size of the NSP@ZIF-71(20) composites in powder forms.

| NSP@ZIF-71(20) | PLQY | |
| --- | --- | --- |
| | Before UV | After UV |
| Single Crystal (SC) | 4.76 | 2.32 |
| Nano Crystal (Nano) | 16.50 | 6.03 |

**Table S7.** The absolute QY of different photochromic composites in powder forms.

| Photochromic materials | PLQY | |
| --- | --- | --- |
| | Before UV | After UV |
| NSP@ZIF-71(20)_Nano | 16.50 | 6.03 |
| NSP@MAF-6(20) | 2.36 | 0.94 |
| NSP@ZIF-11(20) | 1.03 | 0.57 |
| 8NSP@ZIF-71(20) | 1.49 | 1.16 |
| 8NSP@ZIF-11(20) | 8.54 | 7.22 |

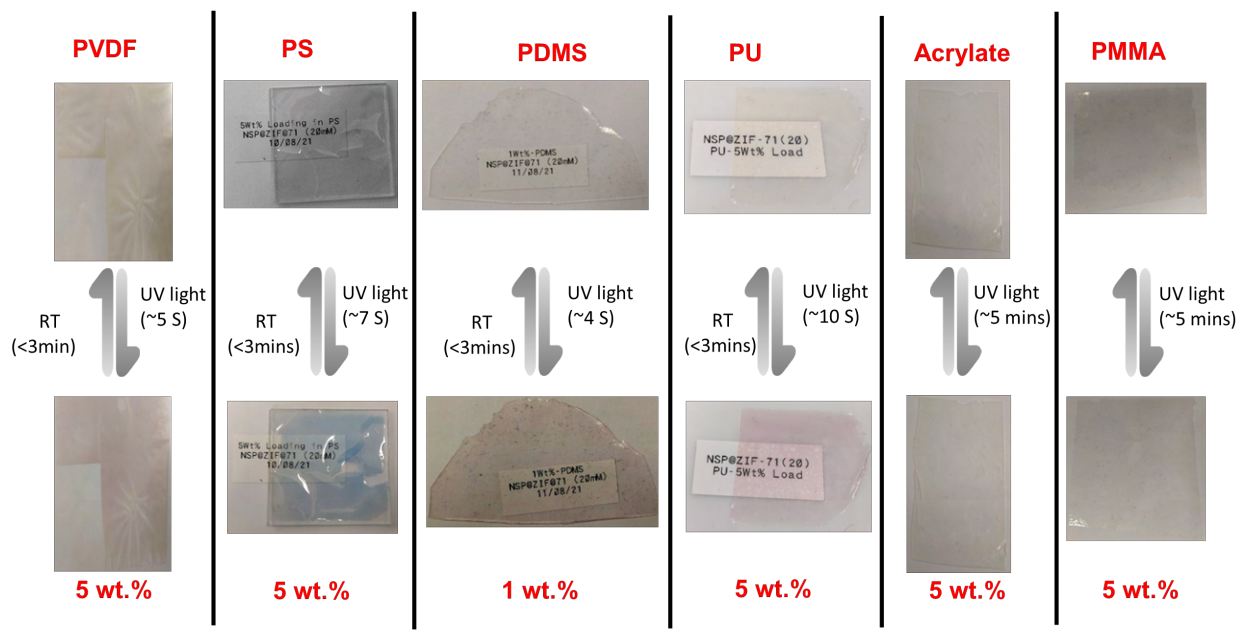

**Fig. S34.** Photographs of different thin films with 5 wt.% loading of NSP@ZIF-71(20) except PDMS film where 1 wt.% of NSP@ZIF-71(20) was loaded. All the films were prepared by using doctor blade technique, where 0.2 mm fixed blade size used for the fabrication of thin films. The thickness of the films falls in the range of 12 to 20 μm. The PS and PDMS films are more optically clear compared to other films. The pictorial images showing PVDF, PS and PDMS switching color upon irradiation of 365 nm (6W) UV light, while acrylate and PMMA films are not exhibiting any color change even after UV exposure for 5 min.

| PVDF alone (0 wt.%) | 16.7 wt.% NSP@ZIF-71 (20) | 5 wt.% NSP@ZIF-71 (20) | 3 wt.% NSP@ZIF-71 (20) |
| --- | --- | --- | --- |
| Thickness: 0.012 mm | Thickness: 0.022 mm | Thickness: 0.018 mm | Thickness: 0.023 mm |
| 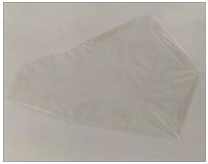 | 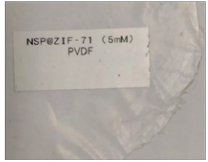 | 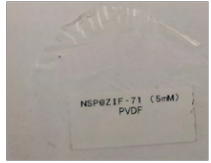 | 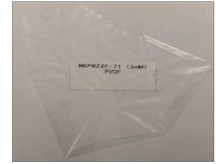 |
| 2 wt.% NSP@ZIF-71 (20) | 1 wt.% NSP@ZIF-71 (20) | 0.5 wt.% NSP@ZIF-71 (20) | 0.1 wt.% NSP@ZIF-71 (20) |
| Thickness: 0.018 mm | Thickness: 0.016 mm | Thickness: 0.018 mm | Thickness: 0.015 mm |
| 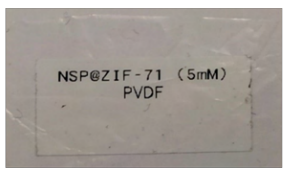 | 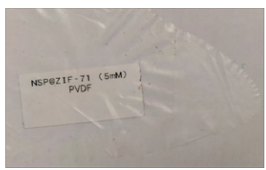 | 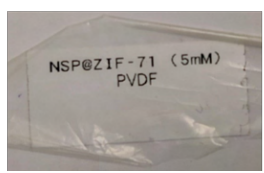 | 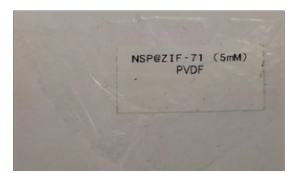 |

**Fig. S35.** Photographs of PVDF thin films with different wt.% loading of NSP@ZIF-71(20). All films were prepared by using doctor blade technique, where a 0.2-mm fixed blade size was used for the fabrication of thin films. The thickness of the films falls in the range of 12 to 20 μm.

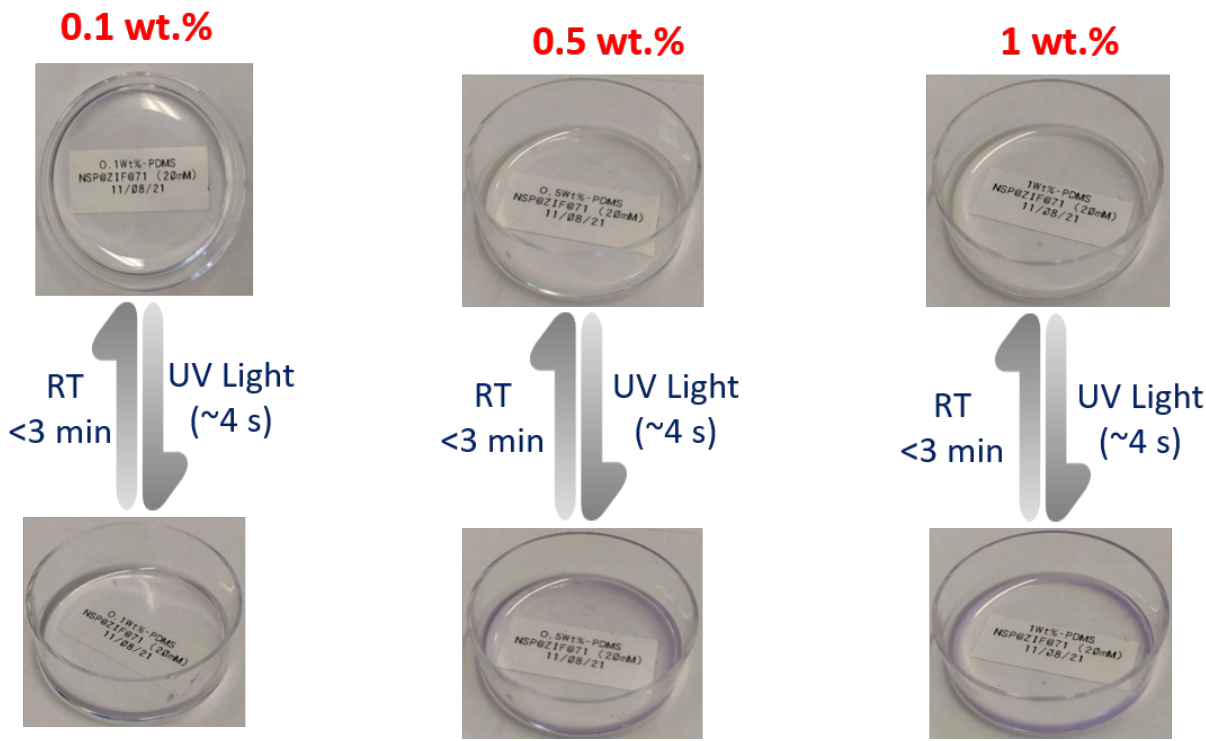

**Fig. S36.** Photographs of PDMS thin films with different wt.% loading of NSP@ZIF-71(20). The optically clear PDMS film changes color after being exposed to 365 nm (6 W) UV light for 4 seconds, returning to its original color after < 10 minutes. The thickness of the films falls in the range of 50 to 100 μm.

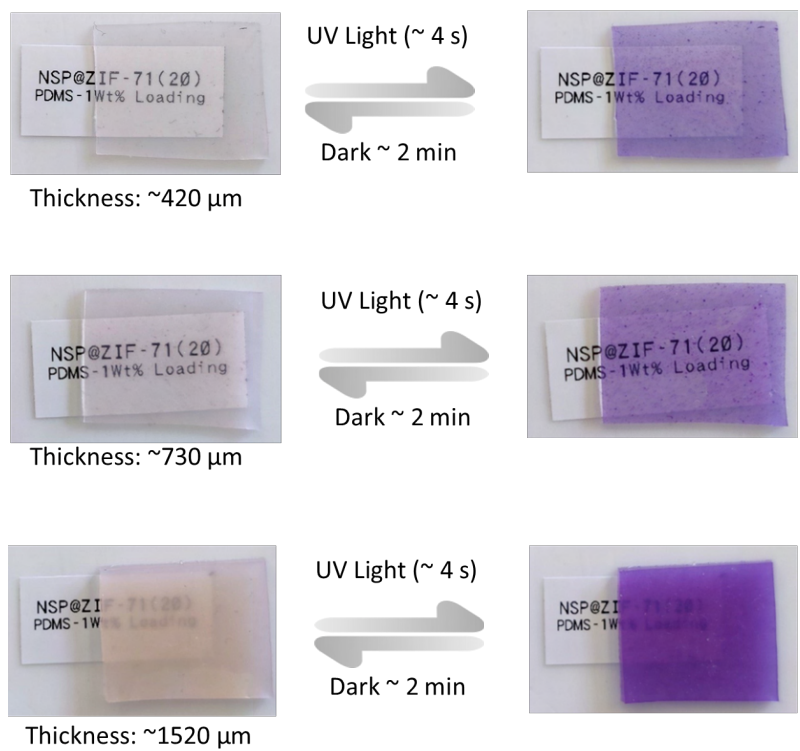

**Fig. S37.** Photographs of different thicknesses of PDMS film with 1 wt.% loading of NSP@ZIF-71(20). The optically clear PDMS film changes color after being exposed to 365 nm (6 W) UV light for 5 seconds, returning to its original color after < 3 minutes. The color switching intensity becomes more prominent with increasing thickness of the films where the filler loading was fixed at 1 wt.%.

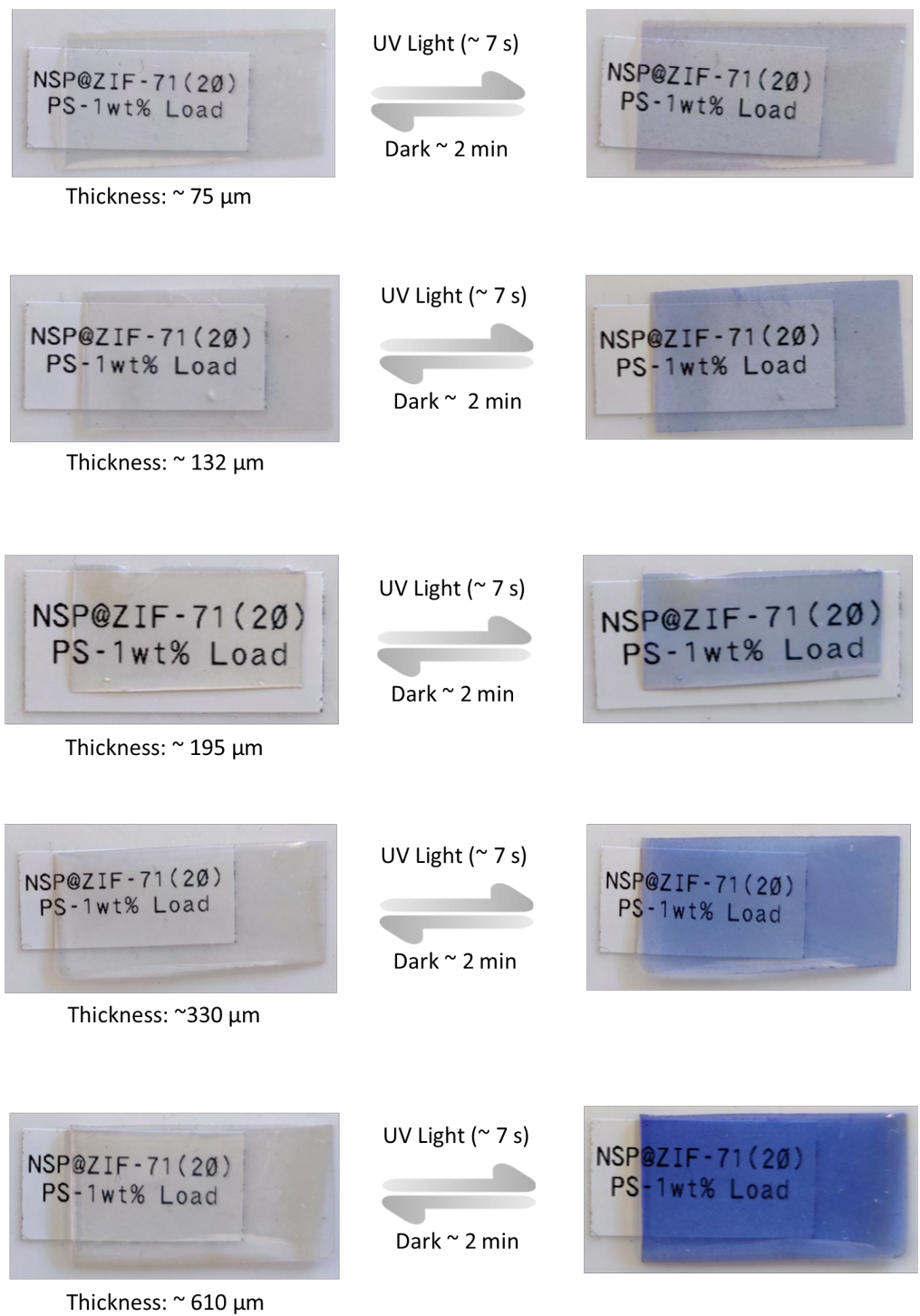

**Fig. S38.** Photographs of different thickness of PS film with 1 wt.% loading of NSP@ZIF-71(20). The optically clear PS film changes color after being exposed to 365 nm (6 W) UV light for 10 seconds, returning to its original color after < 3 minutes. The color switching phenomenon becomes more prominent with increasing thickness of the films where loading of the filler was fixed for each case.

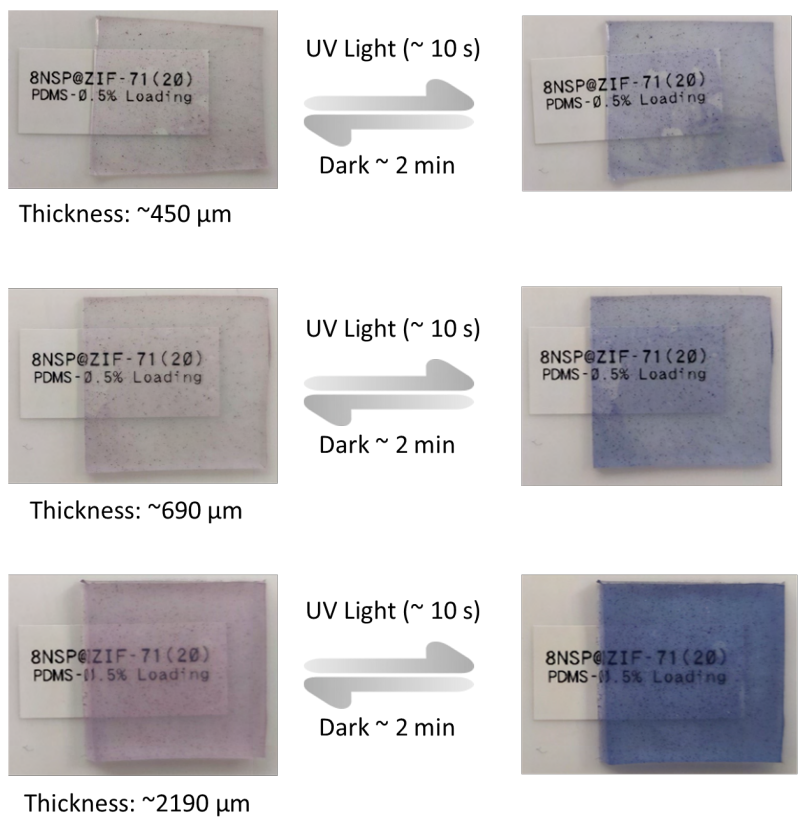

**Fig. S39.** Photographs of different thickness of PDMS film with 0.5 wt.% loading of 8NSP@ZIF-71(20). The optically clear PDMS film changes color after being exposed to 365 nm (6 W) UV light for 15 seconds, returning to its original color after < 10 minutes. The color switching phenomenon becomes more prominent with increasing thickness of the films where loading of the filler was fixed for each case.

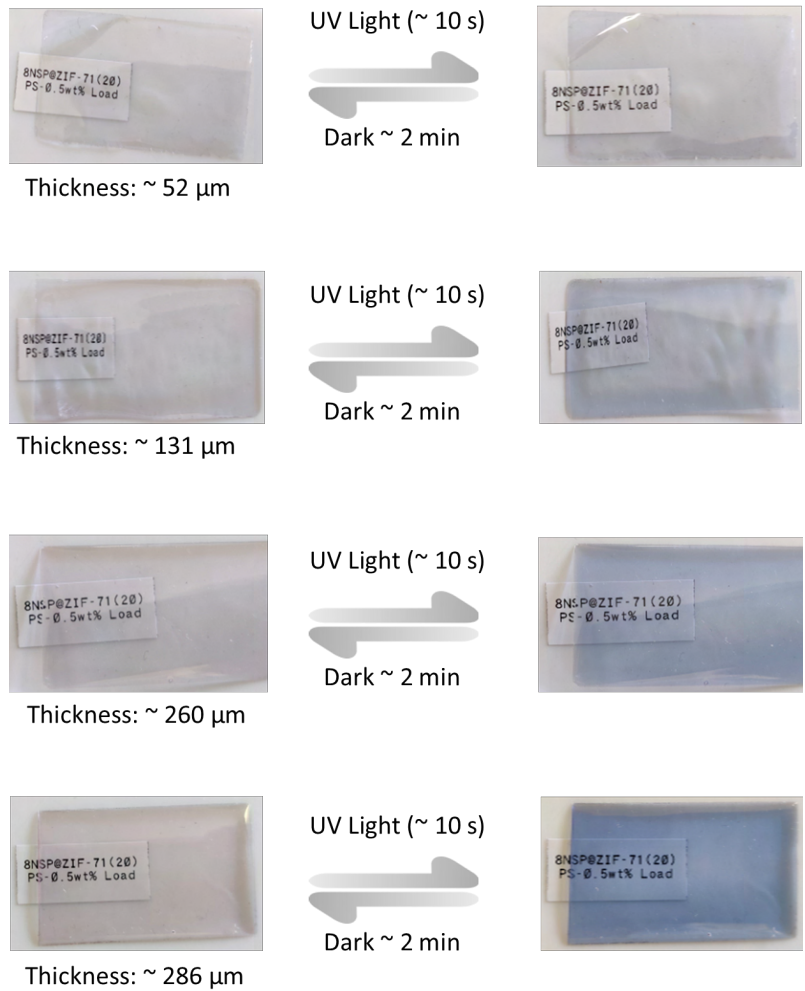

**Fig. S40.** Photographs of different thickness of PS film with 0.5 wt.% loading of 8NSP@ZIF-71(20). The optically transparent PS film changes color after being exposed to 365 nm (6W) UV light for 10 seconds, returning to its original color in less than 5 minutes. The color switching phenomenon becomes more prominent with increasing thickness of the films where loading of the filler was fixed for each case.

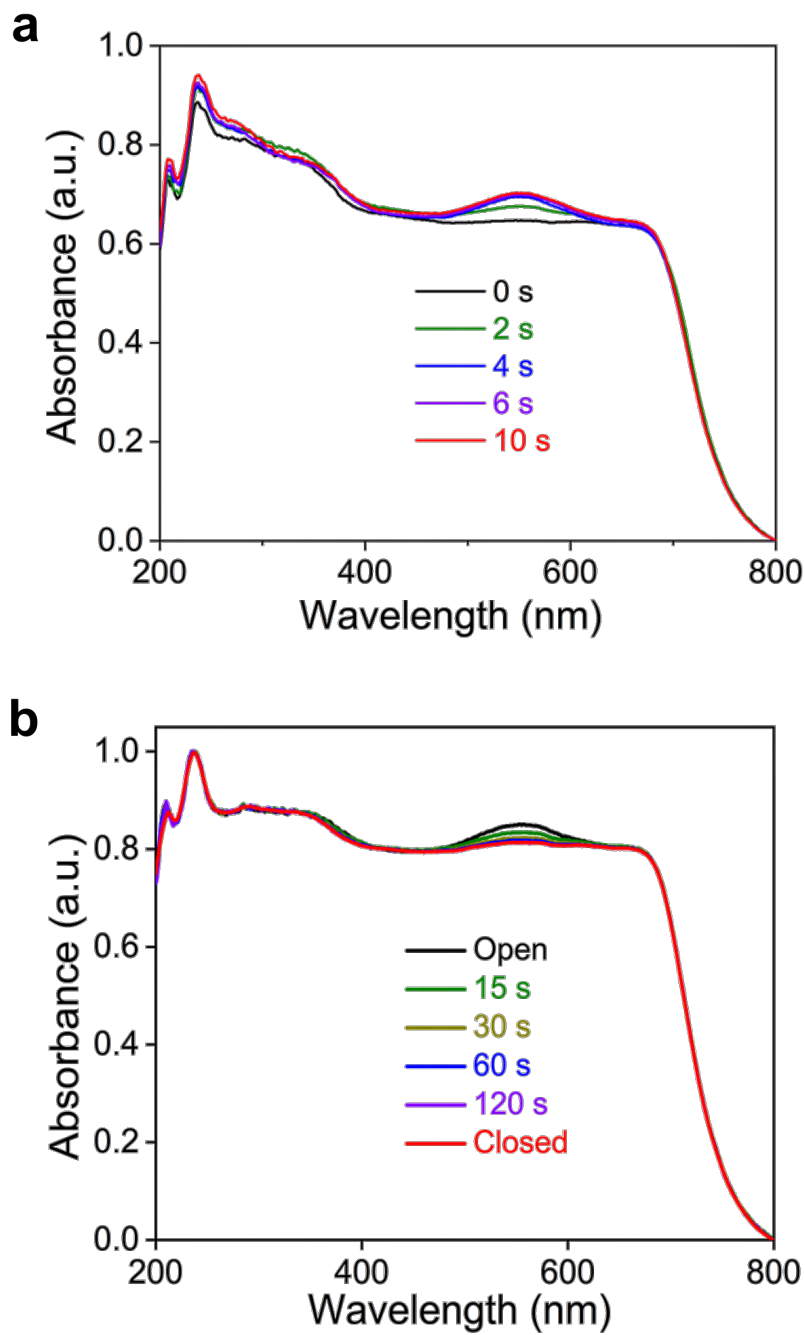

**Fig. S41.** Diffused reflectance spectra of 1 wt.% loading of the NSP@ZIF-71(20)/PDMS film. **(a)** Time-dependent UV irradiation to measure the duration of the color switching process. **(b)** Determination of the time taken to return back to its pristine closed form at room temperature.

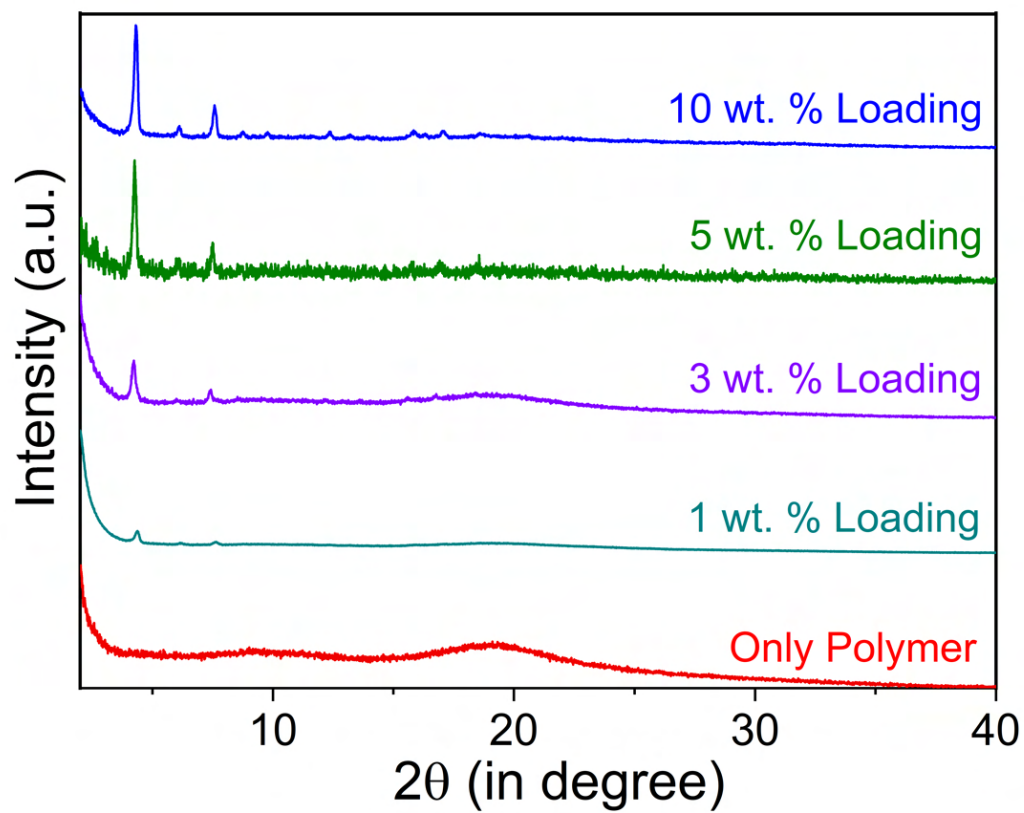

**Fig. S42.** PXRD patterns of the neat polystyrene (PS) polymer and different loading of NSP@ZIF-71(20) composite in the PS films.

**Table. S8.** Comparison of the photochromic features between this work and other related investigations where photochromic molecules are either NSP itself or NSP derivatives.

| Photochromic Material | Type of Materials | Response Time | Time to Revert Back | Refs |
|---|---|---|---|---|
| *NSP@ZIF-71(20)* | *Composite Powder* | *5 s* | *~2.6 min* | *This work* |
| NSP@Cage | Cage | 30 s | 6 h | *Nat. Commun.* **9**, 641 (2018) |
| Zn$_2$(DBTD)(TNDS) | MOF | 3 min | 3 min | *J. Am. Chem. Soc.* **141**, 5350 (2019) |
| PhotoPAF-3.6% | PAF (Porous aromatic framework) | 3 min | Several hours | *Nat. Commun.* **5**, 3588 (2014) |
| WO$_3$/Au film | Inorganic composite | 3 min | - | *Phys. Chem. Chem. Phys.* **4**, 1637 (2002) |
| SP@UiO-67-MOF | MOF | 5 min | - | *Angew. Chem. Int. Ed.* **58**, 1193 (2019) |
| WO$_3$ Nanosheet | Inorganic material | 5 min | 7 min | *J. Mater. Chem. C.* **3**, 7597 (2015) |
| Ag-TiO$_2$ film | Inorganic material | 5 min | ~ 12 hours | *Nat. Mater.* **2**, 29 (2003) |
| BSP/JUC-120 | MOF Film | 20 min | 2.4 h | *J. Mater. Chem.* **22**, 25019 (2012) |
| SP-h | Self-Assembled Monolayers (SAMs) | 20 min | 12 h | *Adv. Mater.* **31**, 1807831 (2019) |
| Lithopone | Inorganic compound | 30 min | Several hours | *Ind. Eng. Chem.* **21**, 348 (1929) |
| MC2⊂MB1 | Cage | 12 h | - | *J. Am. Chem. Soc.* **140**, 7952 (2018) |

**Table S9.** Comparison of the photochromic features between this work and other related investigations where the photochromic parts are different functional groups.

| Photochromic Material | Type of Materials | Photochromic Part | Response Time | Time to Revert Back | Refs |
|---|---|---|---|---|---|
| *NSP@ZIF-71(20)* | *Composite film* | *NSP* | *4 s* | *2 min* | *This work* |
| *NSP@ZIF-71(20)* | *Composite powder* | *NSP* | *5 s* | *2.6 min* | *This work* |
| NSP@Cage | Composite | NSP | 30 s | 6 h | *Nat. Commun.* **17**, 641 (2018) |
| DMOF | MOF | Diarylethene derivative | 1 min | - | *Angew. Chem. Int. Ed.* **53**, 9298 (2014) |
| Mg-NDI | MOF | NDI core | 1 min | 12 h | *Chem. Sci.* **7**, 2195 (2016) |
| PhotoPAF-3.6 | PAF | NSP derivative | 3 min | Several hours | *Nat. Commun.* **5**, 3588 (2014) |
| $Zn_2$(DBTD)(TNDS) | MOF | NSP derivative | ~ 15 s | 3 min | *J. Am. Chem. Soc.* **141**, 5350 (2019) |
| CD-DP | MON (Metal-organic nanosheet) | Organic building block | 3-5 min | 10 min | *Chem* **4**, 1059 (2018) |
| PSZ-1 | MOF | Diarylethenes derivative | 10 min | - | *J. Am. Chem. Soc.* **139**, 13280 (2017) |
| PC-PCN | MOF | Dithienylethene | 10 min | 1 h | *Angew. Chem. Int. Ed.* **54**, 430 (2015) |
| BSP/JUC-120 | MOF film | NSP | 20 min | 2.4 h | *J. Mater. Chem.* **22**, 25019 (2012) |
| SP-h | Self-Assembled Monolayers (SAMs) | NSP derivative | 20 min | 12 h | *Adv. Mater.* **31**, 1807831 (2019) |
| PSF-2 | POP (Porous organic polymer) | Organic building block | 30 min | 15 h | *Nat Chem.* **12**, 595 (2020) |
| ($[Zn_2$(bdc)$_2$(LO)]n) | MOF | Dithienylethene | 30 min | - | *Nat. Commun.* **8**, 100 (2017) |
| PCN-123 | MOF | Azo moiety | 1-2 h | 1 d | *J. Am. Chem. Soc.* **134**, 99 (2012) |

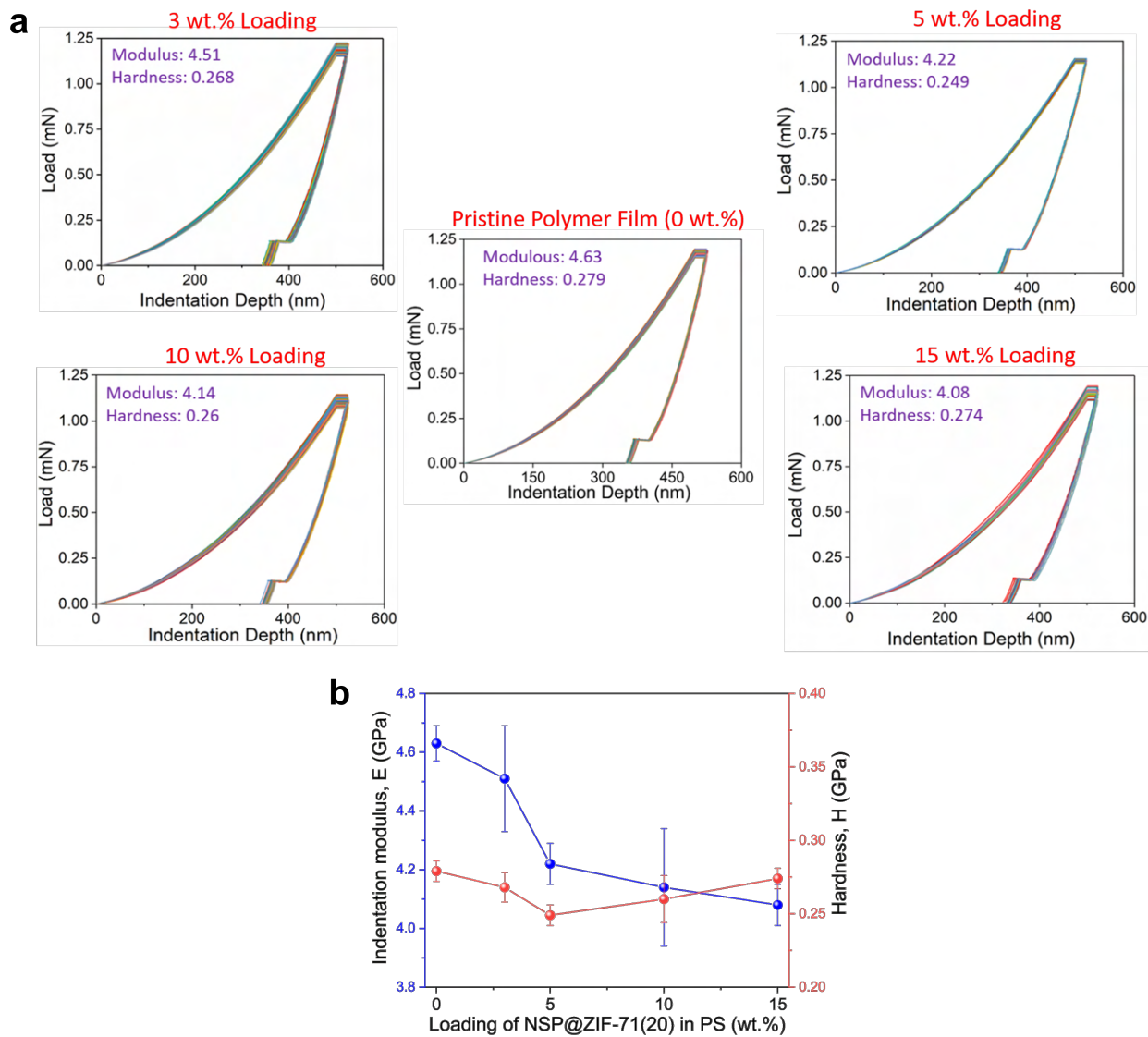

**Fig. S43. (a)** Load-depth curves produced by nanoindentation tests on films made of polystyrene (PS) with various NSP@ZIF-71(20) loadings. The thickness of the films remained in the range of 20 to 25 μm in each case. A total of 16 tests for each sample was performed and the load-depth graphs demonstrated a high degree of repeatability. The Young's modulus ($E$) slightly drops when the NSP@ZIF-71(20) nanocomposite loading percentage in the PS polymer increases, but the hardness ($H$) of the films remains almost constant. **(b)** Average modulus and hardness graph for the different loading of NSP@ZIF-71(20) into the PS films.

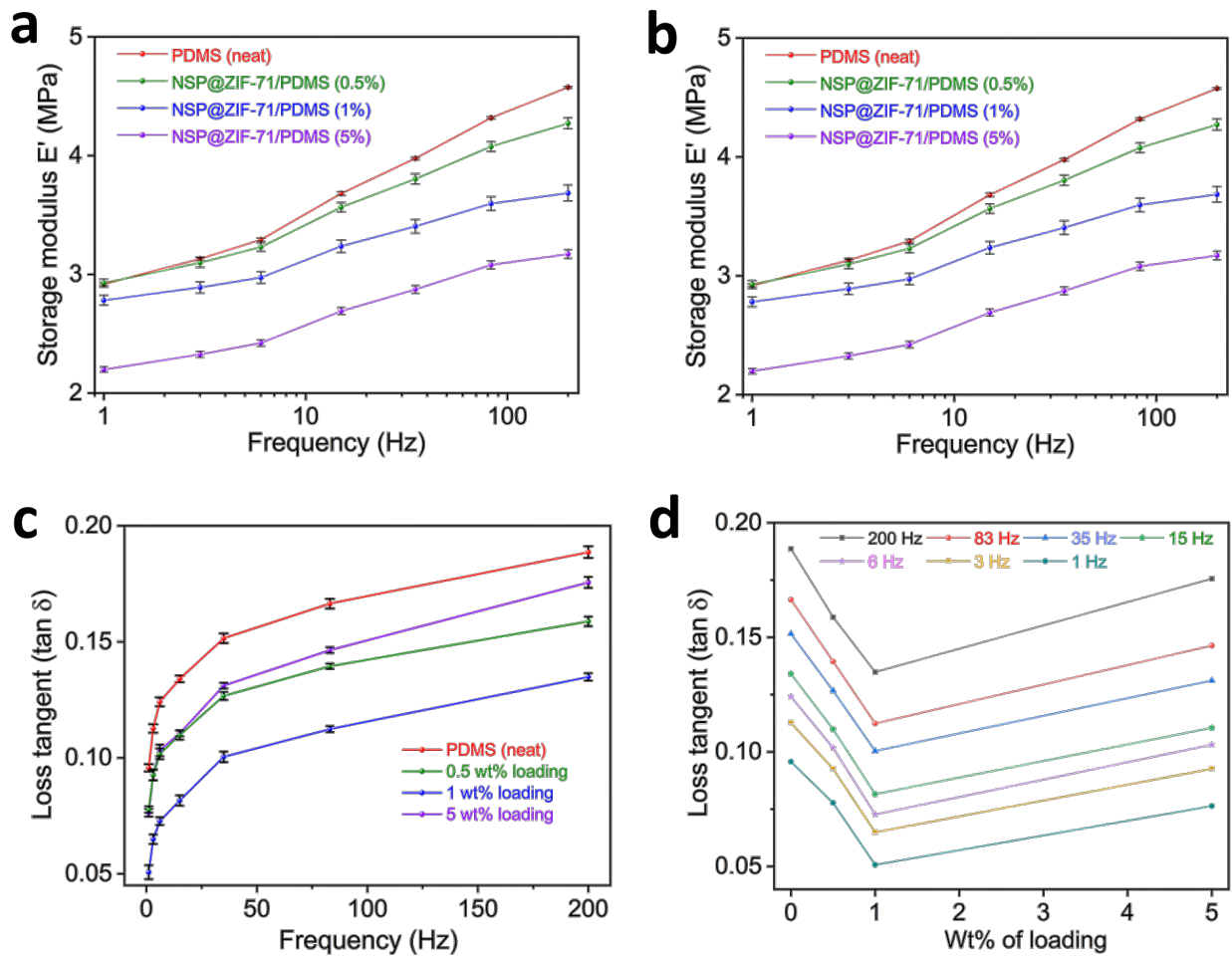

**Fig. S44.** Nanoscale dynamic mechanical analysis (nanoDMA) curves measured using a flat punch indenter probe (Ø 50 µm) for the neat PDMS films and different wt.% loading of NSP@ZIF-71(20) into PDMS films as a function of frequency. **(a)** Storage modulus $E'$ and **(b)** loss modulus $E''$. Both storage and loss modulus decrease upon loading of NSP@ZIF-71(20) filler. The ratio of loss modulus ($E''$) and storage modulus ($E'$) is termed the loss tangent (tan $\delta$ = $E''/E'$). Loss tangent plotted against **(c)** frequency and **(d)** wt.% loading of NSP@ZIF-71(20) composite.

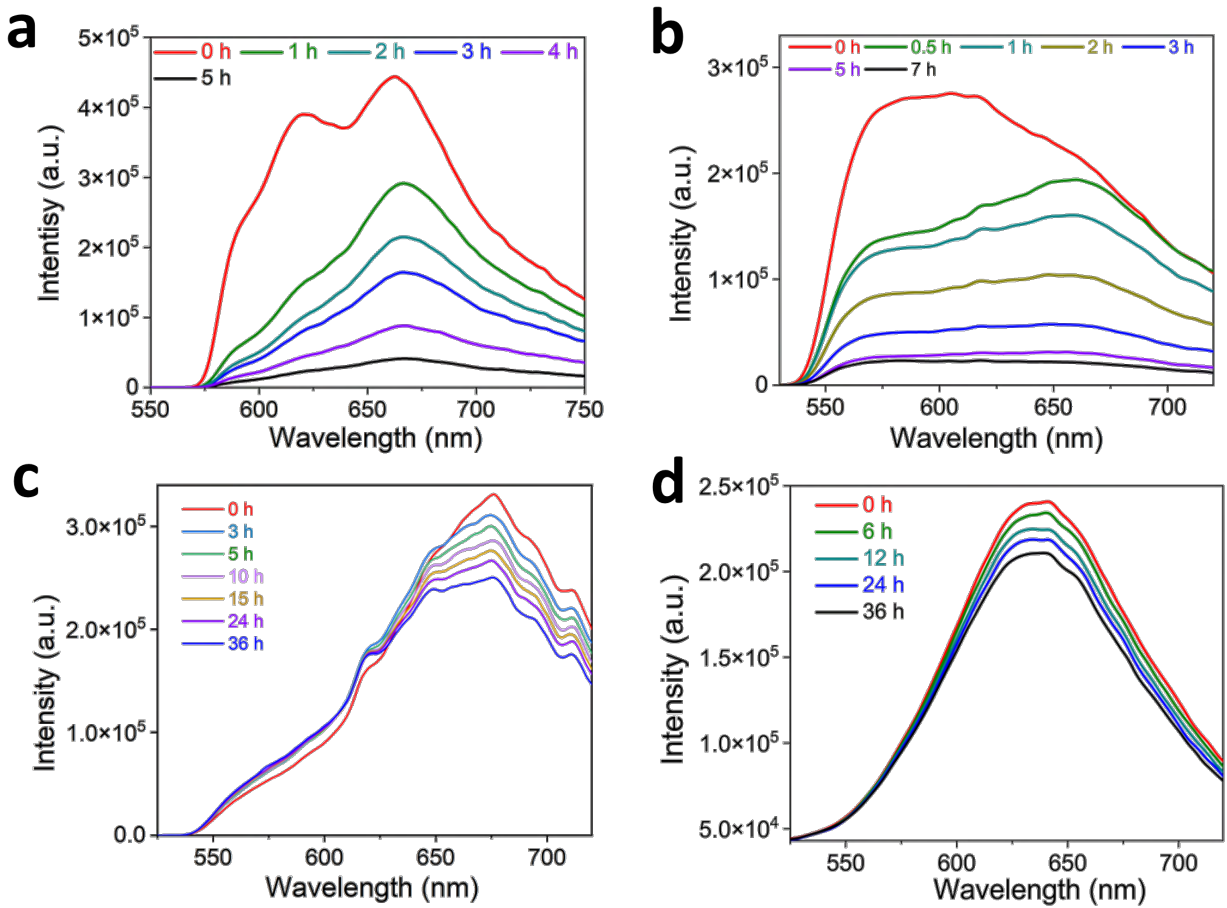

**Fig. S45.** Photostability tests for **(a)** the pristine NSP powder, **(b)** physical mixing of the NSP and ZIF-71 powder, and **(c)** NSP@ZIF-71(20) powder. **(d)** NSP@ZIF-71(20)/PDMS film.

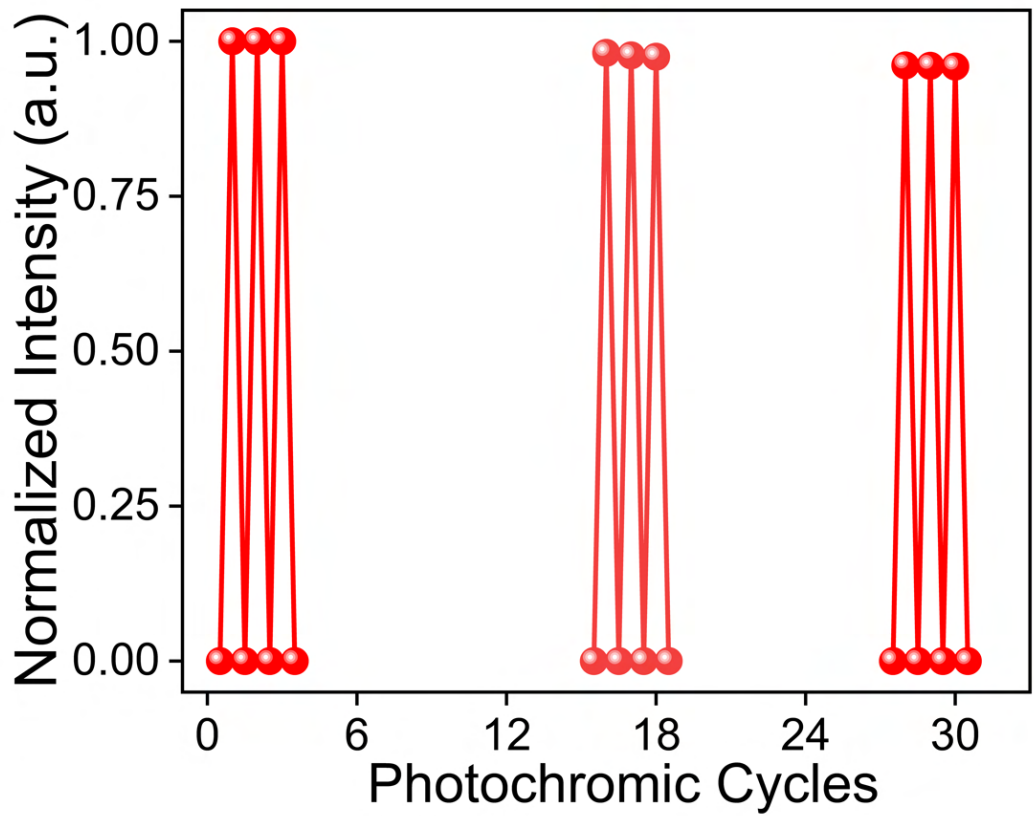

**Fig. S46.** Photochromic cycles showing the photo-fatigue resistance of the powder forms of NSP@ZIF-71(20).

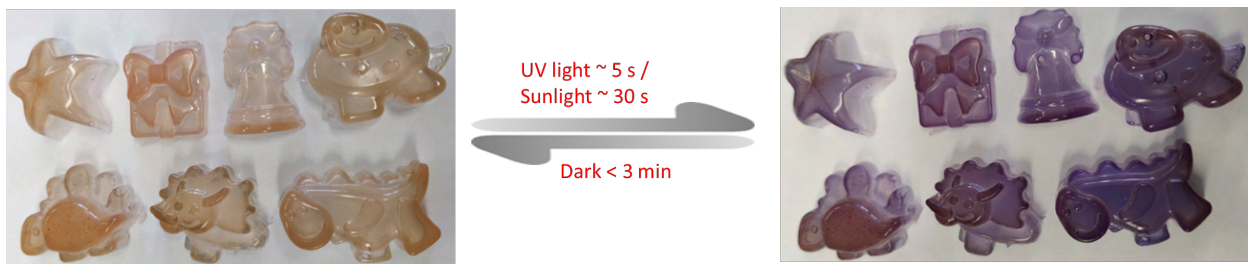

**Fig. S47.** Photographs of various shaped sculptures fabricated from 1 wt.% loading of NSP@ZIF-71(20) in the PS polymer matrix. The photographs show the visual color switching upon exposure to 365 nm UV light or sunlight.

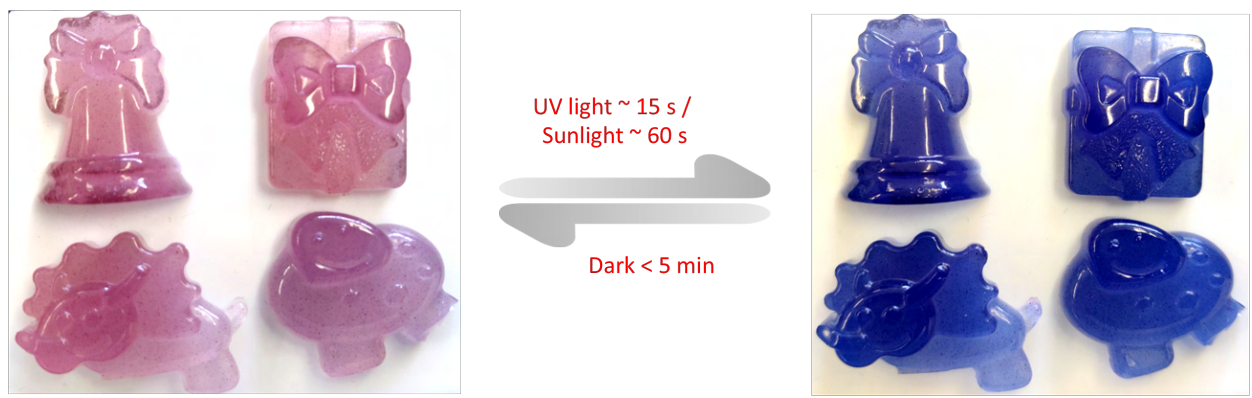

**Fig. S48.** Photographs of various shaped sculptures fabricated from 0.5 wt.% loading of 8NSP@ZIF-71(20) in PDMS polymer matrix. The photographs show the visual color switching upon exposure to 365 nm UV light or sunlight.

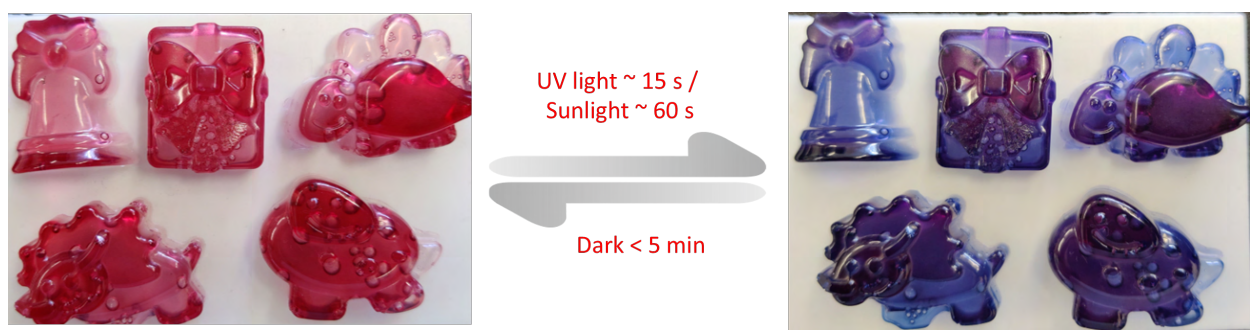

**Fig. S49.** Photographs of various shaped sculptures fabricated from 0.5 wt.% loading of 8NSP@ZIF-71(20) in PS polymer matrix. The photographs show the visual color switching upon exposure to 365 nm UV light or sunlight.

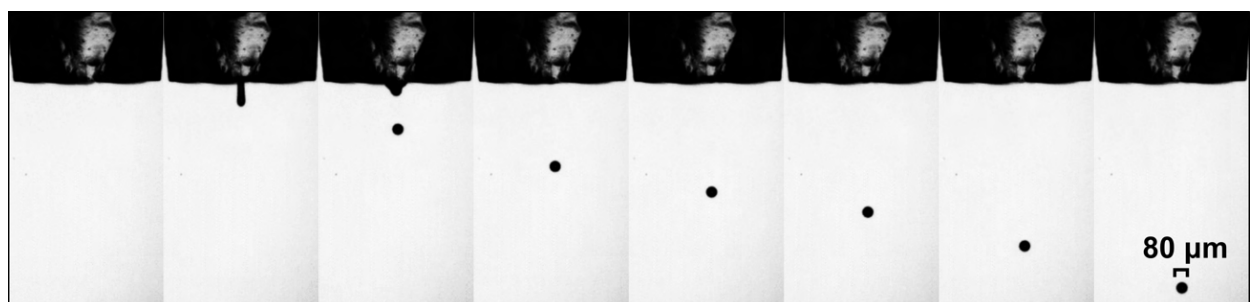

**Fig. S50.** High-speed shadowgraphy images of the printing process of an exemplar NSP@ZIF-71(20) nanocomposite solution. The size of the nozzle orifice was 80 μm. The time interval between two adjacent frames is 48 μs.

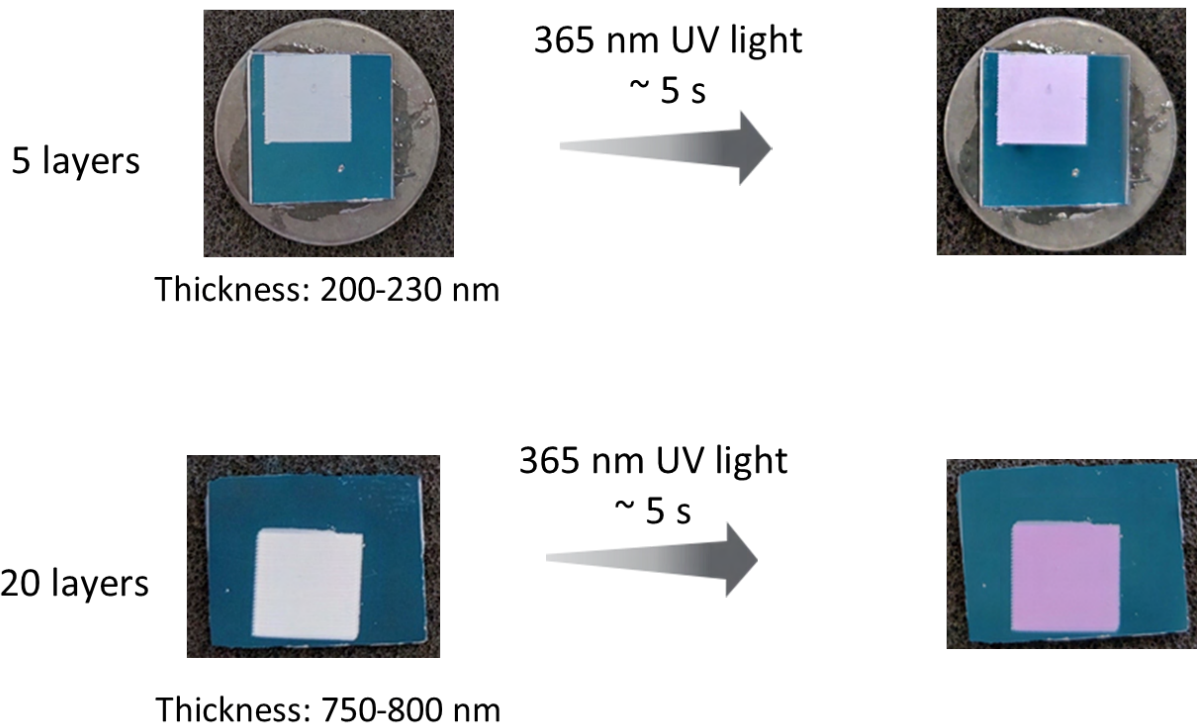

**Fig. S51.** Photographs of inkjet-printed samples. Results for either five or twenty layers printed on a silicon wafer are shown. An IPA solvent was used to prepare the samples.

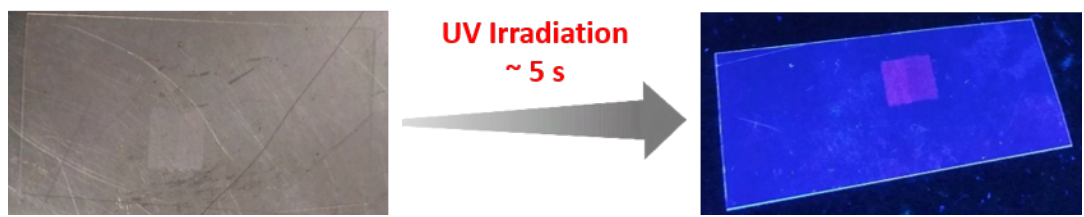

**Fig. S52.** Photographs of inkjet printing of a neat NSP@ZIF-71(20) composite on a flexible acetate sheet. The color of the inkjet-printed samples in the presence of 365 nm UV light. An IPA solvent was used to prepare the sample.

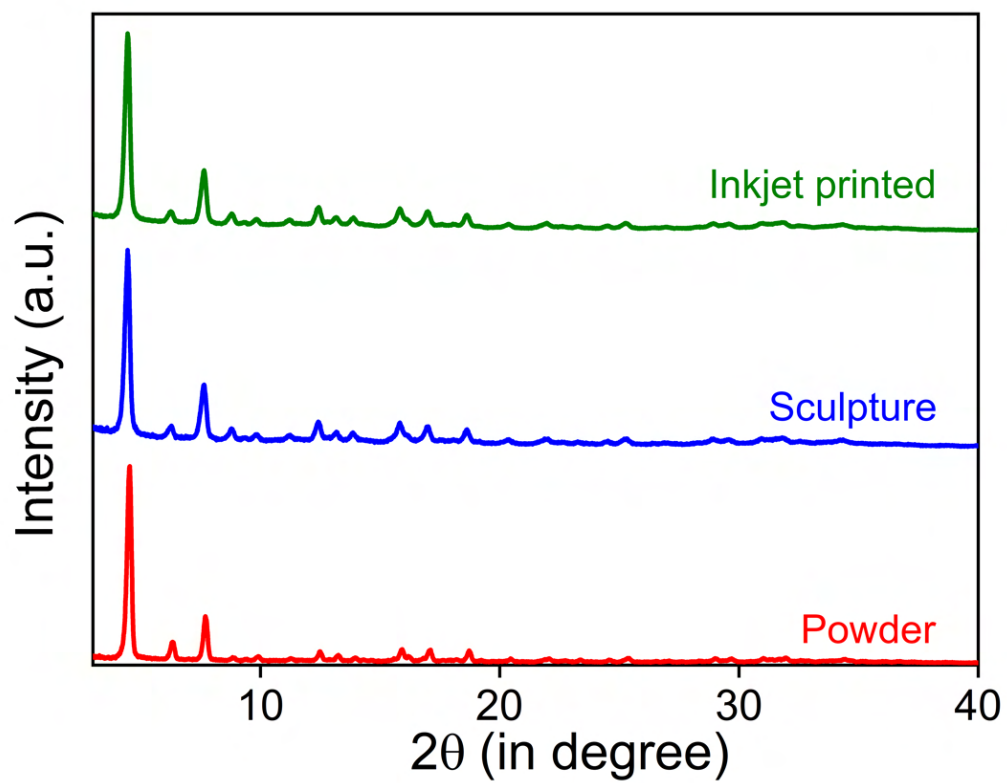

**Fig. S53.** PXRD patterns of NSP@ZIF-71(20) in powder form in PS sculptures (10 wt.% loading) and for an inkjet-printed film (60 layers) on a glass substrate.

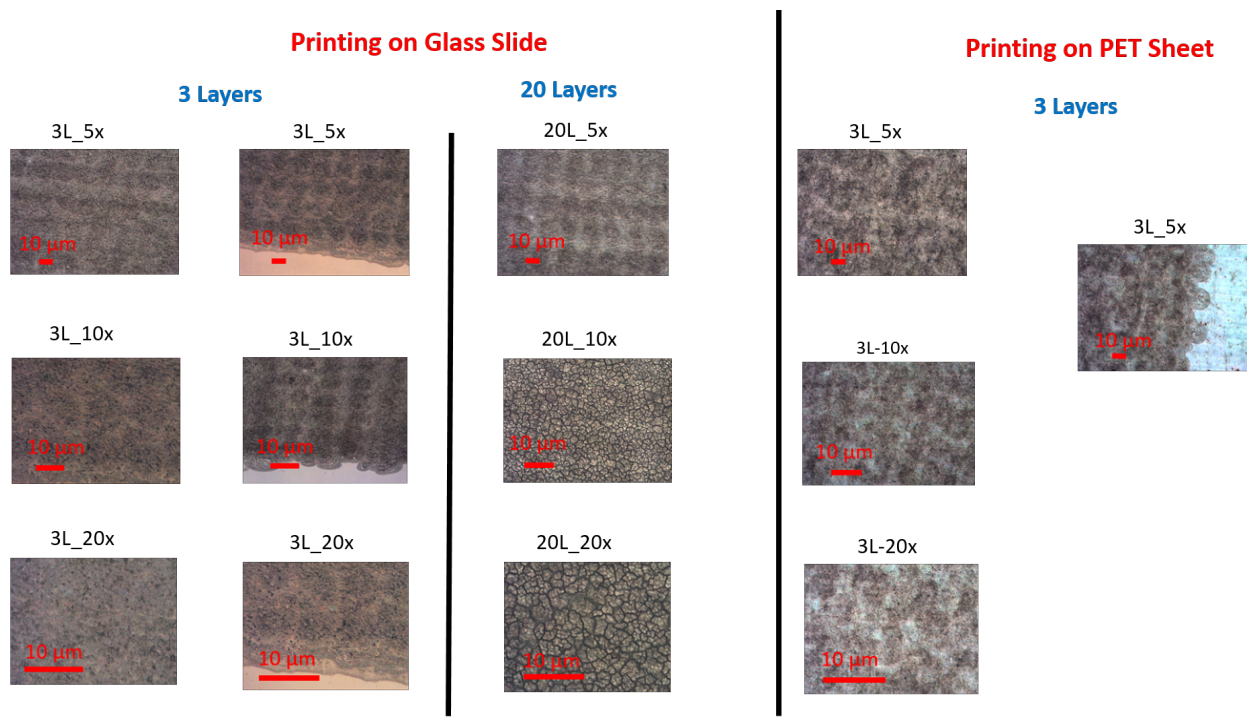

**Fig. S54.** Optical micrographs of inkjet-printed samples on glass slide and PTFE film. Different magnifications demonstrate an enlarged view of the printed films.

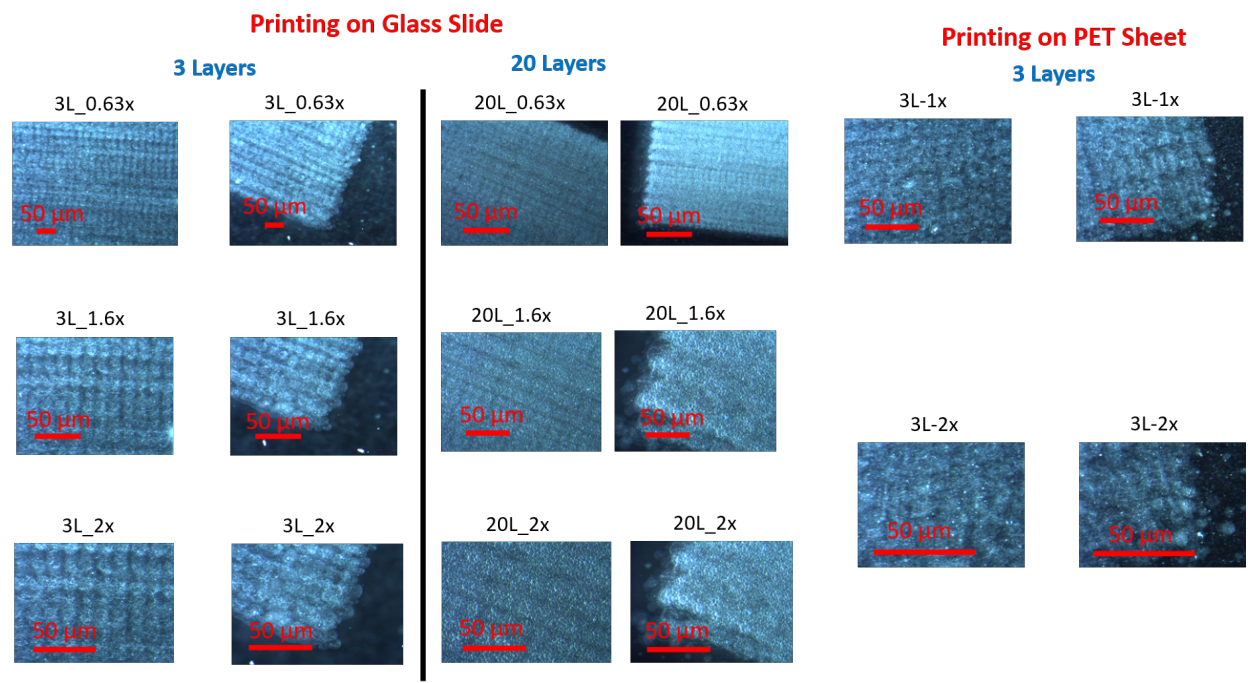

**Fig. S55.** Photographs from a stereo optical microscope of the inkjet-printed samples on a glass slide or PTFE film. Different magnifications demonstrate an enlarged view of the printed samples.

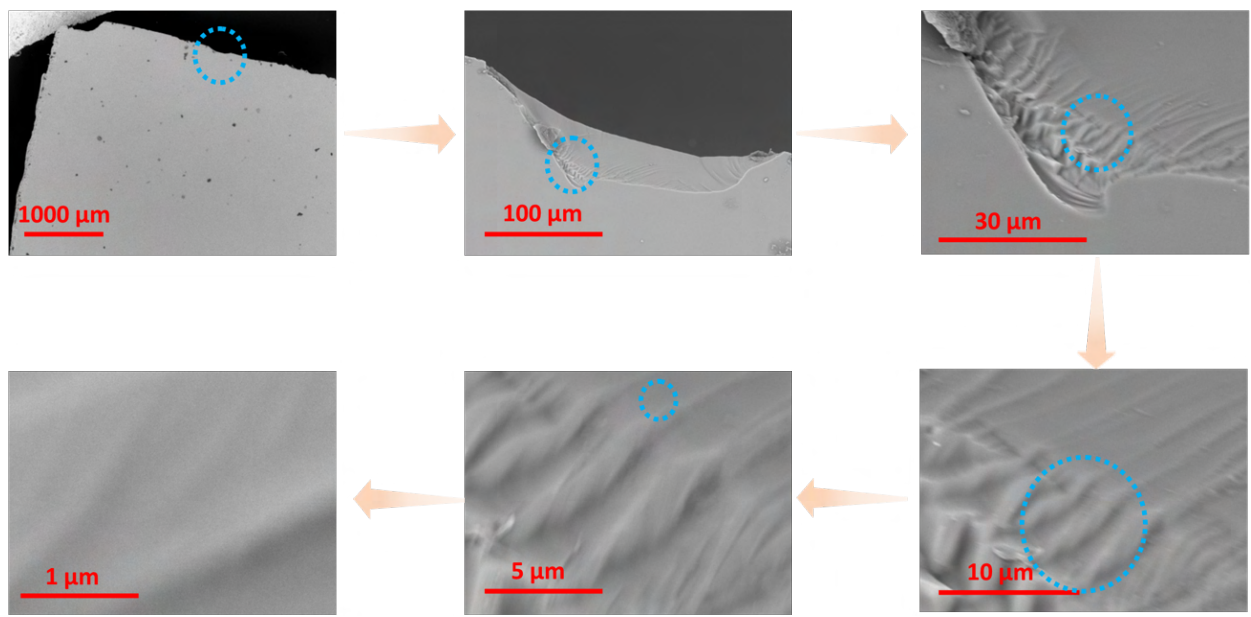

**Fig. S56.** SEM images of 3 layers of the inkjet-printed samples of NSP@ZIF-71(20) on a glass slide. The highlighted part is shown in the subsequent figures.

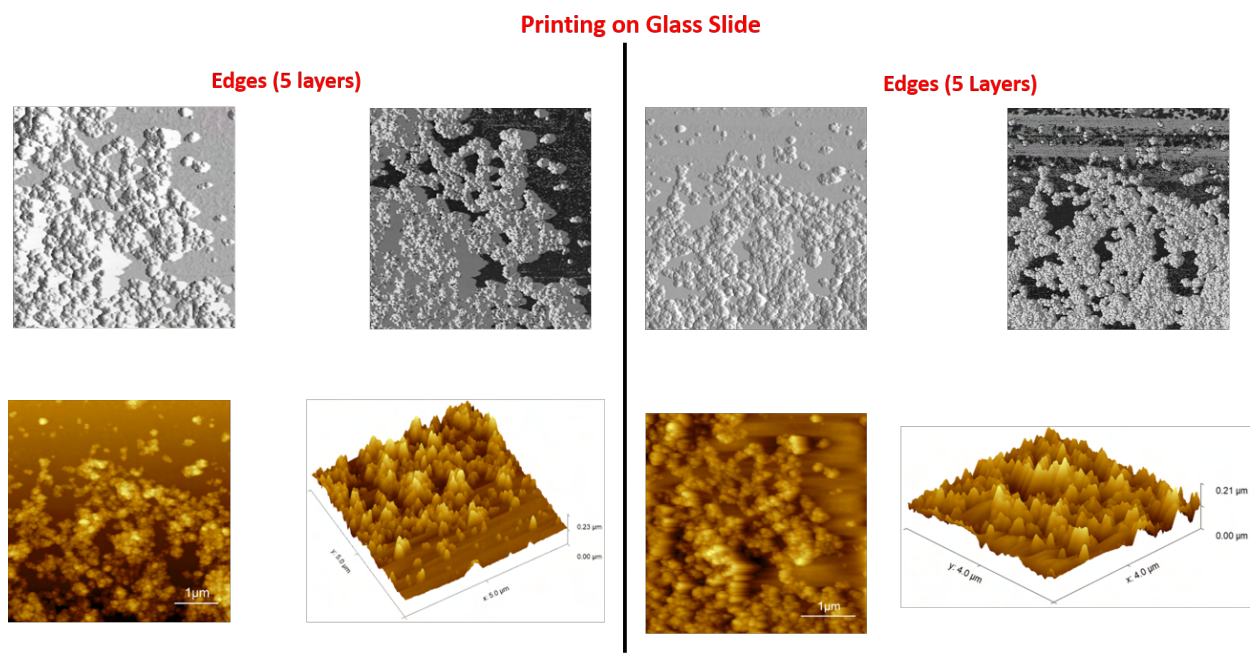

**Fig. S57.** AFM images of different film edges of the inkjet-printed NSP@ZIF-71(20) on a glass slide. The overall thickness of the 5-layer film sample matches the individual crystal size of the composites.

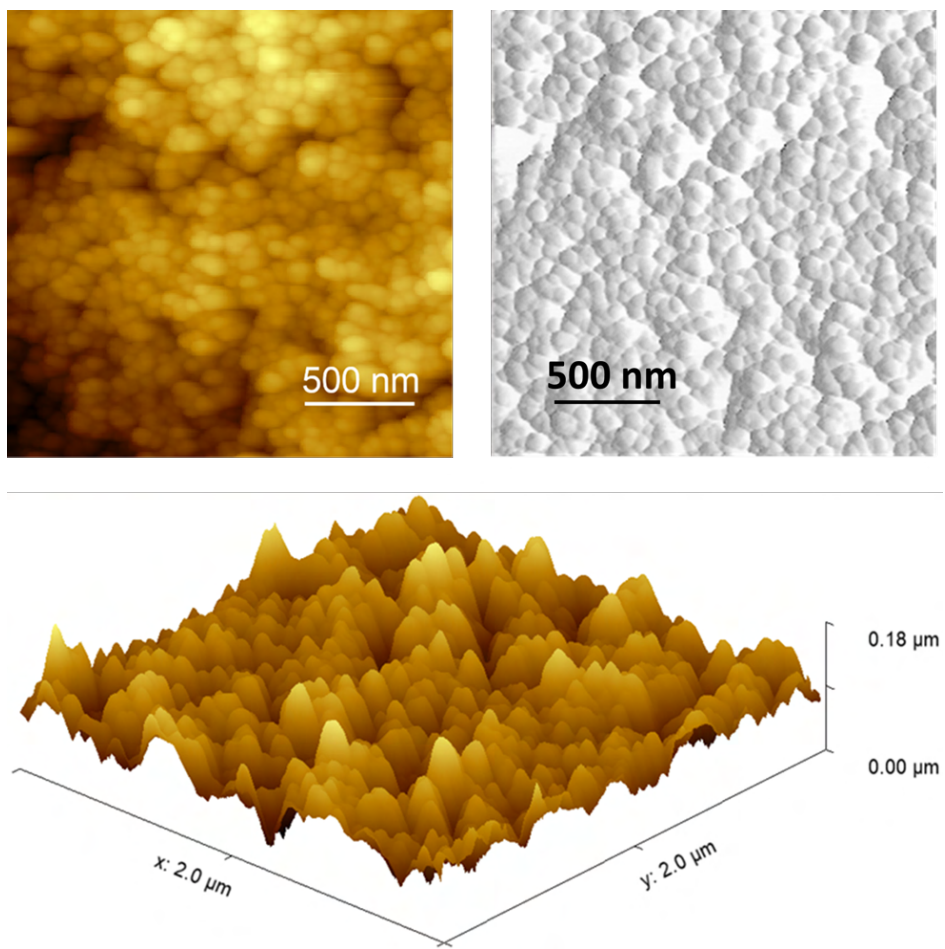

**Fig. S58.** AFM images of 5 layers of the inkjet-printed sample (NSP@ZIF-71(20)) on a glass slide.

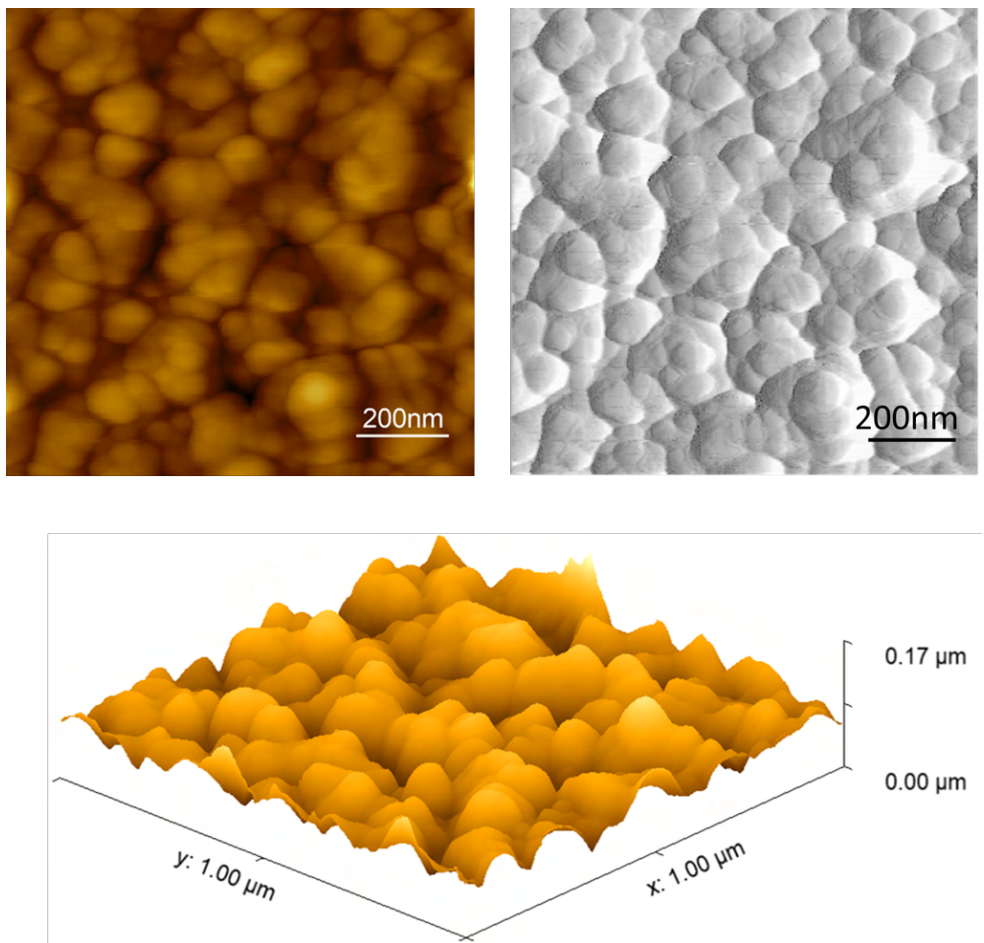

**Fig. S59.** AFM images of 5 layers of the inkjet-printed sample (NSP@ZIF-71(20)) on a glass slide. The AFM images shows the NSP@ZIF-71(20) crystals are nicely coated on the glass slide.

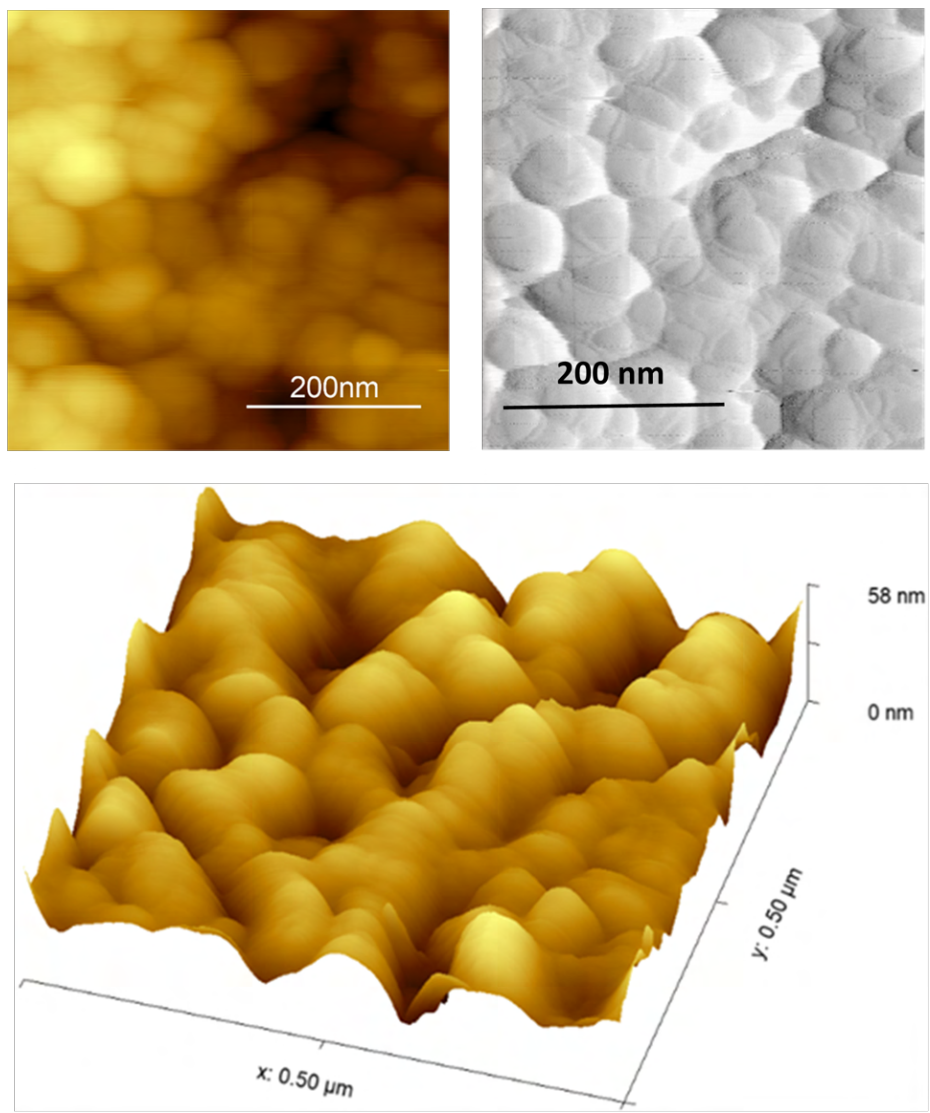

**Fig. S60.** AFM images of 1 layer of the inkjet-printed sample of NSP@ZIF-71(20) on a glass slide. The overall thickness of the 1-layer film matches the size of the individual nanocrystals of the composites.

| Under Ambient light | Under UV light (365 nm) | Under sunlight |
|---|---|---|
| 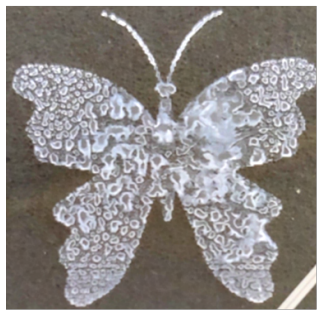 | 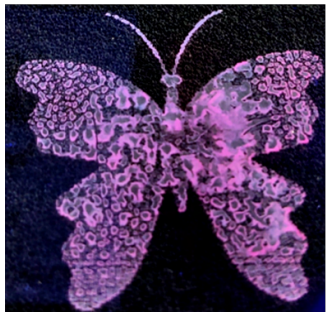 | 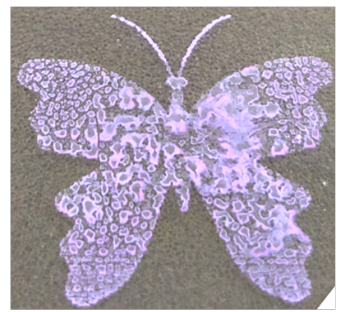 |

**Fig. S61.** Using inkjet printing technology, NSP@ZIF-71(20) was printed to form two layers on a glass substrate to demonstrate the color-changing phenomenon in the presence of various ambient lighting conditions. An IPA solvent was used to prepare the sample.

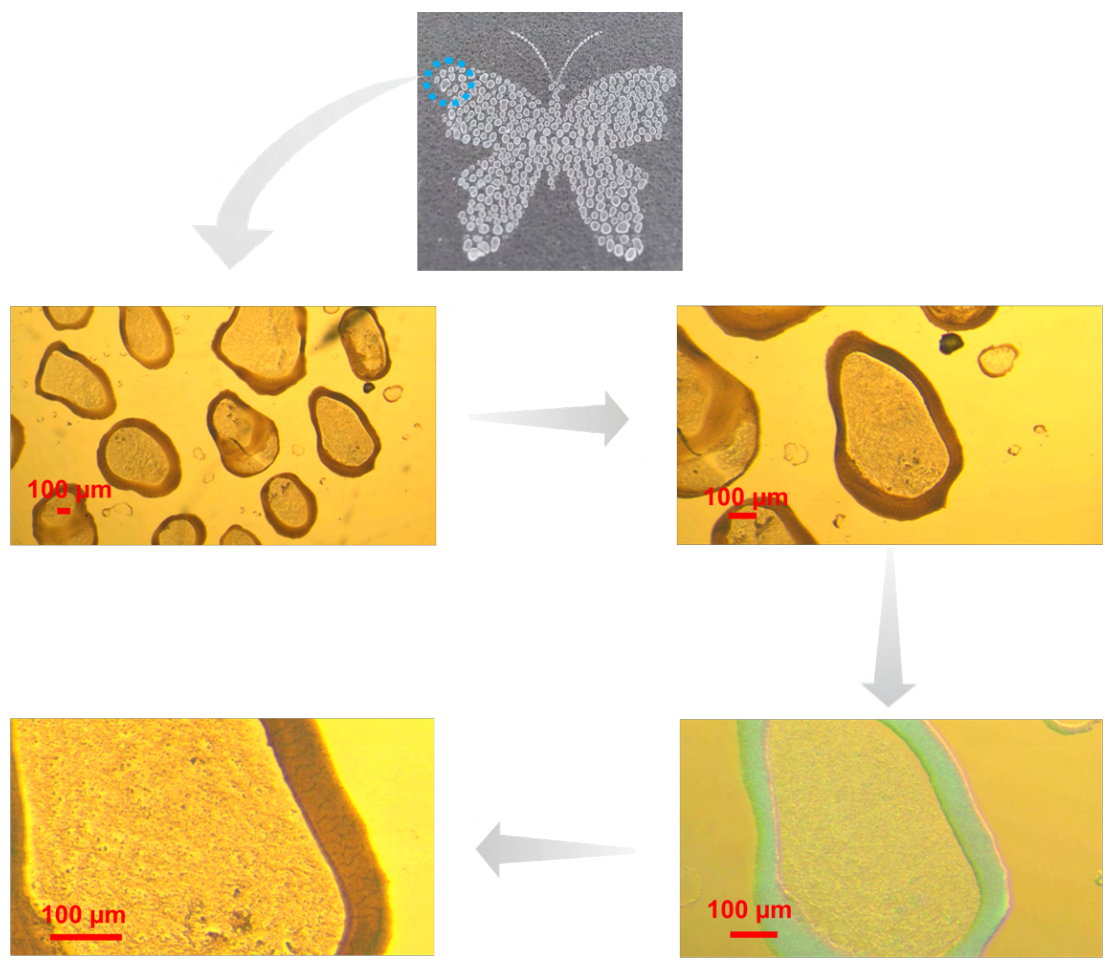

**Fig. S62.** Photographs taken on an optical microscope of a butterfly image printed on a large glass substrate using the inkjet printing method, where the NSP@ZIF-71(20) sample was printed on the glass substrate in two successive layers. An IPA solvent was used to prepare the sample.

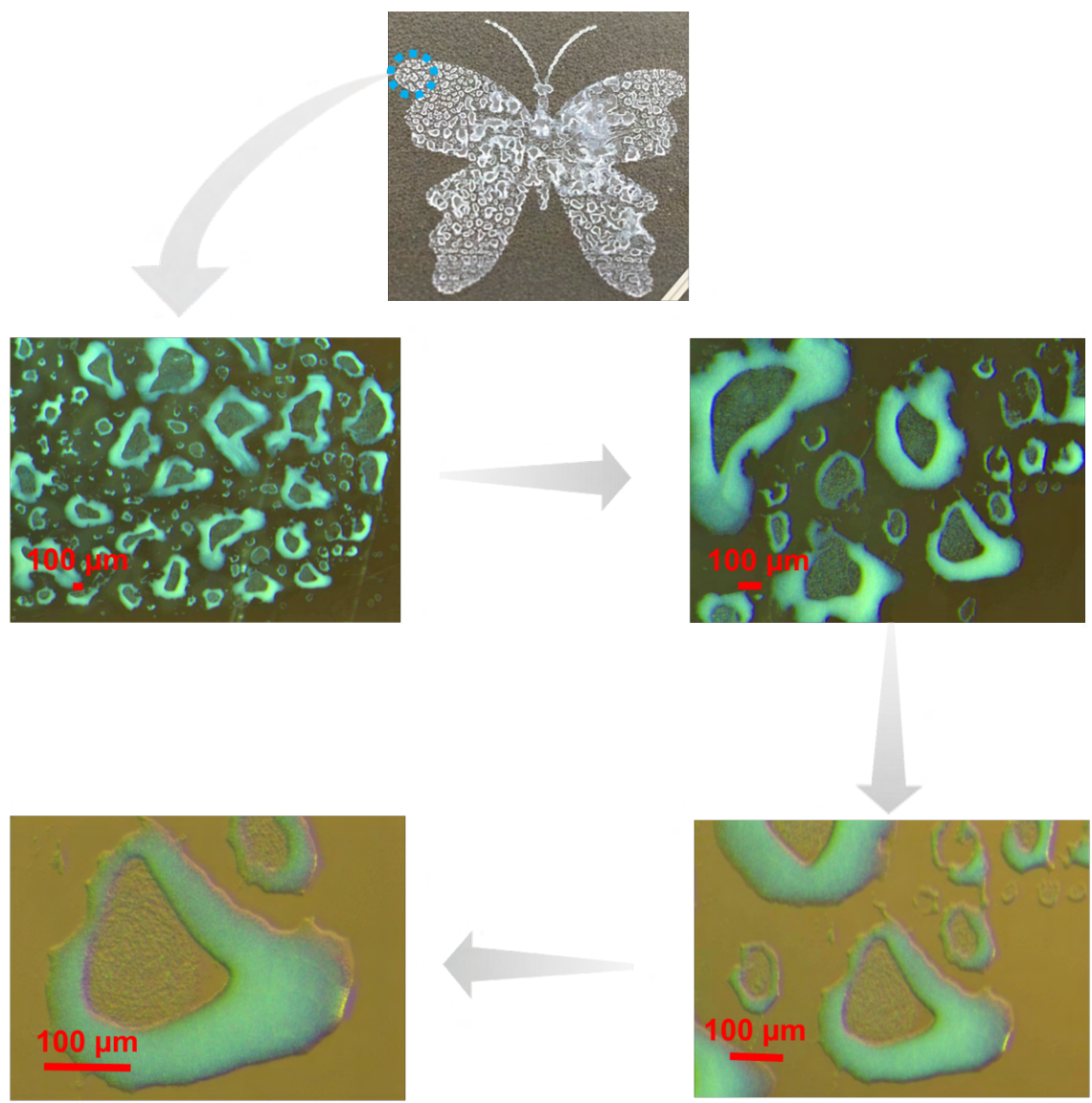

**Fig. S63.** Photographs taken on an optical microscope of a butterfly image printed on a large glass substrate using the inkjet printing method, where the NSP@ZIF-71(20) sample was printed on a glass substrate in five successive layers. IPA solvent was used to prepare the sample.

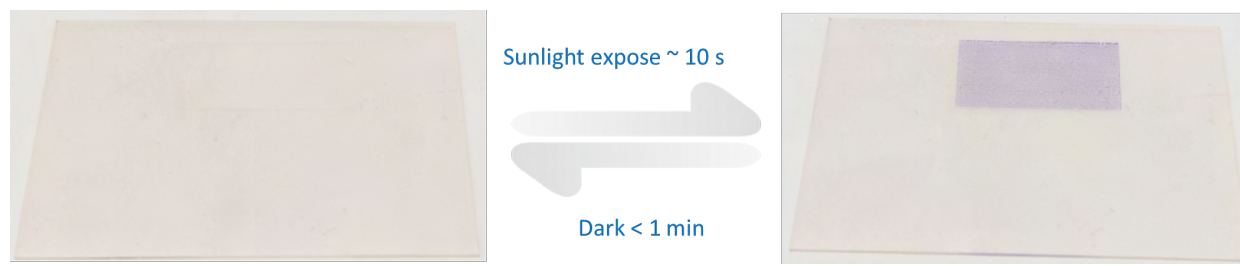

**Fig. S64.** Using inkjet printing technology, NSP@ZIF-71(20) was printed as two layers on an indium tin oxide (ITO)-coated glass substrate to demonstrate the color-changing phenomenon in the presence of sunlight. A DMF solvent was used to prepare the sample. The printed film sample was invisible to the naked eye under ambient room conditions but became visible (against a white background) after exposure to sunlight for ~ 10 s.

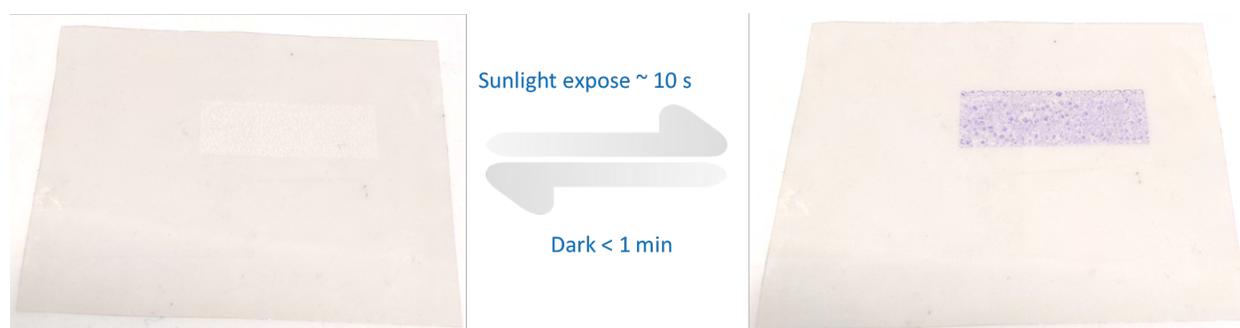

**Fig. S65.** Using inkjet printing technology, NSP@ZIF-71(20) was printed as two successive layers on an ITO-coated flexible acetate sheet to demonstrate the color-changing phenomenon in the presence of natural sunlight. A DMF solvent was used to prepare the sample. Initially, the printed film sample was invisible to the naked eye under ambient room conditions, but it became visible (against a white background) after exposure to sunlight for ~ 10 s.

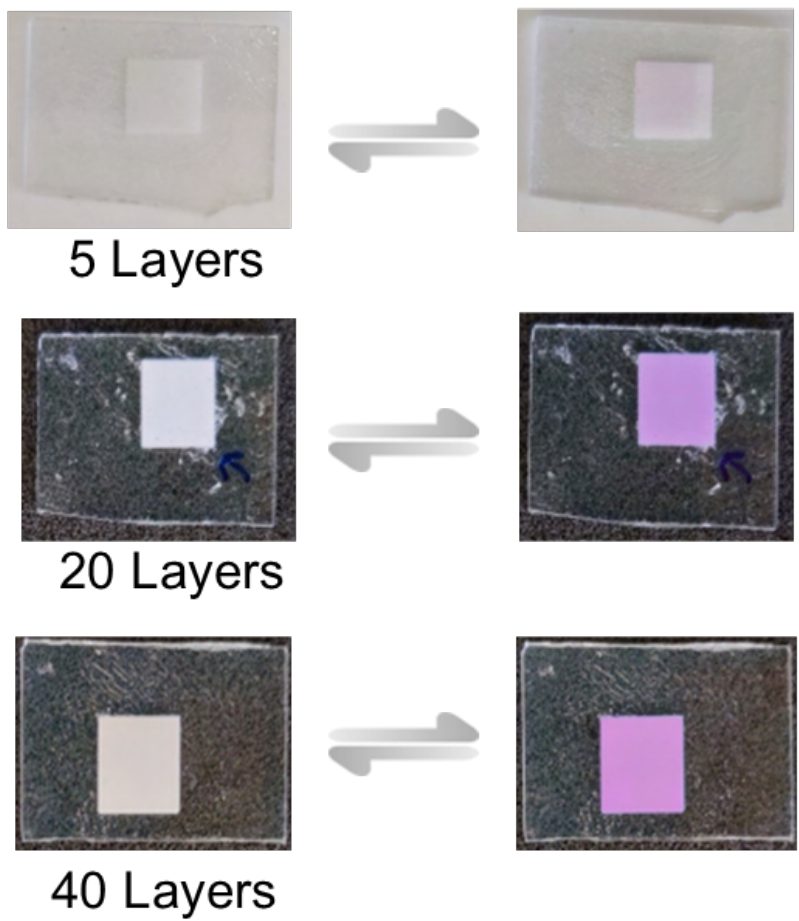

**Fig. S66.** Photographs showing color change after exposure to UV light (365 nm) for a few seconds on different layers of NSP@ZIF-71(20) films inkjet-printed onto a glass substrate. An IPA solvent was used to prepare the sample.

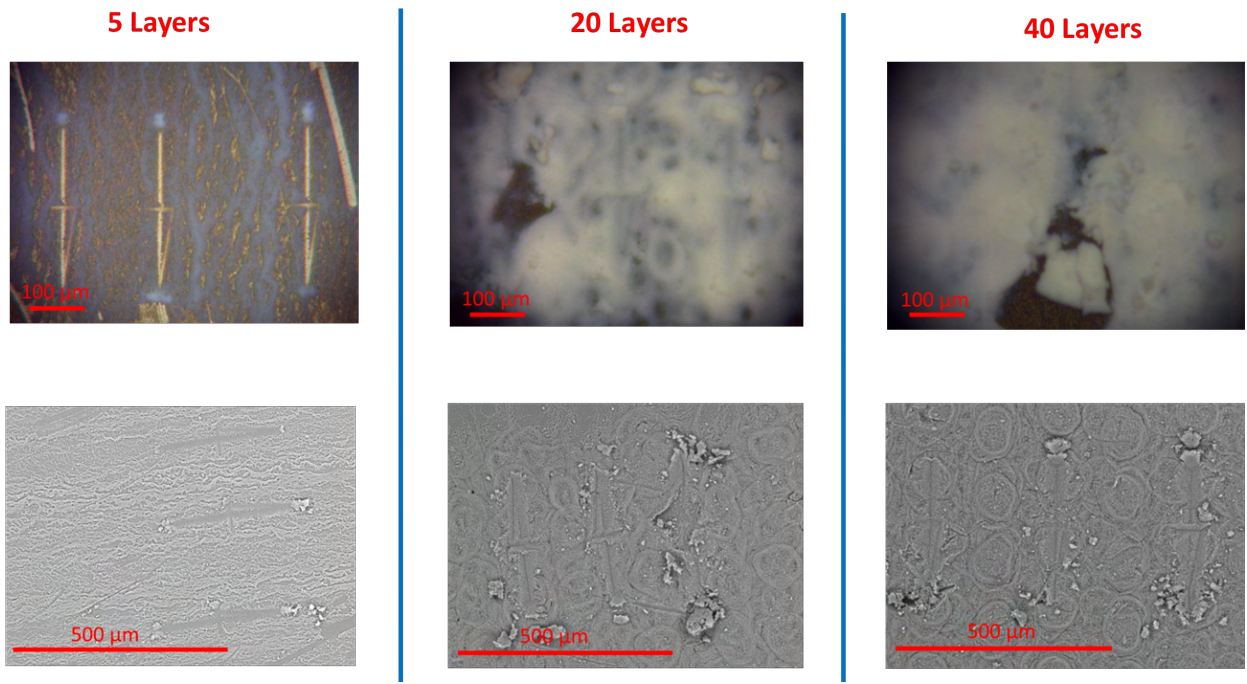

**Fig. S67.** Optical images (top) and SEM micrographs (down) of different inkjet-printed layers after scratch testing.

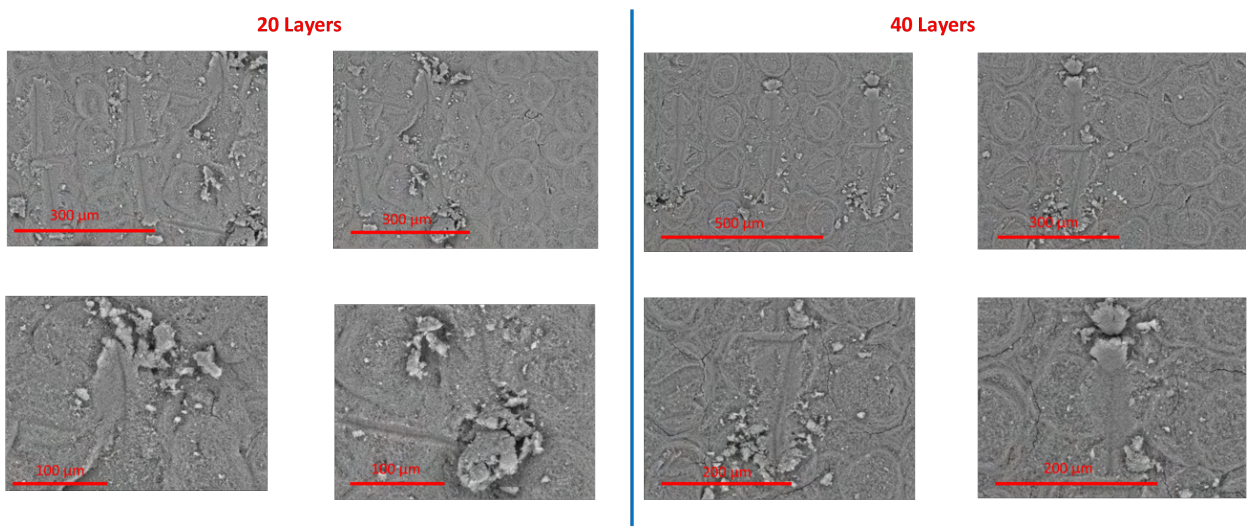

**Fig. S68.** SEM micrographs of different inkjet-printed layers after scratch tests. The build-up of materials at the end of the scratch test indicates the higher cohesion interaction between nanocrystals in the film.

**Movie Clips**

**Movie S1.** Video showing the coloration of the prototype photochromic "window", where the nominal thickness of the window was *ca.* 350 µm and 1 wt.% NSP@ZIF-71(20) was loaded into the PS polymer.

**Movie S2.** Video showing the prototype self-erasing rewritable device where UV exposure was carried out with a 365-nm UV collimated laser source.